%% file: main.tex
\documentclass[sigconf]{acmart}

\usepackage{algorithmic}
\usepackage{graphicx}
\usepackage{textcomp}
\usepackage{xcolor}
\usepackage{xspace}
\usepackage{subfig}
\usepackage{soul}

\usepackage{multirow}
\usepackage{tikz}

\AtBeginDocument{%
  \providecommand\BibTeX{{%
    \normalfont B\kern-0.5em{\scshape i\kern-0.25em b}\kern-0.8em\TeX}}}

\setcopyright{acmlicensed}
\copyrightyear{2023} 
\acmYear{2023} 
\acmDOI{10.1145/3579371.3589064}

\acmConference[ISCA '23]{Proceedings of the 50th Annual International Symposium on Computer Architecture}{June 17--21, 2023}{Orlando, FL, USA}
\acmBooktitle{Proceedings of the 50th Annual International Symposium on Computer Architecture (ISCA '23), June 17--21, 2023, Orlando, FL, USA}
\acmPrice{15.00}
\acmISBN{979-8-4007-0095-8/23/06} \acmDOI{10.1145/3579371.3589064}

\graphicspath{{figs/}}

\begin{document}

\title{\textsc{CamJ}: Enabling System-Level Energy Modeling and Architectural Exploration for In-Sensor Visual Computing}

\author{Tianrui Ma}
\authornote{Both authors contributed equally to the paper.}
\affiliation{
  \institution{Washington University in St. Louis}
  \city{St. Louis}
  \state{MO}
  \country{USA}
}
\email{tianrui.ma@wustl.edu}

\author{Yu Feng}
\authornotemark[1]
\affiliation{
  \institution{University of Rochester}
  \city{Rochester}
  \state{NY}
  \country{USA}
}
\email{yfeng28@ur.rochester.edu}

\author{Xuan Zhang}
\authornote{Both authors are corresponding authors.}
\affiliation{
  \institution{Washington University in St. Louis}
  \city{St. Louis}
  \state{MO}
  \country{USA}
}
\email{xuan.zhang@wustl.edu}

\author{Yuhao Zhu}
\authornotemark[2]
\affiliation{
  \institution{University of Rochester}
  \city{Rochester}
  \state{NY}
  \country{USA}
}
\email{yzhu@rochester.edu}



\input{macro}

\input{abstract}



\maketitle

\setlength{\textfloatsep}{6pt}
\setlength{\floatsep}{6pt}


\input{intro}

\input{background}

\input{framework}

\input{modeling}
\input{eval}

\input{use_case}

\input{related}
\input{conc}


\bibliographystyle{ACM-Reference-Format}
\balance
\interlinepenalty=100000
\bibliography{main}

\end{document}

%% file: macro.tex

\newcommand{\website}[1]{{\tt #1}}
\newcommand{\program}[1]{{\tt #1}}
\newcommand{\benchmark}[1]{{\it #1}}
\newcommand{\fixme}[1]{{\textcolor{red}{\textit{#1}}}}
  \newcommand{\tianrui}[1]{{\color{magenta} [Tianrui: #1]}}

\newcommand*\circled[2]{\protect\tikz[baseline=(char.base)]{
            \protect\node[shape=circle,fill=black,inner sep=1pt] (char) {\textcolor{#1}{{\footnotesize #2}}};}}

\ifx\figurename\undefined \def\figurename{Figure}\fi
\renewcommand{\figurename}{Fig.}
\renewcommand{\paragraph}[1]{\textbf{#1} }
\newcommand{\figline}{{\vspace*{.05in}\hline}}

\newcommand{\Sect}[1]{Sec.~\ref{#1}}
\newcommand{\Fig}[1]{Fig.~\ref{#1}}
\newcommand{\Tbl}[1]{Tbl.~\ref{#1}}
\newcommand{\Equ}[1]{Equ.~\ref{#1}}
\newcommand{\Apx}[1]{Apdx.~\ref{#1}}
\newcommand{\Alg}[1]{Algo.~\ref{#1}}

\newcommand{\specialcell}[2][c]{\begin{tabular}[#1]{@{}c@{}}#2\end{tabular}}
\newcommand{\note}[1]{\textcolor{red}{#1}}

\newcommand{\acomp}{\textsc{A-Component}\xspace}
\newcommand{\acell}{\textsc{A-Cell}\xspace}
\newcommand{\proj}{\textsc{CamJ}\xspace}
\newcommand{\mode}[1]{\underline{\textsc{#1}}\xspace}
\newcommand{\sys}[1]{\underline{\textsc{#1}}}
\newcommand{\PBox}[1]{\vspace*{.05cm}\noindent\fbox{\parbox{.98\columnwidth}{\vspace*{.05cm}{#1}}}\vspace*{.05cm}}

\newcommand{\no}[1]{#1}
\renewcommand{\no}[1]{}
\renewcommand{\hl}[1]{#1}
\newcommand{\RNum}[1]{\uppercase\expandafter{\romannumeral #1\relax}}

\def\cA{{\mathcal{A}}}
\def\cF{{\mathcal{F}}}
\def\cN{{\mathcal{N}}}
\def\bh{{\mathbf{h}}}
\def\bp{{\mathbf{p}}}


%% file: abstract.tex
\begin{abstract}
CMOS Image Sensors (CIS) are fundamental to emerging visual computing applications.
While conventional CIS are purely imaging devices for capturing images, increasingly CIS integrate processing capabilities such as Deep Neural Network (DNN).
Computational CIS expand the architecture design space, but to date no comprehensive energy model exists.
This paper proposes \proj, a detailed energy modeling framework that provides a component-level energy breakdown for computational CIS and is validated against nine recent CIS chips.
We use \proj to demonstrate three use-cases that explore architectural trade-offs including computing in vs. off CIS, 2D vs. 3D-stacked CIS design, and analog vs. digital processing inside CIS. The code of \proj is available at: \color{blue}{\url{https://github.com/horizon-research/CamJ}}.
\end{abstract}

%% file: intro.tex
\section{Introduction}
\label{sec:intro}

Visual computing applications on the horizon such as autonomous machines, computational photography, and space exploration all fundamentally rely on image sensing.
While conventional CMOS Image Sensors (CIS) are responsible only for ``imaging'', i.e., capturing pixels, CIS increasingly integrate \textit{computation} capabilities, ranging from signal preprocessing ~\cite{cheng2008ivisual} to Deep Neural Networks (DNN)~\cite{bong201714, hirata20217} and spanning both the analog and digital domain.
For instance,
a Nikon CIS~\cite{hirata20217} integrates an image processor for per-tile exposure control;
the Sony IMX 500 CIS~\cite{eki20219} integrates a DNN accelerator with the pixel array for edge visual processing.

Computational CIS expand the traditional design space for architects and provide an exciting playground for exploring a diverse range of trade-offs.
For instance, in-CIS processing consumes large volumes of pixel data \textit{in-situ} and reduces the data transmission overhead;
computing inside a sensor, however, is inefficient because CIS tend to be fabricated using older process nodes compared to standard CMOS nodes (limited by the photon sensing sensitivity~\cite{theuwissen20211}), which offsets the gains from reducing the communication cost.
Designers also face a myriad of choices when designing the in-sensor architecture.
For instance, while 3D stacking improves energy efficiency by allowing for hybrid process integration~\cite{xie2015stacking},
it could also increase the power density and, thus, thermal-induced noise, requiring more processing downstream.

This paper presents \proj, a first-of-its-kind energy modeling framework that empowers designers to navigate the large algorithm-hardware co-design space.
\proj provides a component-level energy estimation under a frame-per-second (FPS) target.
To that end, \proj models the interplay across main structures of a computational CIS pipeline:
pixel sensing $\rightarrow$ analog processing $\rightarrow$ digital processing.
Thus, \proj enables end-to-end optimizations of the CIS architecture from photon ingestion to semantic results.

\paragraph{Design.}
\hl{As an energy modeling tool for system-level exploration, \mbox{\proj} has two design principles (\mbox{\Sect{sec:main}}).
First, we use a declarative interface to describe the algorithm and hardware configurations. This is based on the observation that image processing algorithms have regular compute and memory access patterns; thus, a declarative description of the software and hardware is sufficient for accurate estimation of hardware access counts while simplifying users' effort.
The interface also decouples algorithm and hardware descriptions to facilitate iterative architectural explorations.}


Second, we adapt the energy modeling methodology to account for the characteristics of the analog and digital domains (\Sect{sec:model}).
In the digital domain, \proj directly asks users for the per-operation energy of a Processing Element (PE) and per-access memory energy. 
Tools~\cite{balasubramonian2017cacti, mittal2017destiny} and data~\cite{gao2017tetris, han2016eie} are widely available to obtain these statistics, which are routinely used in today’s digital accelerator simulation~\cite{akhlaghi2018snapea, gao2017tetris, han2016eie, feng2019asv}.
The energy of analog components, however, depends on many low-level circuit details (e.g., bias current, capacitance) that designers might not (need to) have access to, and no mature tools exist.
Thus, \proj provides default energy models for common analog components based on designs in classic CIS~\cite{hsu202005,park202151,kaur2020array,young2019data,yang2015}, while exposing an interface to accept custom designs from expert users.

\paragraph{Validation.}
We perform extensive validation of \proj against nine recent CIS chips (\Sect{sec:val}), which cover a diverse range of process nodes, pixel types, memory sizes, and forms of processing (e.g., analog vs. digital and pixel parallel vs. column parallel).
Compared to the measured \textit{absolute} energy consumption of the chips that span several orders of magnitude, \proj achieves a Pearson Correlation Coefficient of 0.9999 and a Mean Absolute Percentage Error of 7.5\%.

\paragraph{Use-cases.}
We use three use-cases to demonstrate the architectural explorations that \proj enables (\Sect{sec:usecases}).
First, we show that a conventional 2D CIS design offers little energy benefit for computation-dominated algorithms.
Second, moving toward a 3D design reduces energy due to the ability to use advanced process nodes for computation---at a cost of increasing power density.
Third, analog processing in CIS could significantly reduce memory energy even when the compute energy reduction is modest.

In summary, this paper makes the following contributions:

\begin{itemize}
    \item We provide a comprehensive survey of the design and scaling trends of CIS, pointing out new opportunities for architectural exploration.
    \item \hl{We propose the first component-level energy modeling tool for CIS and validate it against real silicon.}
    \item We use \proj to show three use-cases, demonstrating architectural trade-offs of inside-vs-off CIS, 2D-vs-3D, and analog-vs-digital processing.
\end{itemize}

%% file: background.tex
\section{Motivation: Computational CIS Design and Scaling Trends}
\label{sec:bck}

We first discuss the main design trend of CIS that underlies this paper: CIS are becoming increasingly computational (\Sect{sec:bck:trend}).
We then explain the energy benefits of such computational CIS, a main driving force behind this design trend (\Sect{sec:bck:benefits}).
Finally, we discuss the challenges of reaping the energy benefits, which motivates this work (\Sect{sec:bck:space}).

\begin{figure}[t]
    \centering
    \includegraphics[width=\columnwidth]{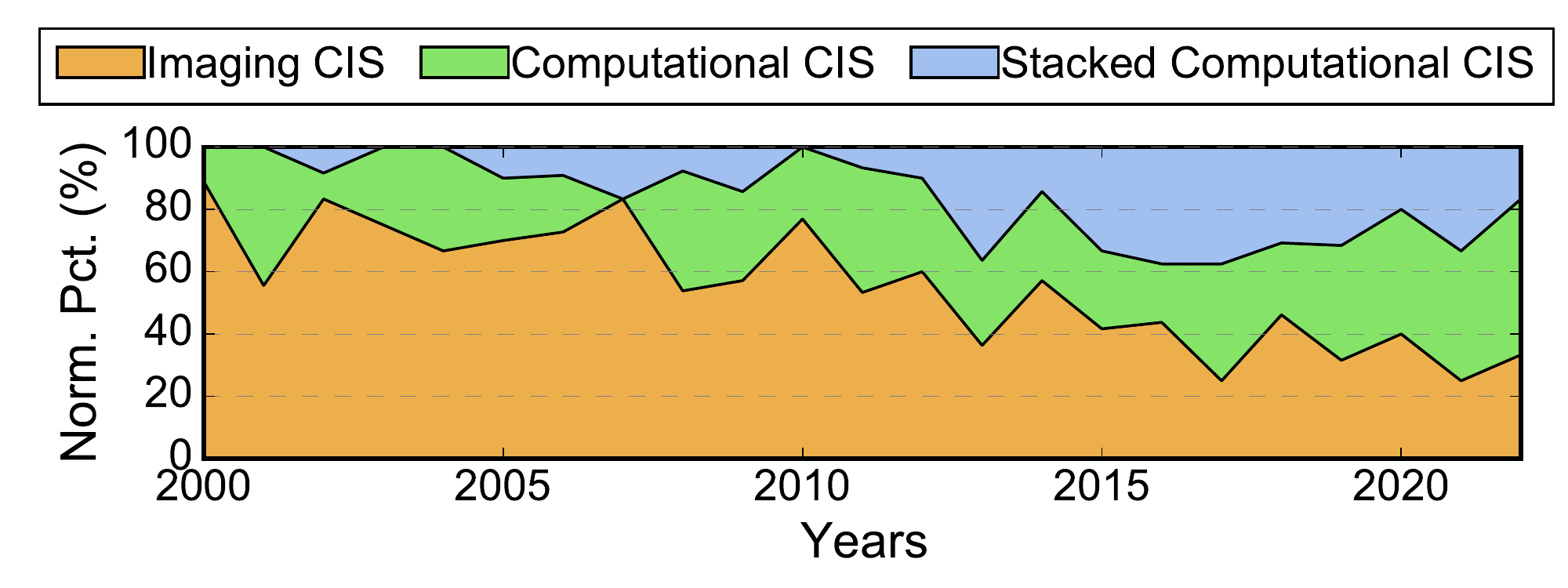}
    \vspace{-5pt}
    \caption{Percentage of conventional CIS, computational CIS, and stacked computational CIS designs from surveying all ISSCC and IEDM papers published between Year 2000 and 2022. Increasingly more CIS designs are computational.}
    \label{fig:trend}
\end{figure}

\begin{figure}[t]
  \centering
    \subfloat[Traditional 2D imaging CIS with photodiode array and ADCs.]
    {
      \label{fig:2d}
      \includegraphics[width=.45\linewidth]{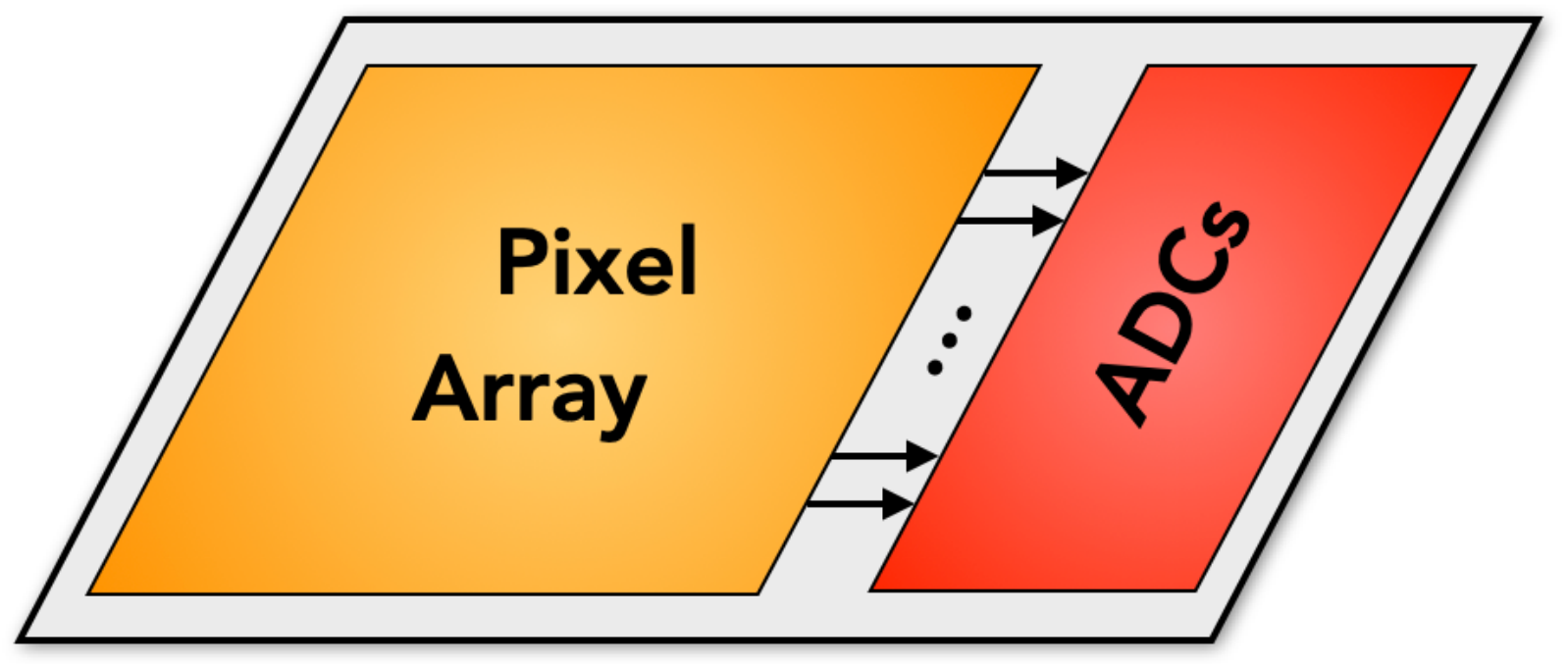}
    }
    \hfill
    \subfloat[Computational CIS with analog processing capabilities.]
    {
      \label{fig:2d-analog}
      \includegraphics[width=.45\linewidth]{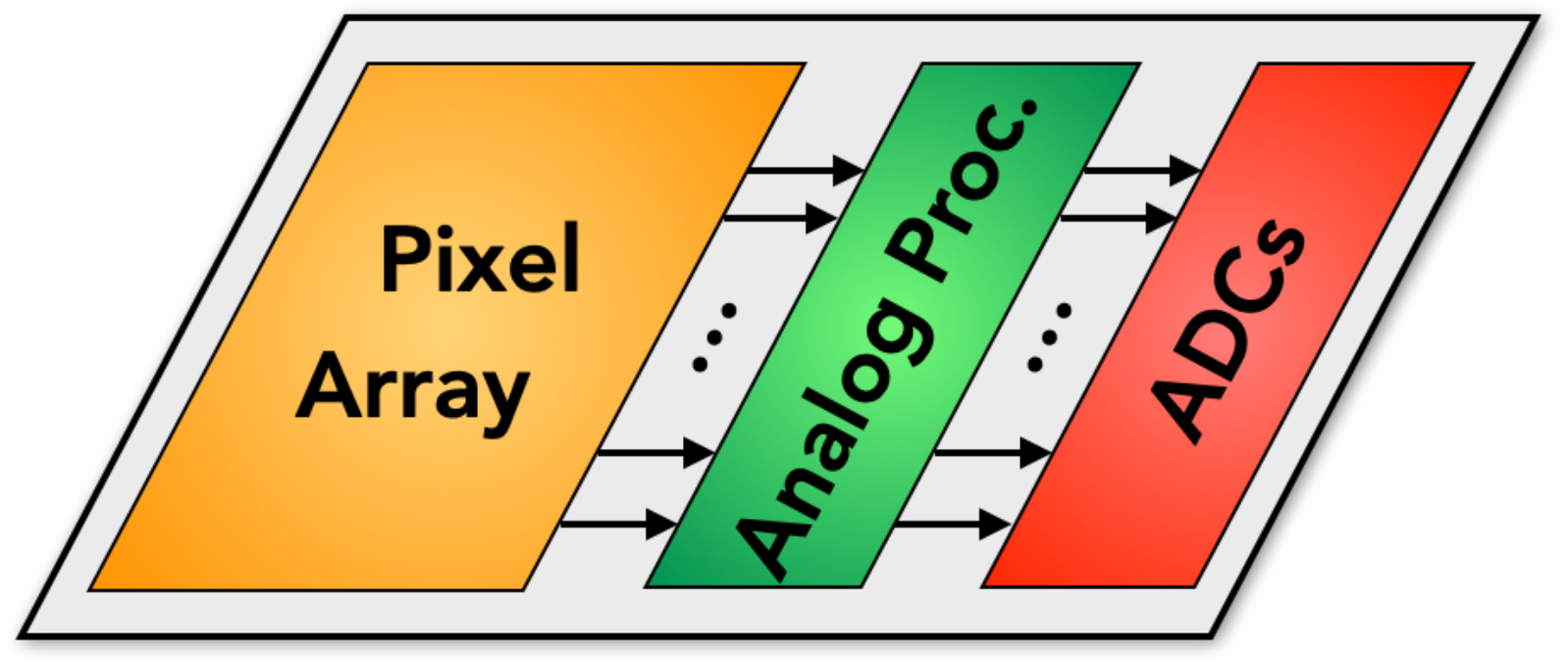}
    }
    \\
    \vspace{-5pt}
    \subfloat[Computational CIS with digital accelerator (ISP here).]
    {
      \label{fig:2d-isp}
      \includegraphics[width=.45\linewidth]{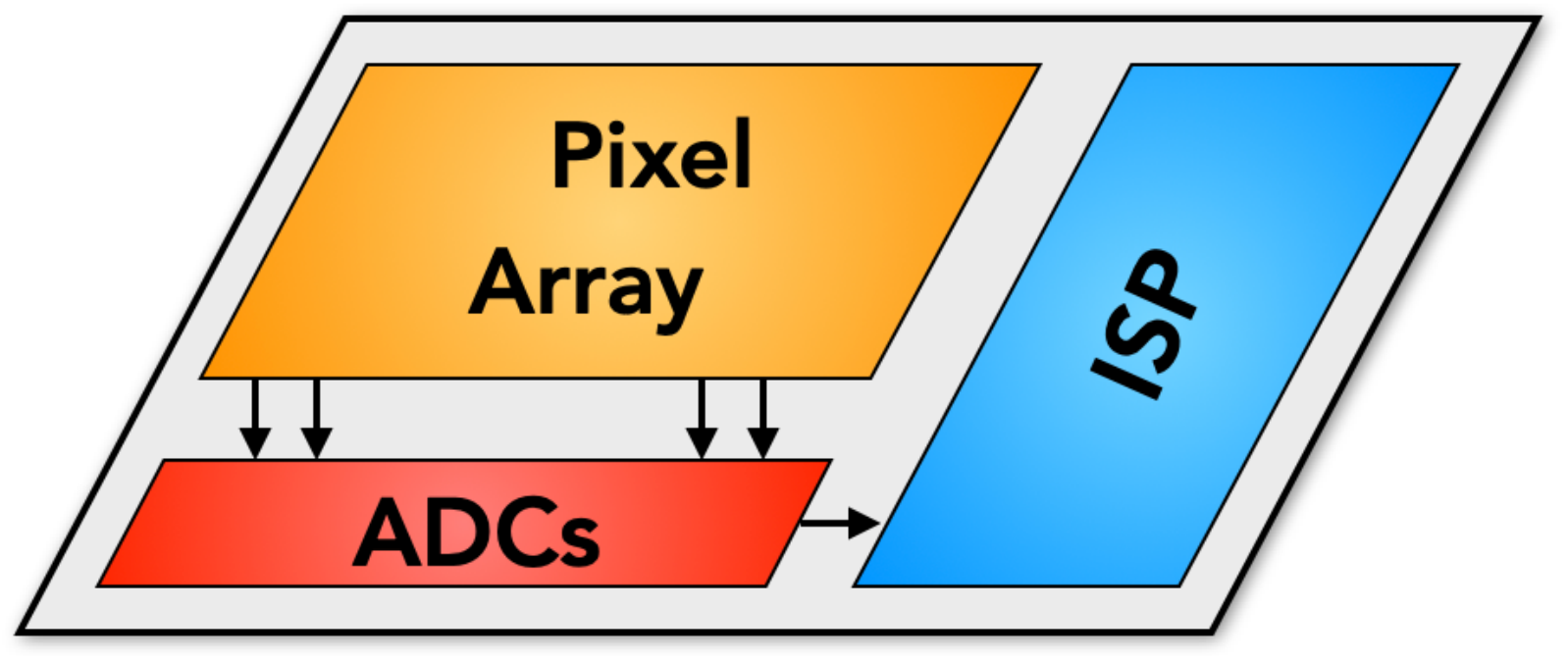}
    }
    \hfill
    \subfloat[Stacked computational CIS with digital accelerators in a separate layer.]
    {
      \label{fig:3d-npu}
      \includegraphics[width=.47\linewidth]{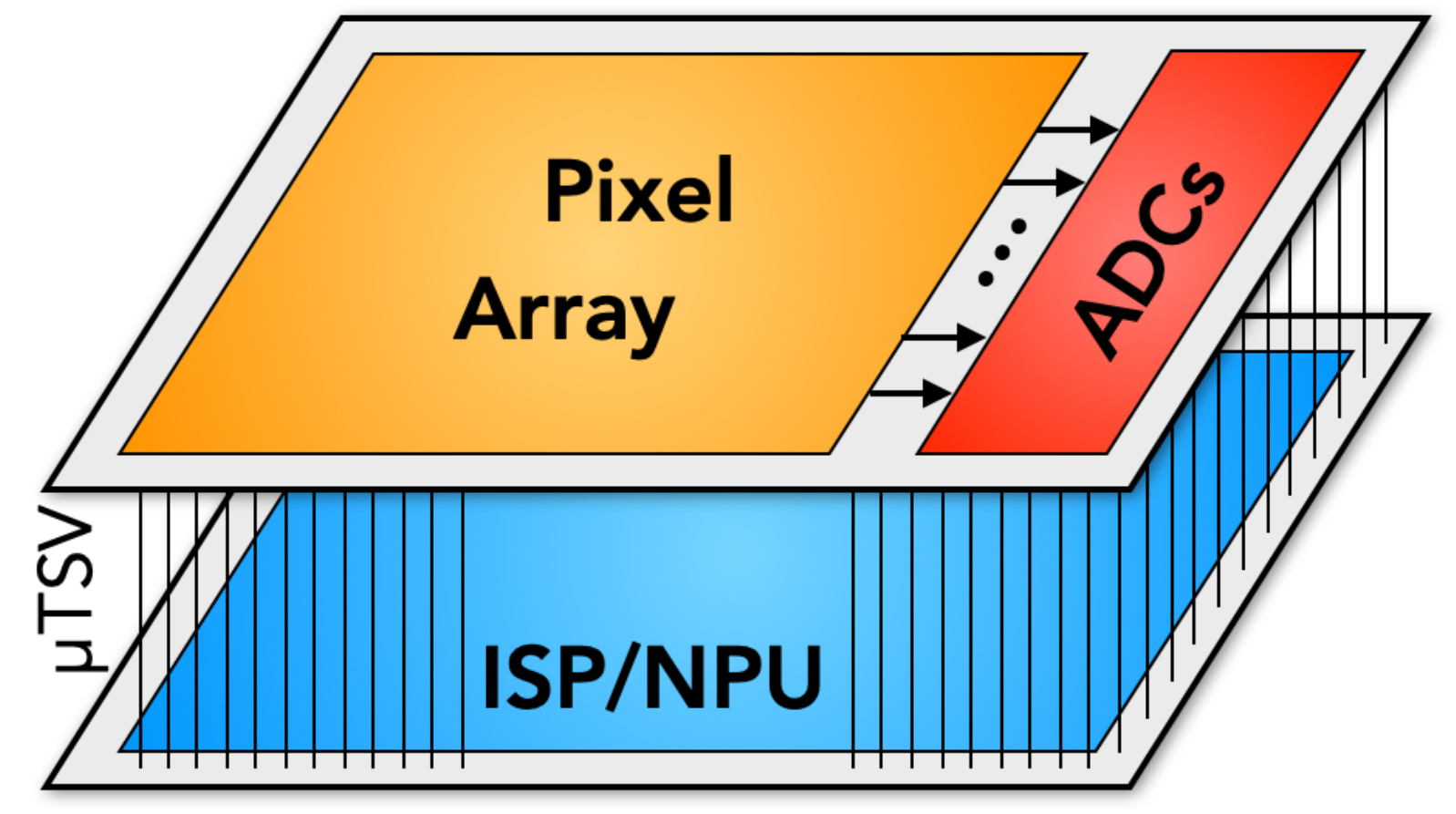}
    }
  \caption{CIS architecture evolution. CIS is moving away from a purely imaging device \protect\subref{fig:2d} to integrate computation capabilities \protect\subref{fig:2d-analog}\protect\subref{fig:2d-isp}, sometimes in a 3D stacking fashion \protect\subref{fig:3d-npu}.}
\label{fig:cis-arch}
\end{figure}

\subsection{Design Trends}
\label{sec:bck:trend}

\paragraph{CIS Primer.} Fundamentally, a CIS consists of two basic components as illustrated in \Fig{fig:2d}: a light-sensitive photodiode array that converts photons to charges and a read-out circuit that converts charges to digital values (i.e., raw pixels) through the analog-to-digital converters (ADC). 
Traditionally, raw pixels are transferred to the host, e.g., a Systems-on-a-Chip (SoC) on a smartphone, through the MIPI CSI-2 interface~\cite{mipi_csi}.
The Image Signal Processor (ISP) in the SoC removes sensing artifacts (e.g., denoising) and prepares pixels for computer vision tasks and/or for visual display.

\paragraph{CIS Design Trend.} A clear trend in CIS design is to move into the sensor computations that are traditionally carried out outside the sensor, which gives rise to the notion of \textit{Computational CIS}.
\Fig{fig:trend} shows the percentage of computational CIS papers in ISSCC and IEDM from Year 2000 and Year 2022 with respect to all the CIS papers during the same time range. Increasingly more CIS designs integrate compute capabilities.

The computations inside a CIS could take place in both the analog and the digital domain.
\Fig{fig:2d-analog} illustrates one example where analog computing is integrated into a CIS chip. Analog operations usually implement primitives for feature extraction~\cite{bong201714, bong2017low}, object detection~\cite{young2019data}, and DNN inference~\cite{hsu202005, xu2021senputing}.
\Fig{fig:2d-isp} illustrates another example that integrates digital processing, such as ISP~\cite{murakami20224}, image filtering~\cite{kim2005200} and DNN~\cite{bong2017low}.

As the processing capabilities become more complex, CIS design has embraced 3D stacking technologies, as is evident by the increasing number of stacked CIS in \Fig{fig:trend}.
\Fig{fig:3d-npu} illustrates a typical stacked design, where the processing logic is separated from, and stacked with, the pixel array layer.
The different layers communicate through hybrid bond or micro Through-Silicon Via ($\mu$TSV)~\cite{liu2022augmented,Tsugawa2017tsv}.
The processing layer typically integrates digital processors; such as ISP~\cite{kwon2020low}, image processing~\cite{hirata20217, kumagai20181}, and DNN accelerator~\cite{eki20219, likamwa2016redeye}.
Three-layer stacked designs have been proposed. Sony IMX 400~\cite{haruta20174} integrates a pixel array layer, a DRAM layer, and a digital layer with an ISP.
Meta conceptualizes a three-layer design~\cite{liu2022augmented} with a pixel array layer, a per-pixel ADC layer, and a digital processing layer that integrates a DNN accelerator.

\subsection{Benefits of Computational CIS}
\label{sec:bck:benefits}

\begin{figure}[t]
  \centering
  \includegraphics[width=\columnwidth]{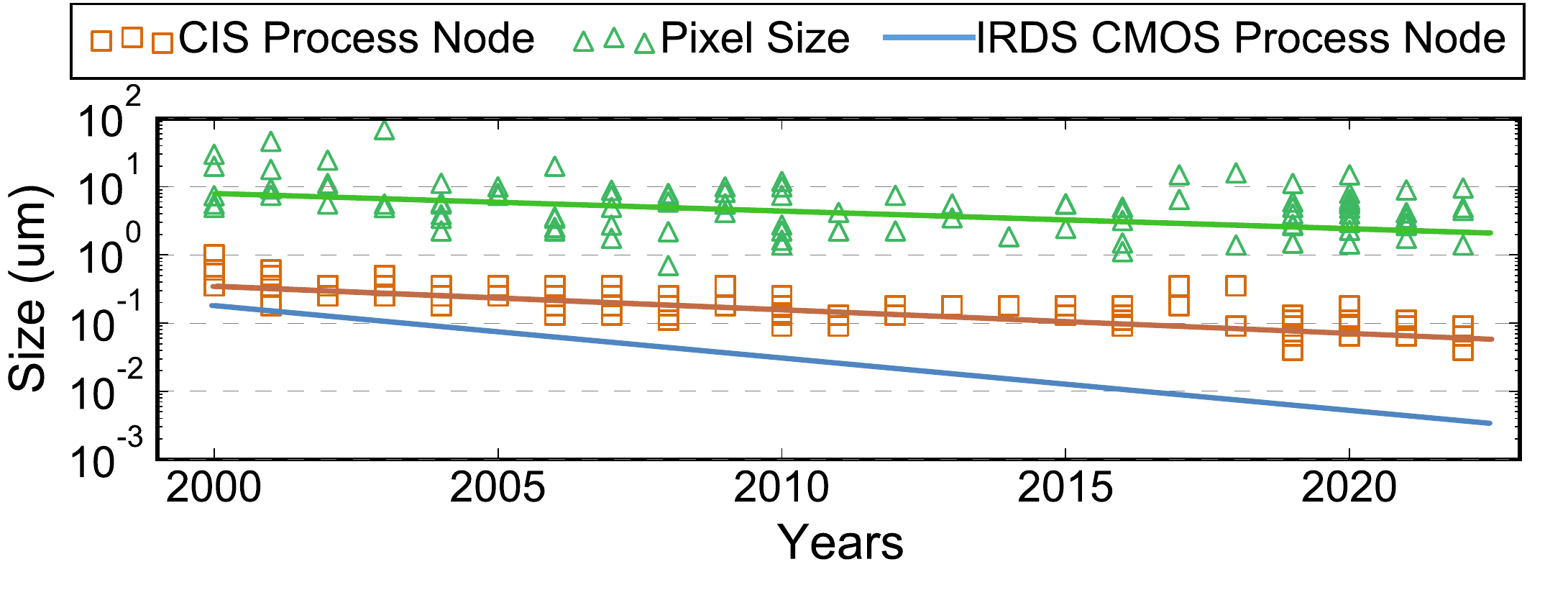}
  \caption{CIS process node always lags behind conventional CMOS process node. This is because CIS node scaling tracks the pixel size scaling, which does not shrink aggressively due to the fundamental need of maintaining photon sensitivity.}
  \label{fig:scaling}
\end{figure}

\begin{figure*}
    \centering
    \includegraphics[width=\textwidth]{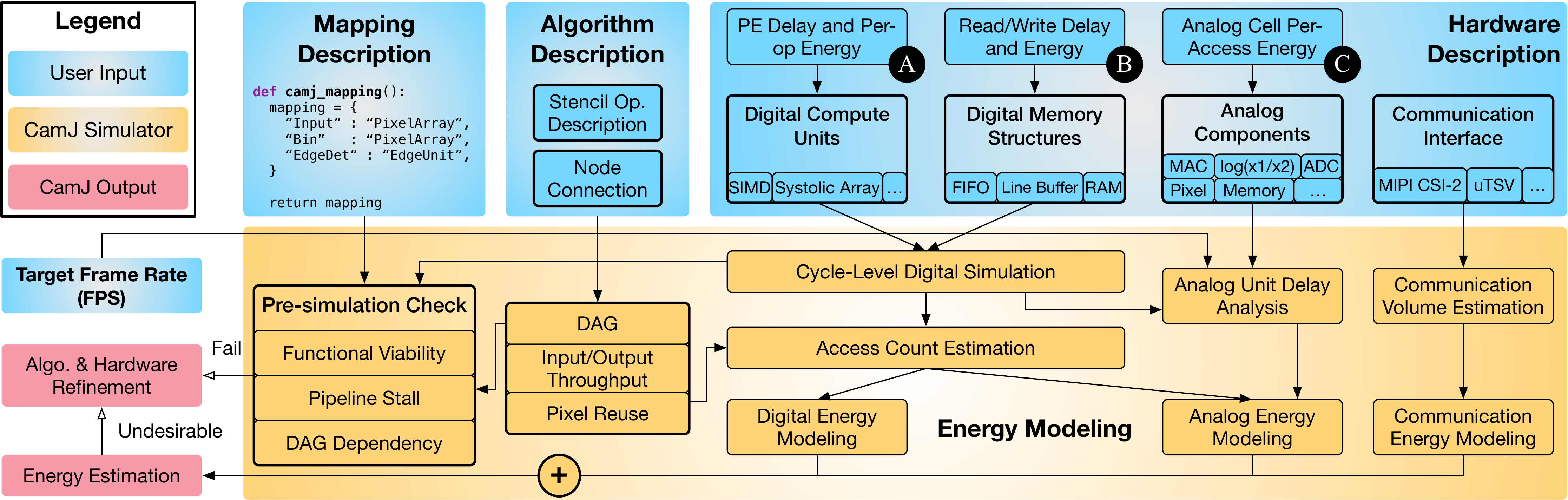}
    \caption{Overview of the \proj framework, which provides a component-level energy estimation under a target frame rate (FPS). Users provide an algorithm and hardware description and the mapping between the two. \circled{white}{A}, \circled{white}{B}, and \circled{white}{C} denote the per-access energy of digital computation, digital memory, and analog units, respectively; the first two are obtained from external tools while the last one is provided within \proj. See \Tbl{tab:hardware} for a list of digital compute/memory and analog components supported in \proj. \proj then estimates the access count to each hardware component to obtain the overall energy estimation.}
    \label{fig:sys_overview}
    \vspace{-10pt}
\end{figure*}

It is no coincidence that computational CIS emerge when energy efficiency is critical.
From an architecture perspective, computational CIS provides two main energy benefits.
First, moving computation inside the sensor allows the pixel data to be consumed closer to where they are generated. Doing so reduces the data transmission energy, which could dominate the overall energy consumption.

Specifically, data communication inside a CIS using a $\mu$TSV consumes about 1 pJ/B, whereas the energy cost of transmitting one Byte out of the CIS through the MIPI CSI-2 interface consumes about 100 pJ of energy~\cite{liu2022augmented}.
As an example, if a CIS is capable of executing an object detection DNN directly, the data volume that has to be transmitted out of the sensor is simply a few Bytes (object location and label), as opposed to, say, 6 MB, for a 1080p image.

Second, computational CIS also provides a natural platform for analog acceleration, since the pixel data originate from the analog domain to begin with, obviating the need for energy-intensive digital to analog converters that often dominate the hardware overheads in conventional analog accelerators.
Compare to digital processing, analog processing minimizes energy-intensive data conversion~\cite{ma2022hogeye, cao2022} and can reduce both the computation and memory energy consumption.

\subsection{Challenges and Design Space}
\label{sec:bck:space}

Moving computation inside a CIS, however, is not without challenges.
Most importantly, processing inside the sensor is far less efficient than that outside the sensor, fundamentally because the CIS process node significantly lags behind that of the conventional CMOS.
\Fig{fig:scaling} illustrates this difference, where square markers show the process nodes used in CIS designs from all ISSCC papers appeared during Year 2000 and Year 2022, which include leading industry CIS designs at different times.
We overlay a trend line regressed from these CIS designs to better illustrate the scaling trend.
As a comparison, the blue line at the bottom represents the conventional CMOS technology node scaling laid out by International Roadmap for Devices and Systems (IRDS)~\cite{irds}.

At around Year 2000, the CIS process node started lagging behind that of the conventional CMOS node, and the gap is increasing.
CIS design today commonly use 65nm and older process nodes.
This gap is not an artifact of the CIS designs we pick; it is fundamental: there is simply no need to aggressively scale down the process node because the pixel size does not shrink much.
The triangles in \Fig{fig:scaling} represent the pixel sizes of all the CIS designs we surveyed. The slope of CIS process node scaling almost follows exactly that of the pixel size scaling.
The reason that pixel size does not shrink is to ensure light sensitivity: a small pixel reduces the number of photons it can collect, which directly reduces the dynamic range and the Signal-to-Noise ratio  (SNR)~\cite{bigas2006review}.

Inefficient in-sensor processing can be mitigated through 3D stacking technologies~\cite{xie2015stacking}, which allows for heterogeneous integration: the pixel layer and the computing layer(s) can use their respective, optimal process node.
Stacking, however, could increase power density especially when future CIS integrate more processing capabilities.
Therefore, harnessing the power of (stacked) computational CIS requires exploring a large design space and address key challenges, some of which we list below.
Providing a tool to easily navigate the design space is the goal of our \proj framework.

\begin{itemize}
    \item Whether and what to compute in vs. off CIS?
    \item How to architect each layer in stacked CIS to achieve energy reduction without increasing power density?
    \item What to compute in the analog vs. digital domains?
\end{itemize}

%% file: framework.tex
\section{\proj Framework}
\label{sec:main}

No framework to date allows designers to explore the complicated design space of computational CIS at a system level. Our \proj framework is designed to fill this void (\Sect{sec:main:when}).
We first outline the design principles of \proj, followed by an overview of the \proj design internals (\Sect{sec:main:ov}). We use a concrete example to demonstrate from a designer perspective how \proj is used (\Sect{sec:main:prog}).

\subsection{When is \proj Used in the Design Cycle?}
\label{sec:main:when}

\hl{\mbox{\proj} is meant to be used for \textit{system-level} exploration after each component design is sketched out; an analogy would be Systems-on-a-Chip (SoC) vs. accelerator design. Before system-level exploration, a team usually has at hand a range of component-level designs, which could be licensed Intellectual Property (IP) blocks, reference designs from the literature, or earlier designs from other teams in the organization (e.g., using a synthesis flow or High-Level Synthesis tools); in all cases the component-level energy behavior is known or can be modeled using external tools like Aladdin~\mbox{\cite{shao2014aladdin}} and OpenRAM~\mbox{\cite{guthaus2016openram}}.

\mbox{\proj} helps designers make design decisions when assembling the individual (digital and analog) components into an optimal system.
Ideally, a designer uses \mbox{\proj} to estimate the system energy given initial designs of individual components; using the estimation, a designer can \textit{iteratively} refine the components/system design. For instance, \mbox{\proj} can identify energy bottlenecks and guide the re-design of corresponding components. Orthogonally, a designer can use \mbox{\proj} to explore optimal mapping and partitioning of the algorithms between analog vs. digital domains or in vs. off CIS to minimize overall system energy under performance targets.}

\hl{\mbox{\proj} is \textit{not} a synthesis tool; it does not generate (nor estimate the energy of) an accelerator. Rather, \mbox{\proj} can be used in conjunction with HLS: one could use HLS to first generate an accelerator and then use \mbox{\proj} to explore, in the bigger system, how/whether that accelerator would fit in a computational CIS to maximize end-to-end application gains.
}

\subsection{Design Principles and Overview}
\label{sec:main:ov}


\hl{As with any energy modeling tool~\mbox{\cite{brooks2000wattch, leng2013gpuwattch, kandiah2021accelwattch}}, the total CIS energy is sum of the product of 1) the access count to each hardware unit and 2) the per-access energy consumption.
Therefore, the central objective of \mbox{\proj} is to develop a modeling methodology that accurately estimates those two statistics using a programming interface that, critically, only requires user inputs for information that cannot be automatically inferred.
\mbox{\Fig{fig:sys_overview}} shows an overview of the \mbox{\proj} framework. \mbox{\Sect{sec:main:prog}} and \mbox{\Sect{sec:model}} discuss the programming interface and modeling details, respectively.
Here, we provide an overview of the interface and modeling methodology.}

\hl{\mbox{\paragraph{Interface.}} \mbox{\proj} observes that CIS deal with stencil-based image processing, which has regular computation and memory access patterns (i.e., little to no control flow) that are statically mapped to hardware units.
The access count statistics can, thus, be inferred with only the high-level algorithm description and hardware configuration without knowing the implementation details.
Therefore, \mbox{\proj} exposes a \textit{declarative} interface, in which an algorithm is described by a DAG and the input/output/stencil dimensions at each node and the hardware is abstracted as a set of basic units, each performing a specific sensing, computation, or memory operation.}

\hl{\mbox{\proj}'s interface also \textit{decouples} the description of an algorithm, the underlying hardware, and the mapping between the two. 
A decoupled interface facilitates an iterative system design process, during which algorithm, hardware, and algorithm to hardware mapping can change independently.
For instance, one can evaluate algorithmic changes by re-writing the algorithm description without touching the hardware design, or explore different algorithm-to-hardware mappings (e.g., split between analog vs. digital and between in vs. off sensor) by describing a new mapping.}

\hl{\mbox{\paragraph{Internal Modeling.}}
\mbox{\proj} judiciously uses different methodologies to obtain the per access energy in the digital vs. analog domains.}
For digital structures, \mbox{\proj} directly asks users to provide the per-cycle energy of computation PEs (\mbox{\circled{white}{A}}) and the per-access energy of memory units (\mbox{\circled{white}{B}}).
These statistics are usually obtained by an ASIC synthesis flow or from commonly used tools (e.g., CACTI~\mbox{\cite{balasubramonian2017cacti}} and OpenRAM~\mbox{\cite{guthaus2016openram}}), and are routinely used in today's digital accelerator simulators for energy estimation~\mbox{\cite{gao2017tetris, kodukula2021rhythmic, akhlaghi2018snapea}}.

One might be surprised to find that 
\proj directly asks for the per-cycle/access energy of the digital structures. 
This is because of the design philosophy of \proj (\Sect{sec:main:when}): it is \textit{not} used to generate digital accelerators; rather, it helps assess how an accelerator fits in the entire computational CIS system. For that reason, \proj expects designers to have a preliminary design of the digital accelerators (whether it's your manual design, HLS generated, or a licensed IP), in which case one will have the per-cycle/access energy statistics.

\hl{Unlike digital structures, few energy modeling tool exists for analog structures, whose energy consumption (\mbox{\circled{white}{C}}) depend on many low-level circuit details (e.g., load capacitance, gain, bias current) that are cumbersome and perhaps unreasonable to ask system-level designers for.
Our design decision is to expose a low-level interface to accept these parameters from expert users, but also provide default energy models based on
classic implementations of analog components and delay analyses of these components (\mbox{\Sect{sec:model:delay}}).}

Finally, \proj performs a series of pre-simulation design checks to ensure that the algorithm and hardware combination 1) is functionally viable (e.g., ADCs must exist between the analog and digital domain), 2) does not have pipeline stalls (to avoid accumulating long frame latency), and 3) has well-formed dependencies in the algorithm DAG (e.g., no circle).
We provide feedback upon check failures and a detailed energy breakdown, which helps designers iteratively refine the algorithm and/or hardware.

\begin{figure}[t]
    \centering
     \includegraphics[width=\columnwidth]{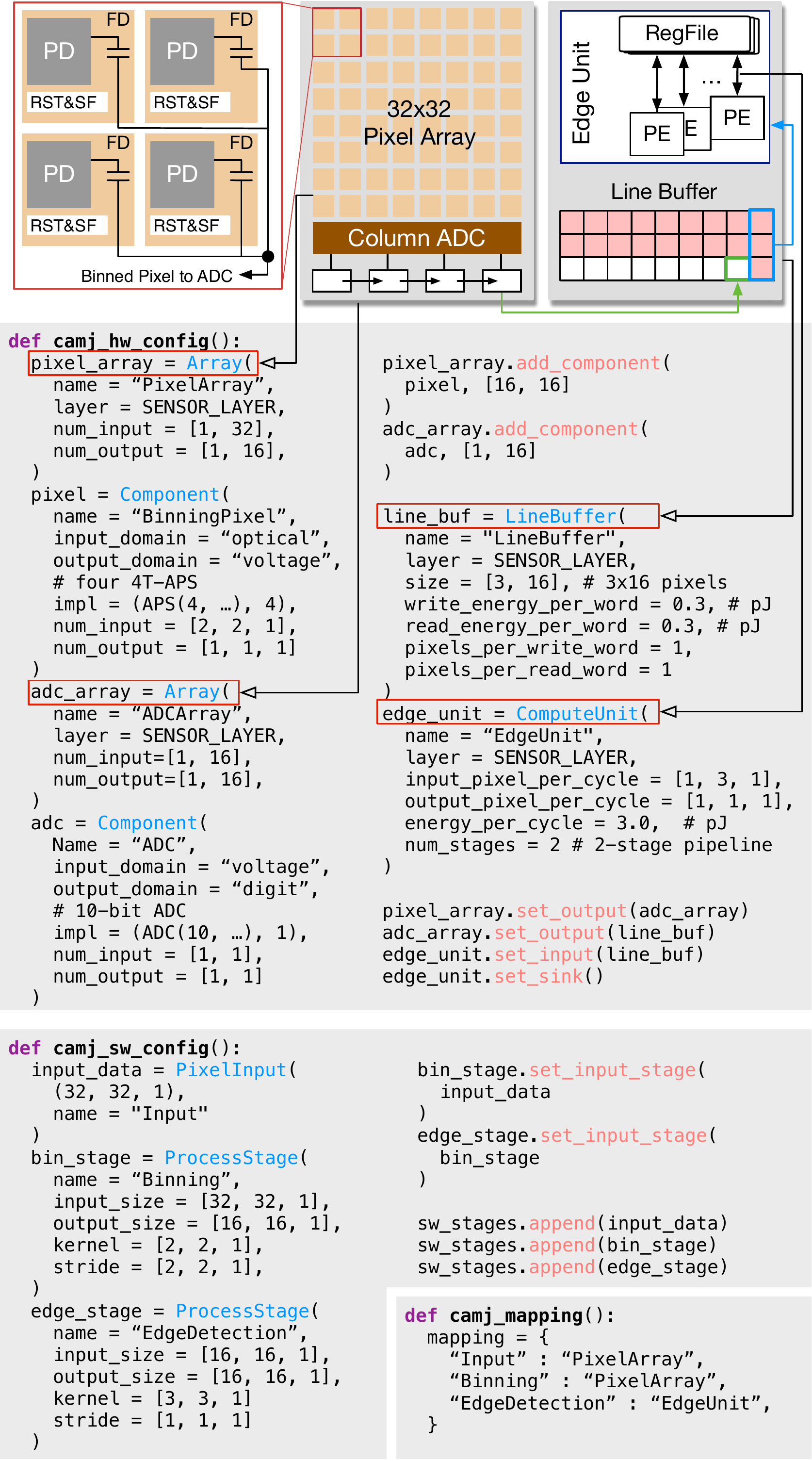}
    \caption{An example of defining a simple CIS using the \proj programming interface in Python. The hardware architecture of the simulated CIS is illustrate at the top.}
    \label{fig:code_example}
\end{figure}

\subsection{Programming Interface}
\label{sec:main:prog}

We use a running example in \Fig{fig:code_example} to introduce the programming interface and illustrate the main design decisions.

\paragraph{An Example.} The Python code in \Fig{fig:code_example} shows a concrete example to use the programming interface of \proj. In this conceptual CIS design with a $32 \times 32$ pixel array, every $2 \times 2$ pixel tile is first averaged (i.e., ``binned'') to produce a $16 \times 16$ image. The sensor then performs a digital edge detection on the image before sending the edge data through the MIPI CSI interface.
The \texttt{camj\_hw\_config} function and the \texttt{camj\_sw\_config} function describe the hardware components and the algorithm, respectively. The \texttt{camj\_mapping} maps each algorithm stage to a hardware component. We explain each part next.

\paragraph{Algorithm Description.}
The code in \texttt{camj\_sw\_config} describes the DAG of the entire processing pipeline, starting from the raw pixels generated by the pixel array (\texttt{PixelInput}), which go through two processing stages: \texttt{bin\_stage} for pixel binning and \texttt{edge\_stage} for edge detection. The \texttt{set\_input\_stage} method connects the stages together to form a DAG.

Notice how the algorithm description does not require the actual arithmetic details; we observe that image processing algorithms can be abstracted as stencil operations that operate on a local window of pixels at a time~\cite{hegarty2014darkroom, qadeer2013convolution} --- convolution (or image filtering in conventional image processing parlance) being a prime example.
\hl{This observation holds in all the ISSCC/IEDM papers since Year 2000 we surveyed. Irregular computations complicate hardware design and increase energy, defeating the purpose of in-CIS computing.}

Therefore, users express only the input/output image dimensions (\texttt{input\_size}, \texttt{output\_size}) along with the stencil window (\texttt{kernel}) and stride size (\texttt{stride}).
Given the regular computation and data access pattern of stencil operations, \proj could accurately estimate the access counts to different hardware structures for energy estimations.
\hl{Nonetheless, \mbox{\proj} does accept as input a memory trace offline collected for an irregular algorithm, which can then be integrated with external tools such as DRAMPower~\mbox{\cite{chandrasekar2012drampower}} to estimate the energy consumption for irregular algorithms.}

\paragraph{Hardware Description.}
\texttt{camj\_hw\_config} describes the hardware architecture, which we illustrate at the top of \Fig{fig:code_example}.
The hardware description consists of two components: analog processing units and digital processing units.

\circled{white}{1} \underline{Analog Units.}
CIS hardware necessarily starts from analog units, which, at a high level, are described as a set of Analog Functional Arrays (AFA),
which is in turn is composed of a set of Analog Functional Components (\acomp{s}).
The most important AFA in a CIS is the pixel array (\texttt{pixel\_array}), in which each \textsc{A-Component} is a pixel, which is added to the pixel array through the \texttt{add\_component} method.
In the example of \Fig{fig:code_example}, the pixel array is followed by another AFA, i.e., the ADC array (\texttt{adc\_array}), where each \acomp is an ADC.

\begin{table}
\caption{A list of hardware units supported in \proj.  APS: Active Pixel Sensor; DPS: Digital Pixel Sensor; PWM: Pulse Width Modulation; MAC: Multiply-Accumulate.}
\resizebox{\columnwidth}{!}{
\renewcommand*{\arraystretch}{1}
\renewcommand*{\tabcolsep}{6pt}
\begin{tabular}{ccc} 
\toprule[0.15em]

& \specialcell{\textbf{Analog}\\(\acomp)} & \textbf{Digital} \\
\midrule[0.05em]
\textbf{Memory} & \specialcell{Passive/Active,\\Sample-and-Hold} & \specialcell{FIFO, Line Buffer,\\Double-Buffered SRAM}\\
\midrule[0.05em]
\textbf{Compute} & \specialcell{Pixel (APS, DPS, PWM),\\ ADC, MAC, Max, Scaling, \\ Add, log, Abs, Comparator} & \specialcell{Systolic Array, Generic\\ Pipelined Accelerator} \\
\bottomrule[0.15em]
\end{tabular}
}
\label{tab:hardware}
\end{table}

From users' perspective, each \acomp performs a particular kind of (arithmetic) operation.
In addition to a pixel or an ADC, \proj provides other common \acomp{s} used in CIS such as MAC or logarithmic operations.
The complete list of analog \acomp{s} is in \Tbl{tab:hardware}.
The energy consumption of each \acomp, which is dictated by its circuit-level implementation, is abstracted away from the users by the \texttt{impl} method.
\Sect{sec:model:analog} will later describe how we model the energy of each \acomp by mapping it to its analog circuit implementation.

What users do have to provide, however,
is the signal dimension (\texttt{num\_input} and \texttt{num\_output}) and signal domain (\texttt{input\_domain} and \texttt{output\_domain}) of an AFA's input and output data. These parameters allow \proj to check whether the simulated CIS is functionally viable.
Specifically, the \texttt{input\_domain} of a consumer unit and the \texttt{output\_domain} of a producer unit must match.
If, for instance, the producer is in the charge domain and the consumer is in the voltage domain, \proj will ask designers to insert a charge-to-voltage conversion component\footnote{unless the output of the consumer is in the voltage domain, where the inherent capacitor of the consumer naturally acts as an analog buffer.}, which has energy implications.
Similarly, if the \texttt{num\_input} of a consumer unit and the \texttt{num\_output} of a producer unit do not match, the hardware must have an analog buffer in-between, which, again, could have energy implications.



\circled{white}{2} \underline{Digital Units.}
The digital part of the hardware is described by specifying a set of compute units that communicate through memory structures.
In this example, the compute unit is the edge detection accelerator (instantiated through \texttt{ComputeUnit}), which reads from the line buffer (\texttt{LineBuffer}), a pre-defined memory structure, that stores data from the pixel array, an analog unit as described before.

Column 2 of \Tbl{tab:hardware} lists the memory structures and compute units available in \proj. We support three memory structures commonly found in image/vision processing: FIFO (\texttt{FIFO}), line buffer~\cite{hegarty2014darkroom, whatmough2019fixynn} (\texttt{LineBuffer}), and double-buffered SRAM (\texttt{DoubleBuffer}).
The compute units are abstracted as pipelined accelerators through the \texttt{ComputeUnit} interface. We also provide a \texttt{SystolicArray} class to describe a systolic array due to its importance in executing DNNs.

With the generic pipelined accelerator interface (\texttt{ComputeUnit}), users can model a wide range of (image processing) accelerators.
To describe a pipelined accelerator, CamJ requires three main parameters:
the shape of pixels read per cycle (\texttt{input\_pixel\_per\_cycle}), the shape of pixels generated per cycle (\texttt{output\_pixel\_per\_cycle}), and the pipeline depth (\texttt{num\_stages}).
Using these statistics, \proj performs cycle-level simulation for two purposes.
First, \proj can check whether the accelerator will stall the CIS pipeline and, if so, asks for a re-design of the accelerator.
Second, \proj can estimate the total latency of the digital domain, which is critical for analog energy estimation.
Stall checking and latency estimation are critical for analog energy estimation as we will discuss in \Sect{sec:model:delay}.




\paragraph{Mapping.}
The \texttt{camj\_mapping} function maps each algorithm stage to a hardware unit.
The code is self-evident.
Users can simply remap the algorithm to hardware to explore a different system design.
The decoupling of algorithm and hardware description through the mapping function also enables easy expression of hardware reuse---by simply mapping different algorithm nodes to a hardware component.

%% file: modeling.tex
\section{Energy Modeling Methodology}
\label{sec:model}

The energy consumption per frame of a CIS sensor is the sum of that of the analog, digital, and data communication:
\begin{align}
    E^{\text{frame}} = E_\text{a}^{\text{frame}} + E_\text{d}^{\text{frame}} + E_\text{c}^{\text{frame}}
\end{align}

Before we describe how the analog component $E_\text{a}^{\text{frame}}$ (\Sect{sec:model:analog}), digital component $E_\text{d}^{\text{frame}}$ (\Sect{sec:model:digital}), and communication component $E_\text{c}^{\text{frame}}$ (\Sect{sec:model:comm}) are modeled separately, we first discuss a prerequisite of energy modeling: delay estimation (\Sect{sec:model:delay}).

\subsection{Delay Estimation}
\label{sec:model:delay}

The energy consumption, both analog and digital, is correlated with the circuit speed.
For example, in the analog domain an operational amplifier (OpAmp) with higher response speed requires larger bias current, which increase the energy consumption (assuming the OpAmp is active over a fixed duration, e.g., when used for an analog frame buffer).
The latency of digital units is estimated through cycle-level simulation as described in the previous section.
\hl{The delay of an analog unit, in contrast, depends on many parameters specific to a fully-designed circuit.
We find it cumbersome and error-prone to ask users for input: users often find themselves tuning low-level parameters only to end up with a design that misses the target frame rate.
Instead, \mbox{\proj}'s insight is that each analog unit's delay can be automatically inferred from \textit{the prescribed frame rate}.}

Specifically, the fundamental observation that \proj relies on is that \textit{the CIS pipeline is designed to never stall}.
This is because the input data to the pipeline is generated at a constant rate as the pixel array is exposed to light at the constant speed.
If the pipeline ever stalls in a later stage, the frame latency would gradually accumulate, leading to excessively long responsive latency or frame drops.
Therefore, CIS designers ensure that the hardware pipeline never stalls.
In a fully-pipelined hardware, each pipeline stage must have roughly the same delay; this is the basis of our delay estimation.

\begin{figure}[t]
    \centering
    \includegraphics[width=\columnwidth]{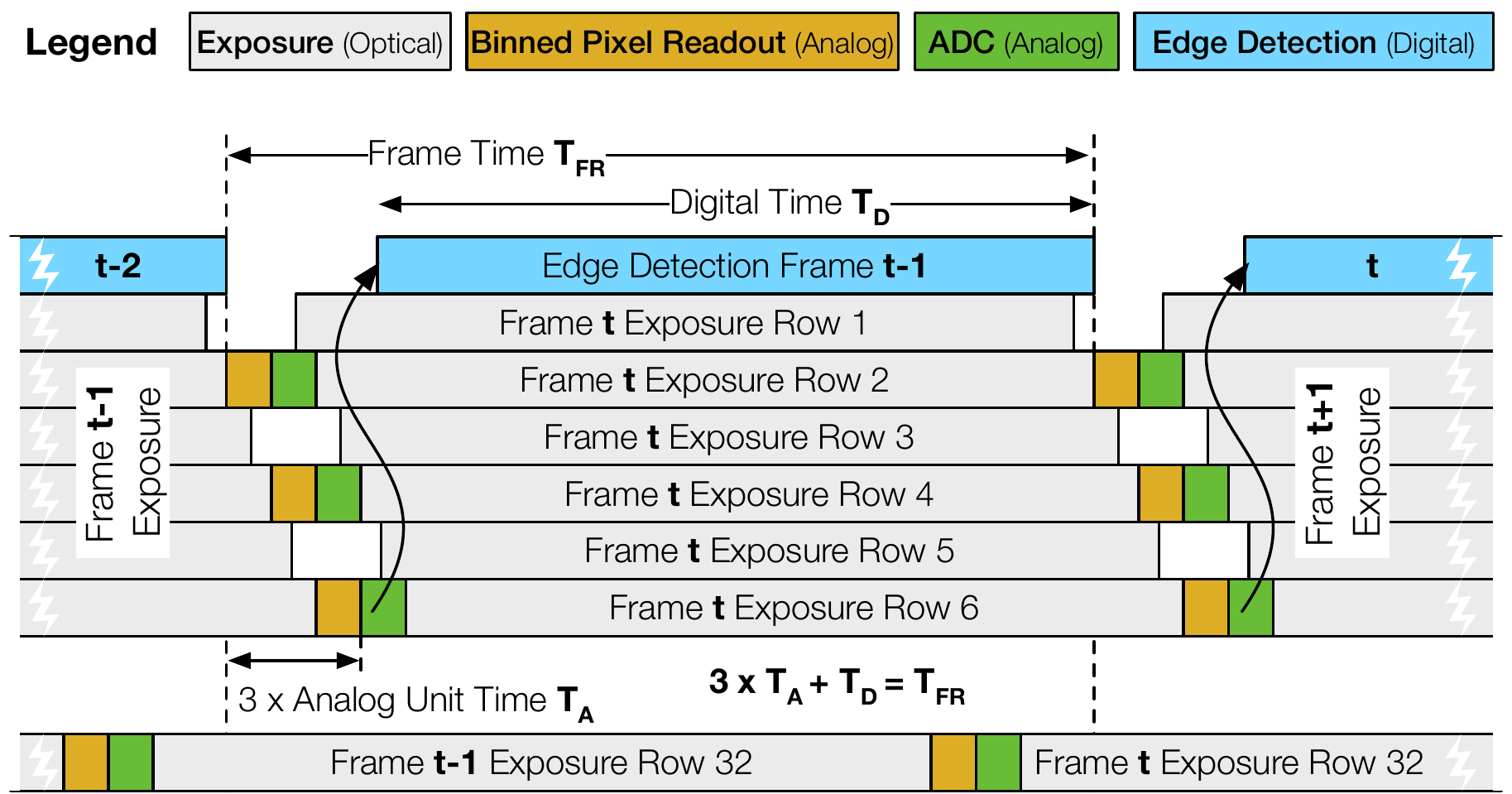}
    \caption{The (not-to-scale) pipeline diagram of the example in \Fig{fig:code_example} when there is no pipeline stall.
    }
    \label{fig:timing}
\end{figure}

\paragraph{Example.}
\Fig{fig:timing} shows the pipeline timing for the example in \Fig{fig:code_example}.
The frame time $T_{FR}$ is 1/FPS, where FPS is the target frame rate.
In the diagram, the frame time is the delay between when the pixels of the current frame can be read-out to when the computation of the current frame finishes.
The ``Binned Pixel Readout'' and ``ADC'' are the two analog units, who share the same delay $T_A$ (i.e., balanced pipeline) to be estimated. The ``Edge Detection'' is the digital unit, which starts once the second line has been written to the line buffer.

To estimate $T_A$, we first simulate the digital domain to estimate the latency of the entire digital domain ($T_D$ here). Given the frame time $T_{FR}$, we can then estimate how much time is left for the analog units. In this example, $T_A = \tfrac{T_{FR} - T_{D}}{3}$.

\proj will analyze the hardware description and, upon detecting potential stalls, asks the user to re-design the hardware to avoid stall.
Specifically, \proj checks to avoid three scenarios:
1) pixel required is not generated by the producer yet, 2) the memory in-between two stages is full, or 3) the number of access ports in the memory structures is not enough.


\subsection{Analog Energy Modeling}
\label{sec:model:analog}



The analog energy per frame, $E_\text{analog}^\text{frame}$, is the sum of the energy consumption per access of each \acomp weighted by the access count to that component. 
Refer to \Tbl{tab:hardware} for a list of \acomp{s} that \proj supports.

\begin{equation}
\begin{aligned}
    E_\text{a}^\text{frame} =\sum_i (E_\text{a}^{\text{component}[i]}\times \text{Num}_\text{access}^{\text{component}[i]})
\end{aligned}
\label{eq:analog_energy_component_1}
\end{equation}

\paragraph{Modeling \acomp{s} Access Count.}
The access count to a \acomp is the number of times the \acomp is used per frame.
Recall from \Sect{sec:main:prog} that each \acomp is part of an Analog Functional Array (AFA).
\proj leverages the fundamental observation that stencil operations in image processing have regular computation and memory access patterns and, thus, the access count to each \acomp in the same AFA is the same.

As a result, the access count of a component $i$ is simply the ratio between the total number of operations mapped to the AFA $j$ that contains the component ($\text{Num}_\text{ops}^{\text{AFA}[j]}$) and the number of components in that AFA ($\text{Num}_{\text{component}[i]}^{\text{AFA}[j]}$):
\begin{equation}
    \text{Num}_\text{access}^{\text{component}[i]}=
    \frac{\text{Num}_\text{ops}^{\text{AFA}[j]}}{\text{Num}_{\text{component}[i]}^{\text{AFA}[j]}}
\end{equation}
The numerator is easily derived from the algorithm description of a stencil operation (e.g., calculating the number of MAC operations in a convolution).
The denominator is the \texttt{num\_component} attribute of the AFA (see \Fig{fig:code_example}). 

%
%

\paragraph{Modeling \acomp{s} Access Energy.}
Internally, each \acomp is built from a set of analog cells, which we call \acell{s}.
Modeling per-access energy of an \acomp requires knowing its cell-level implementation.
Expert users can define new cell parameters and/or cell-level implementation of an \acomp.
Absent those, each \acomp has a default implementation, surveyed from classic and recent CIS designs~\cite{hsu202005,park202151,kaur2020array,young2019data,yang2015}.
For instance,
a 4T-APS pixel \acomp consists of a photodiode (PD) \acell, a floating diffusion node (FD) \acell, and a source follower (SF) \acell;
a multiplier implemented by switched-capacitor charge re-distribution~\cite{lee2017} consists of a capacitor array \acell and an OpAmp \acell.

We now describe our energy modeling of \acomp{s}, but keep in mind that these design details are abstracted away from typical users. $E_\text{a}^\text{component}$ is the weighted sum of the energy consumption of each constituting \acell in the component and the access counts to the \textsc{A-Cell}:
\begin{equation}
\begin{aligned}
    E_\text{a}^{\text{component}[i]} =\sum_j E_\text{a}^{\text{cell}[j]}\times \text{Num}_\text{access}^{\text{cell}[j]}
\end{aligned}
\label{eq:analog_energy_cell_1}
\end{equation}

Despite large varieties of high-level analog circuits, the \acell used for analog in-sensor computing can be categorized to three classes according to circuit characteristics: dynamic \acell, static-biased \acell, and non-linear \acell.
They each consume energy in a different way.

\circled{white}{1} \underline{Dynamic \acell.}
The energy of a dynamic circuit comes from the charging and discharging of the total capacitance in the circuit:
\begin{equation}
    E_\text{a}^\text{cell,d} = \sum_i^{N_{c}} C[i] \times V_\text{VS}[i]^2
\label{eq:dynamic-acell}
\end{equation}
where $N_{c}$ represents the total number of capacitance nodes in the dynamic circuit, and $C[i]$ and $V_\text{VS}[i]$ are the capacitance and the voltage swing at $i^{th}$ capacitance node, respectively.
Typical dynamic \acell{s} include capacitive digial-to-analog converter (CDAC) and passive analog memory.

In \Equ{eq:dynamic-acell}, $V_\text{VS}$ is determined by the analog supply $V_\text{DDA}$ and the number of transistors placed between the analog supply and the ground.
The nodal capacitance $C$ is determined by its thermal noise and the computation precision.
To guarantee the accuracy of analog computing, the maximum thermal noise should be kept below $\frac{1}{2}$LSB of the data resolution:
\begin{equation}
    \sigma_\text{thermal}=\sqrt{\frac{kT}{C}},\quad 3\sigma_\text{thermal}<\frac{1}{2}\text{LSB}
    \label{eq:thermal}
\end{equation}
where $\text{LSB}=V_\text{VS}/2^\text{data resolution}$.
Data resolution is algorithm dependent. For example, if $V_\text{VS}=1V$ and the required resolution is 8-bit, the thermal noise should be less than $\frac{1}{3}\frac{1}{2}\frac{1\text{V}}{2^{8\text{-bit}}}=2.6\text{mV}$, from which $C$ is obtained.

\circled{white}{2} \underline{Static-biased \acell.}
The energy of a static-biased circuit comes from the integration of the bias current over a specific time period under the analog supply $V_\text{DDA}$:
\begin{equation}
    E_\text{a}^\text{cell,s} = V_\text{DDA}\times I_\text{bias} \times t_\text{static}
    \label{eq:staticcell}
\end{equation}
where $I_\text{bias}$ is the bias current and $t_\text{static}$ is the time during which the \acell is statically biased.

We provide two ways to estimate $I_\text{static}$ based on circuit details.
For \acell{s} where $I_\text{bias}$ directly drives the load capacitance (e.g. static-biased SF in a pixel), $I_\text{bias}$ is determined by charging up the load within the given time:
\begin{equation}
    I_\text{bias,1} = \frac{C_\text{load}\times V_\text{VS}}{t_\text{static}}
\label{eq:bias_1}
\end{equation}
where $C_\text{load}$ is the load capacitance.
The energy is reduced to:
\begin{equation}
    E_\text{a}^\text{cell,s} = C_\text{load}\times V_\text{VS}\times V_\text{DDA}
\end{equation}

For \acell{s} where $I_{bias}$ does not directly drive the load capacitance (e.g. differential operational amplifier in analog memory or discrete-time integrator), $I_\text{bias}$ can be determined by the classic $\frac{g_\text{m}}{I_\text{d}}$ method~\cite{gm_id}:
\begin{equation}
    I_\text{bias,2} = \frac{2\pi \times C_\text{load}\times \text{GBW}}{g_\text{m}/I_\text{d}}
\label{eq:i_bias2}
\end{equation}
where $\frac{g_m}{I_d}$ is a technology-insensitive factor ranging from 10 to 20 depending on the inversion level of the transistors, and GBW is product between gain (G) and bandwidth (BW).

To use \Equ{eq:staticcell} and \Equ{eq:i_bias2}, \proj must estimate BW and $t_\text{static}$, both of which depend on the \acell delay.
Specifically, BW is the reciprocal of the \acell delay and $t_\text{static}$ is:
\begin{equation}
    t_\text{static}=T_A-\sum_i^K t_i^\text{cell}
\label{eq:acell_delay}
\end{equation}
where $T_A$ is the delay of the \acomp containing the \acell and is estimated in \Sect{sec:model:delay}; $K$ is the number of cells before the current \acell on the \acomp critical path, and $t_i^\text{cell}$ is the delay of an \acell.
Absent timing condition from users, we evenly allocate the \acomp delay to each \acell,
based on the fact that the analog signal uni-directionally flows through the \acomp{s} we support so all \acell{s} are on the critical path.




\circled{white}{3} \underline{Non-linear \acell.}
For those circuits with non-linear transfer functions, such as ADCs and comparators (which are essentially 1-bit ADCs), they contain both dynamic/static-biased circuit cells and digital logic so it is difficult to estimate the energy from analytical formulas.
Instead, we use the ADC's Walden Figure-of-Merit (FoM) plot~\cite{adc_fom} surveyed from recently published CIS papers, which shows the ADC's energy-per-conversion vs. its sampling rate.
Specifically, given the ADC sampling rate (the reciprocal of the \acell delay) we, absent detailed user input, use the median energy-per-conversion at that sampling rate as the estimation.
The total energy of non-linear \acell is thus obtained by the product of its estimated FoM and the number of required conversions:
\begin{equation}
    E_\text{a}^\text{cell,nl} = FoM\text{ [J/conversion]}\times \text{Num}_\text{conversion}
\end{equation}

The access counts to a specific \acell are the number of times the \acell is used along both the spatial and temporal scale to generate one \acomp output:
\begin{equation}
\begin{aligned}
    \text{Num}_\text{access}^{\text{cell}[j]}=\text{Num}_\text{spatial}^{\text{cell}[j]}\times \text{Num}_\text{temporal}^{\text{cell}[j]}
\end{aligned}
\label{eq:analog_energy_cell_2}
\end{equation}
For example, if an \acell represents an static-biased SF in a pixel, $\text{Num}_\text{spatial}^{\text{static-SF}}$ would be the number of SFs in the pixel and $\text{Num}_\text{temporal}^{\text{static-SF}}$ would be the number of times the pixel charge is read out (e.g., 2 if correlated double sampling is used to reduce noise~\cite{cds}).
The access counts information for \acell{s} is hard-coded for each \acomp and is abstracted away from typical users, but can be updated for a custom design.

\input{table/validation_examples}
\input{table/fig_validation_examples}

\subsection{Digital Energy Modeling}
\label{sec:model:digital}


The digital energy of a frame $E_{\text{d}}^{\text{frame}}$ is the sum of the computation energy of each compute unit $E_{\text{d}}^{\text{comp}}$ and the energy of each memory structure $E_{\text{d}}^{\text{mem}}$:
\begin{equation} \label{eq:d_inf_energy}
E_{\text{d}}^{\text{frame}} = \sum_{i} E_{\text{d}}^{\text{comp}[i]} + \sum_{j} E_{\text{d}}^{\text{mem}[j]}
\end{equation}


The energy of each compute unit is the product of the energy per cycle $E_{\text{d}}^{\text{cycle}}$ and the number of cycles $\text{Num}_{\text{cycle}}$:
\begin{equation}
\label{eq:E_c_d_2}
E_{\text{d}}^{\text{comp}[i]} = E_{\text{d}}^{\text{cycle}[i]} \times \text{Num}_{\text{cycle}}
\end{equation}

We rely on users to provide $E_{\text{d}}^{\text{cycle}}$, which usually is obtained through HLS/ASIC synthesis flows. The cycle counts, in contrast, are obtained through cycle-level simulation by \proj.


The energy consumption for memory accesses is the sum of leakage energy and the dynamic access energy; the latter is the product of the energy consumption for one memory read $E_{\text{d}}^{\text{read}}$ or write ($E_{\text{d}}^{\text{write}}$) and the total number of memory reads ($\text{Num}_{\text{read}}$) or writes ($\text{Num}_{\text{write}}$):
\begin{align}
E_{\text{d}}^{\text{mem}[j]} &= E_{\text{d}}^{\text{read}[j]} \times \text{Num}_{\text{read}}^{[j]} + E_{\text{d}}^{\text{write}[j]} \times \text{Num}_{\text{write}}^{[j]} \nonumber\\
&+ P_{\text{d}}^{\text{leakage}[j]} \times \frac{1}{\text{FR}} \times \alpha
\end{align}

The leakage energy is the product of the leakage power $P_\text{d}^{\text{leakage}}$ and the memory active time (i.e., not power-gated), which is a fraction $\alpha$ of the frame time $\tfrac{1}{\text{FR}}$.
Users supply the dynamic read/write energy and leakage power; the access counts and the active time are from the \proj simulation.

\subsection{Communication Energy Modeling}
\label{sec:model:comm}

The communication power is dominated by the energy to transfer the data outside the sensor using the energy-hungry MIPI CSI-2 interface and, in the case of 3D-stacking CIS, the energy of $\mu$TSV.
In literature the energy of the two interfaces is usually given for energy per Byte.
Therefore, the communication energy is given by: 
\begin{align}
    E_{\text{c}}^{\text{frame}} = E_{\text{c}}^{\text{mipi}} \times \text{Num}_{\text{Bytes}}^{\text{mipi}} + E_{\text{c}}^{\text{tsv}} \times \text{Num}_{\text{Bytes}}^{\text{tsv}}
\end{align}

$E_{\text{c}}^{\text{mipi}}$ and $E_{\text{c}}^{\text{tsv}}$ are user supplied with represented data reported in the literature~\cite{liu2019intelligent}.
The data volume statistics in both interfaces are generated in \proj simulation (based on the algorithm description and algorithm to hardware mapping).

%% file: table/validation_examples.tex
\begin{table*} 

\caption{Summery of CIS designs for validation, which cover a wide range of design variations. $^\ast$indicates data not reported in the original papers and are based on our educated guess. Unit of analog memory size is ``number of  analog values''.}
\resizebox{\textwidth}{!}{
\renewcommand*{\arraystretch}{1}
\renewcommand*{\tabcolsep}{4pt}
\begin{tabular}{ c|cc|ccccc|cc } 
\toprule[0.15em]
\textbf{\multirow{2}{*}{CIS}} & \textbf{\multirow{2}{*}{Process Node}}& \textbf{\multirow{2}{*}{Stacked}} & \multicolumn{5}{c|}{\textbf{Analog}}  & \multicolumn{2}{c}{\textbf{Digital}} \\ 
 &  & & Pixel & Memory & PE Operation & PE Position & Op Domain  & Memory & PE Size \\
\midrule[0.05em]
ISSCC'17~\cite{bong2017low} & 65nm & No & 3T APS & $20\times80$ & Avg\&Add & Column\&Chip & Charge\&Voltage  & 160KB & $4\times4\times64$ \\
JSSC'19~\cite{young2019data} & 130nm & No & 4T APS & $4\times240$ & Logarithmic Sub. & Column & Voltage  & - & - \\
Sensors'20~\cite{choi2020design} & 110nm & No & 4T APS & No & MAC\&MaxPool & Column & Voltage & - & - \\
ISSCC'21~\cite{eki20219} & 65nm/22nm & Yes & 4T APS$^\ast$ & No & - & - & - & 8MB & $1 \times 2304$\\
JSSC'21-I~\cite{hsu202005} & 180nm & No & PWM & No & MAC & Column & Time\&Current &  - & - \\
JSSC'21-II~\cite{park202151} & 110nm & No & 4T APS & No & MAC & Column & Charge & - & - \\
VLSI'21~\cite{seo2021} & 65nm/28nm & Yes & DPS & No & - & - & - & 6MB & -\\
ISSCC'22~\cite{hsu202208} & 180nm & No & PWM & No & MAC & Column & Time\&Current & 256B$^\ast$ & 1 \\
TCAS-I'22~\cite{xu2021senputing} & 180nm & No & 3T APS & No & Mul.\&Add & Pixel\&Chip & Current & - & - \\
\bottomrule[0.15em]
\end{tabular}
}
\label{tab:validation_summary}
\vspace{-5pt}
\end{table*}

%% file: table/fig_validation_examples.tex
\begin{figure*}[t]
\vspace{-10pt}
\centering
\subfloat[Energy Correlation]{
	\label{fig:overall_validation}
	\includegraphics[width=0.38\columnwidth]{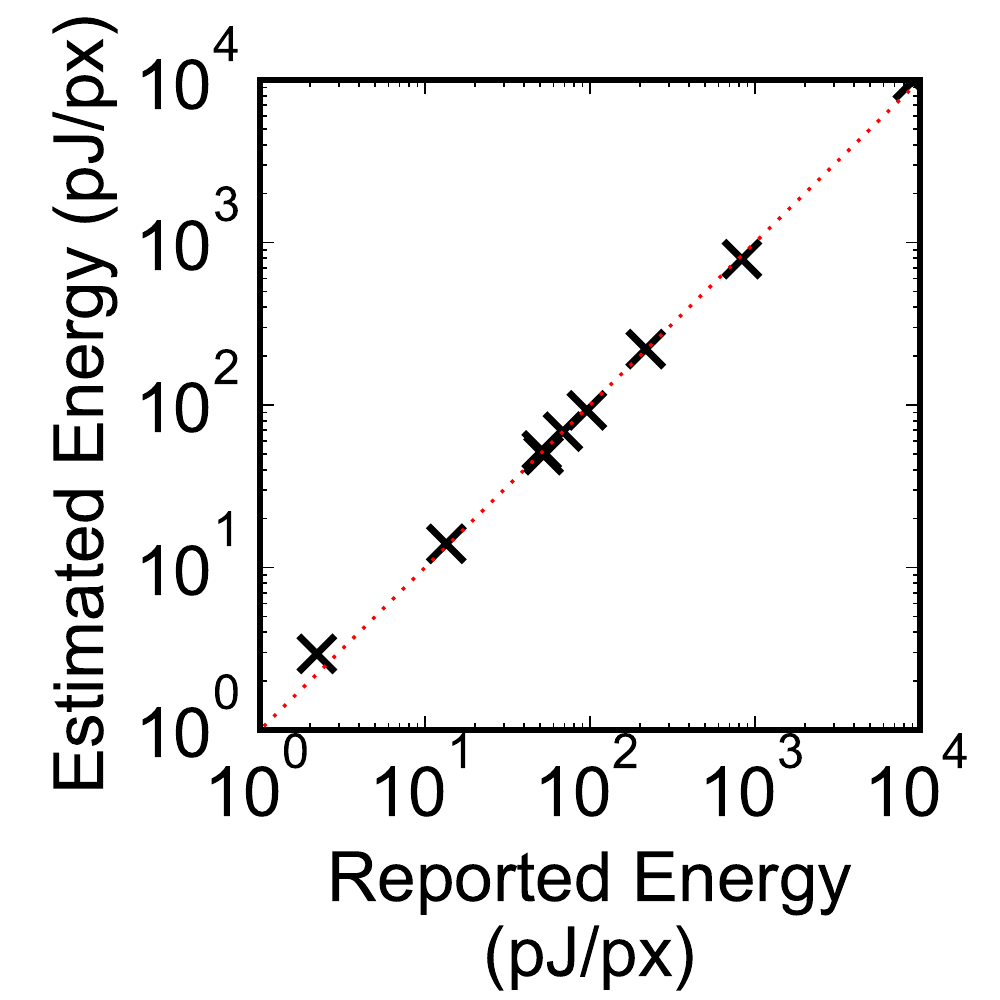} } 
\subfloat[ISSCC'17~\cite{bong2017low}]{
	\label{fig:isscc_17}	
	\includegraphics[width=0.38\columnwidth]{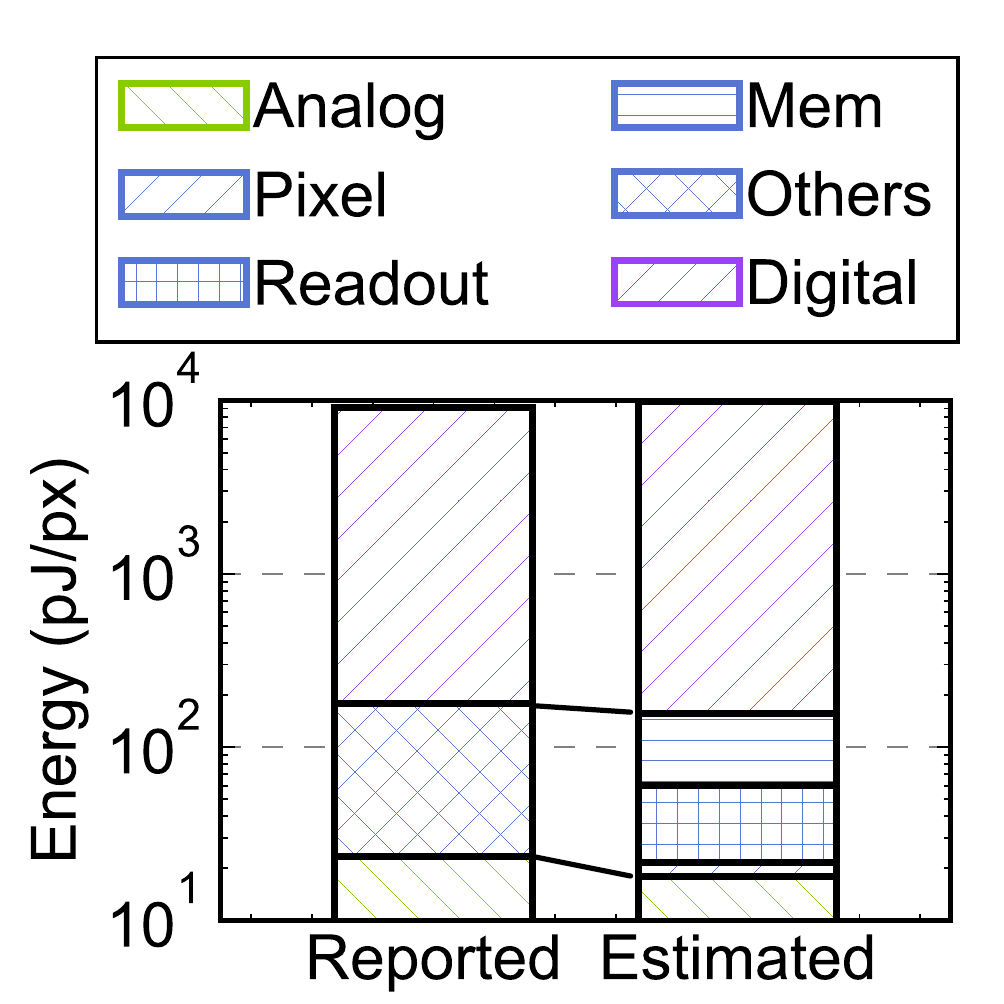} }
\subfloat[JSSC'19~\cite{young2019data}]{
	\label{fig:jssc_19}
	\includegraphics[width=0.38\columnwidth]{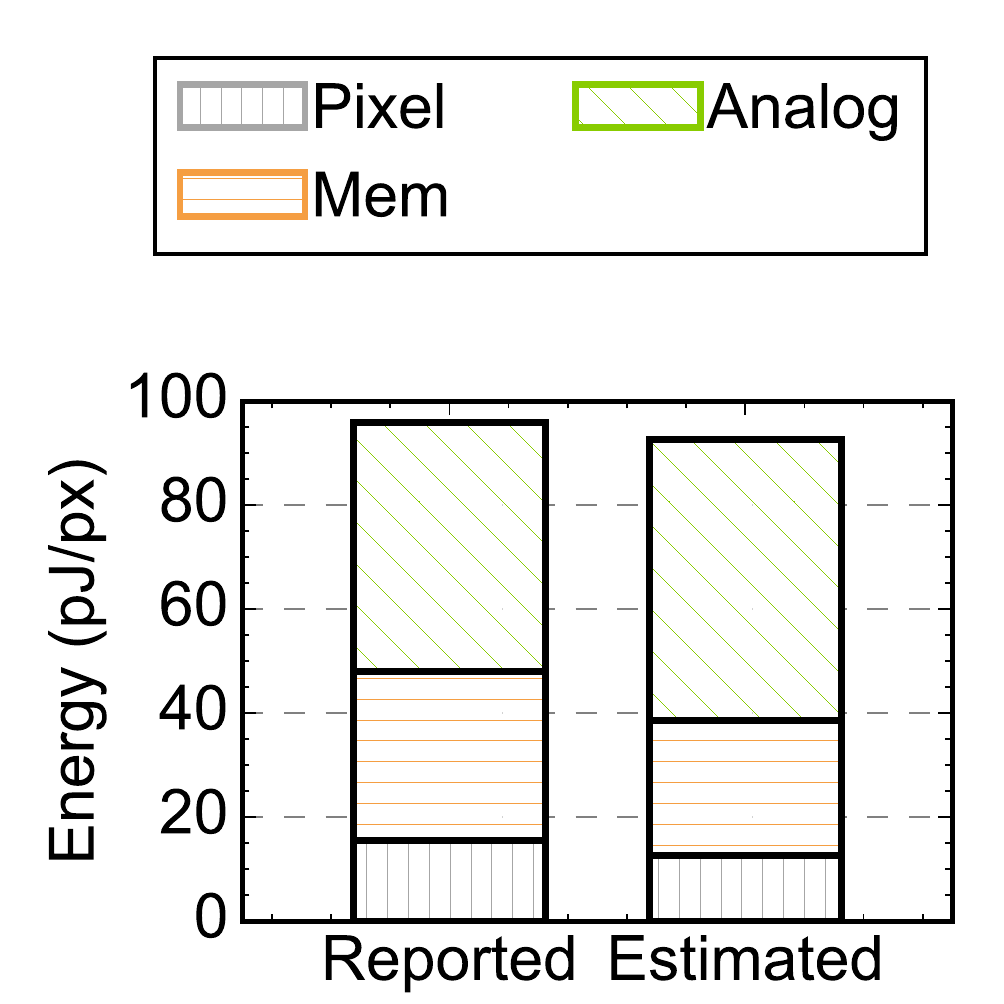} }
\subfloat[Sensors'20~\cite{choi2020design}]{
	\label{fig:sensors_20}	
	\includegraphics[width=0.38\columnwidth]{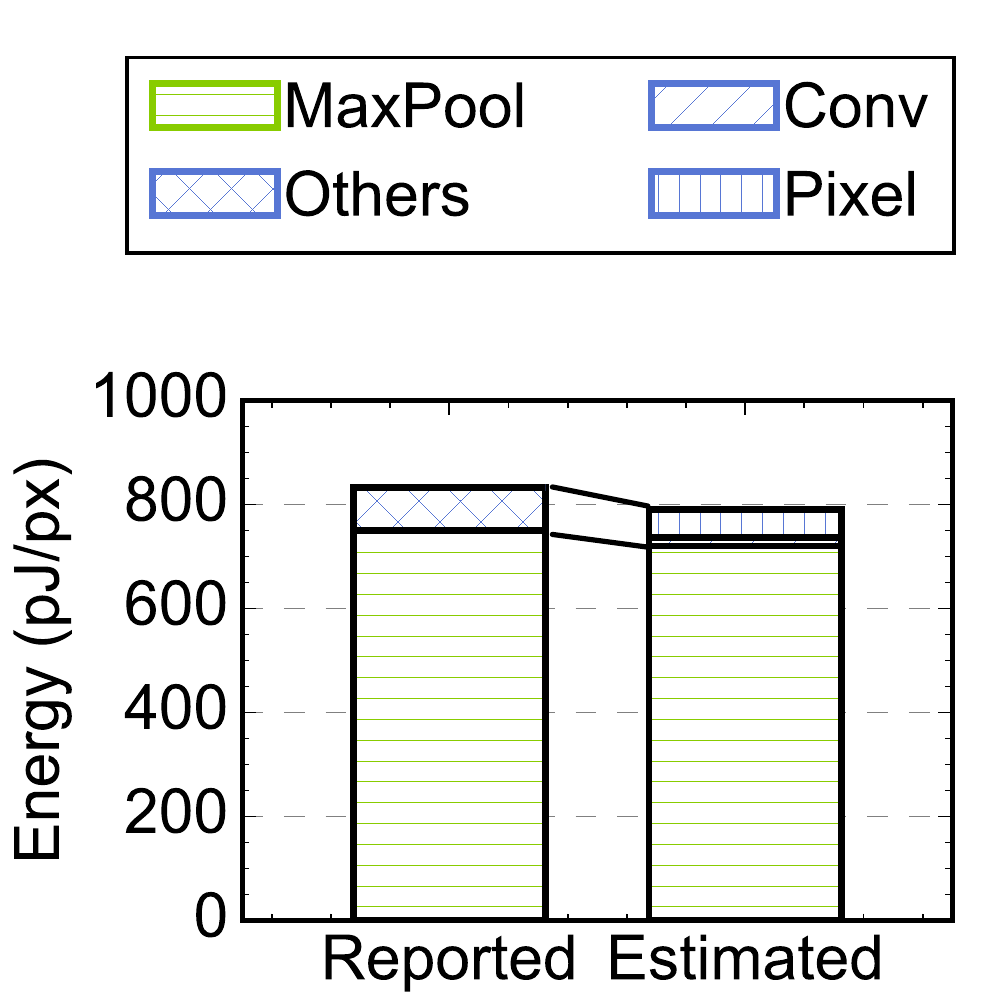} }
\subfloat[ISSCC'21~\cite{eki20219}]{
	\label{fig:isscc_21}
	\includegraphics[width=0.38\columnwidth]{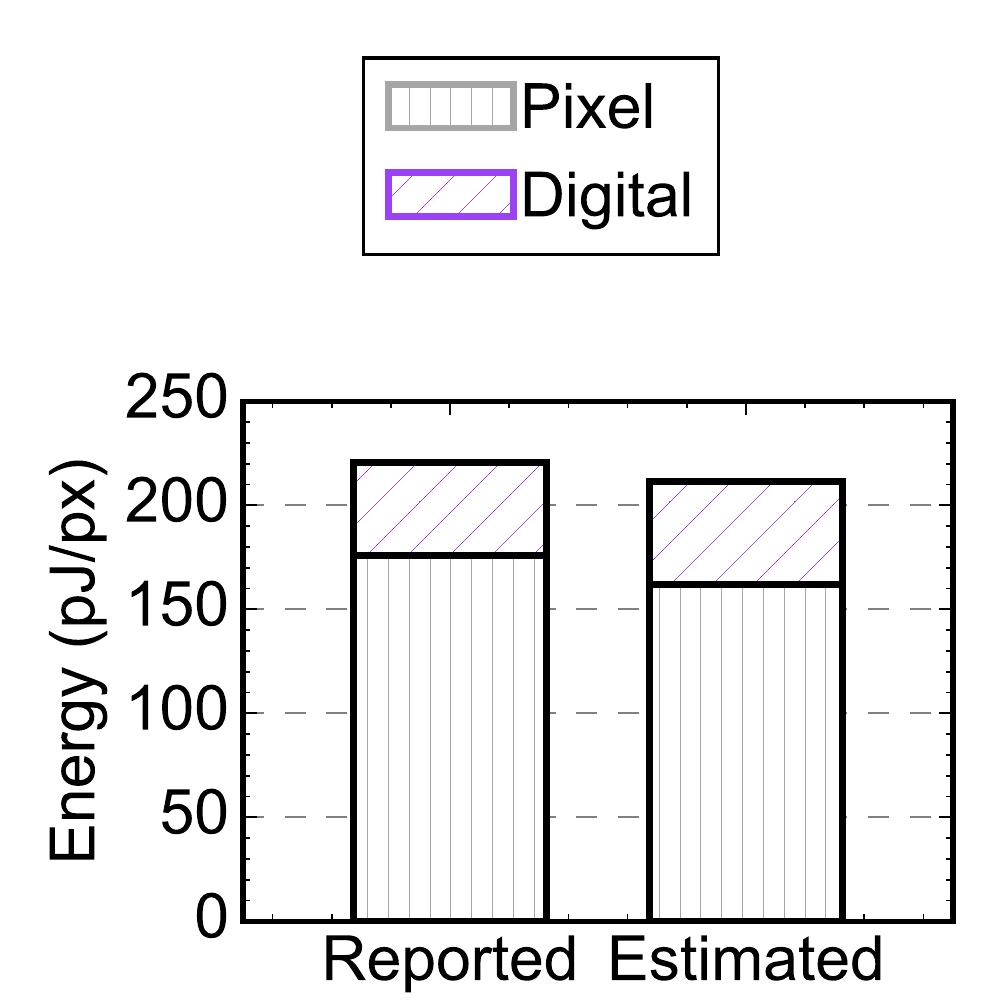}
}
\\
\subfloat[JSSC'21-I~\cite{hsu202005}]{
	\label{fig:jssc_21_1}
	\includegraphics[width=0.38\columnwidth]{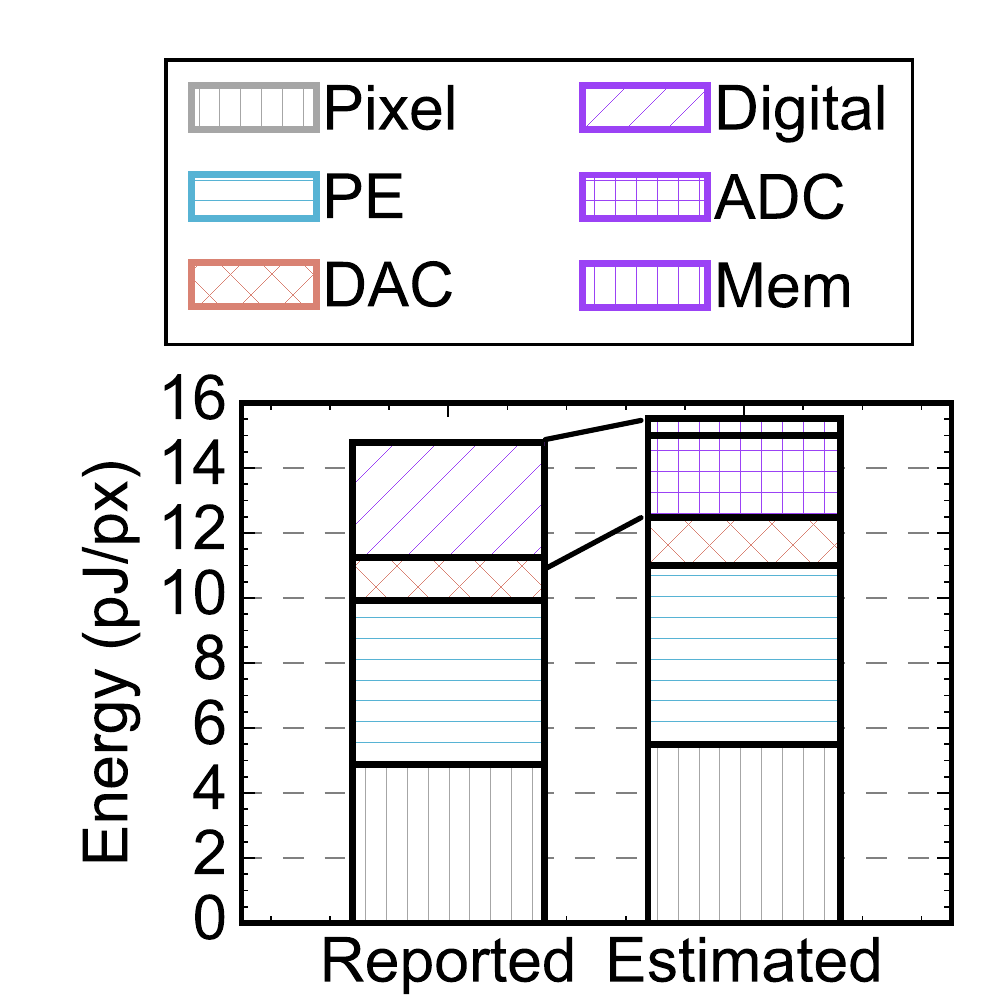}
} 
\subfloat[JSSC'21-II~\cite{park202151}]{
	\label{fig:jssc_21_2}	
	\includegraphics[width=0.38\columnwidth]{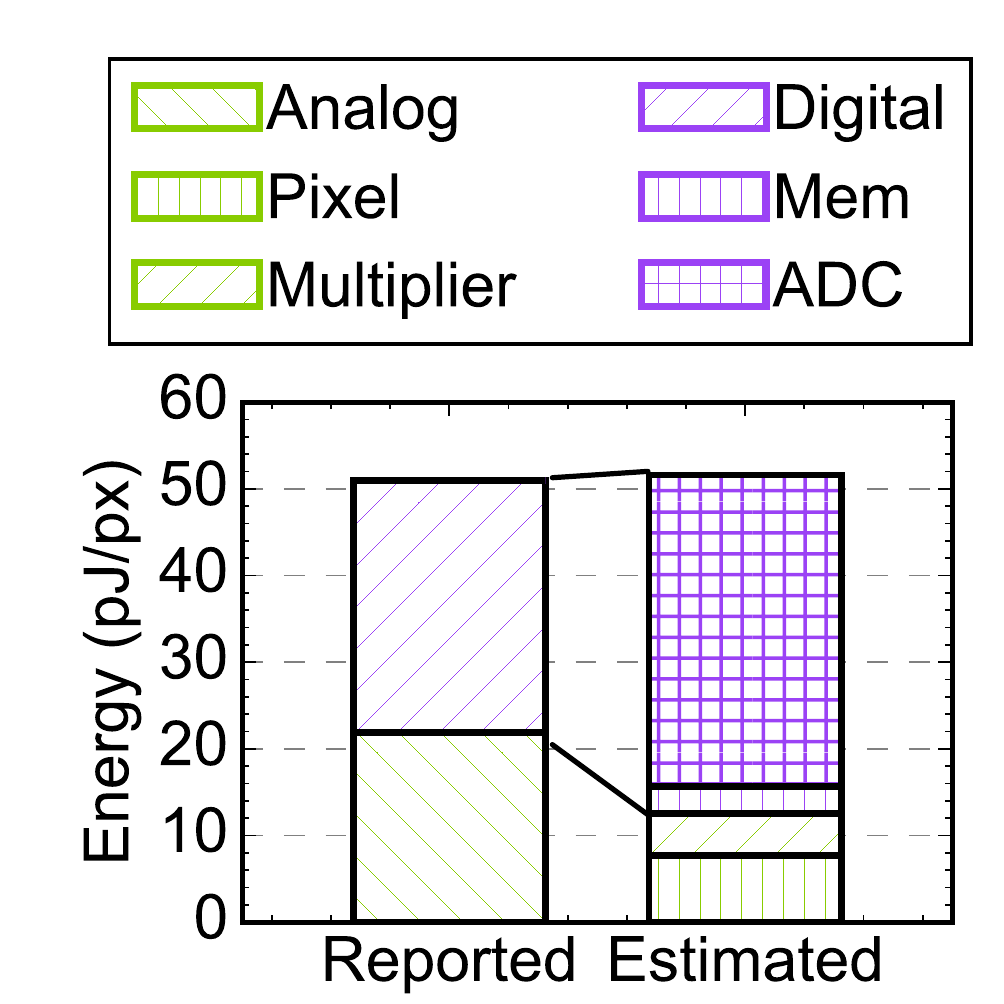} }
\subfloat[VLSI'21~\cite{seo2021}]{
	\label{fig:vlsi_21}	
	\includegraphics[width=0.38\columnwidth]{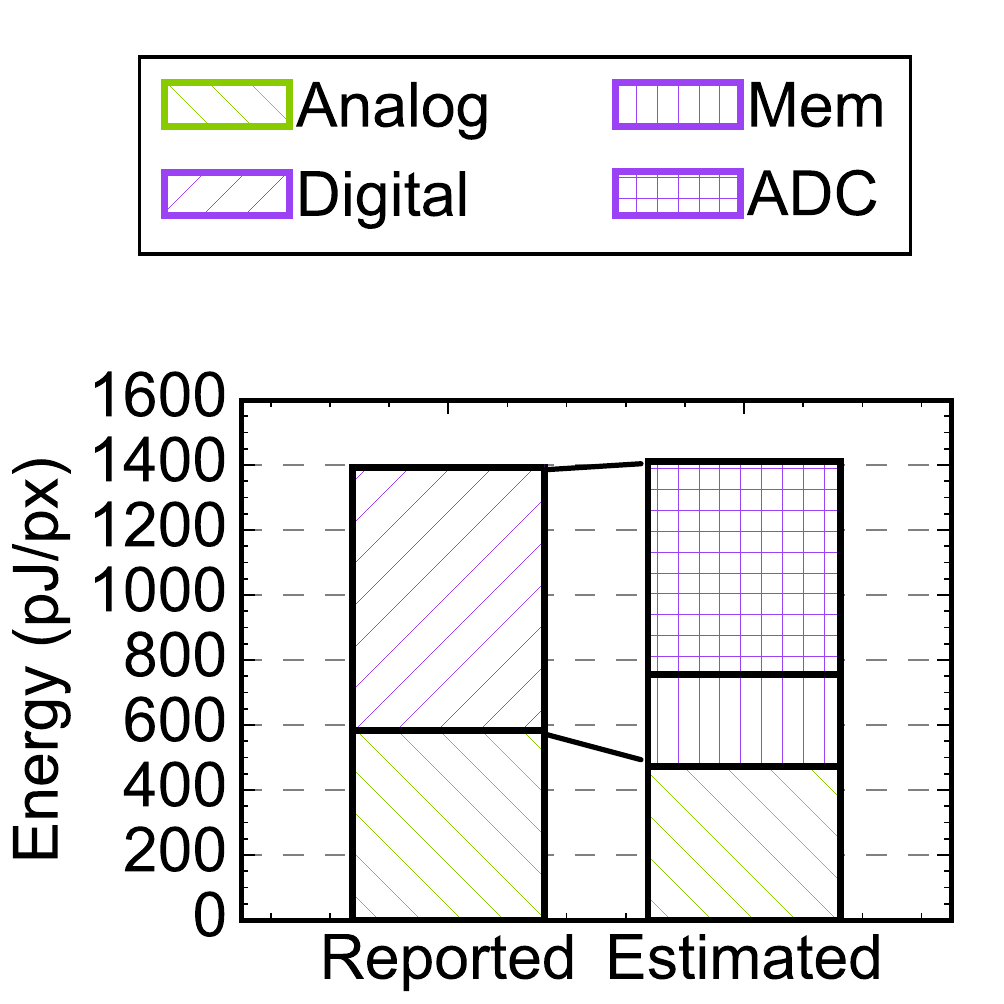} }
\subfloat[ISSCC'22~\cite{hsu202208}]{
	\label{fig:isscc_22}
	\includegraphics[width=0.38\columnwidth]{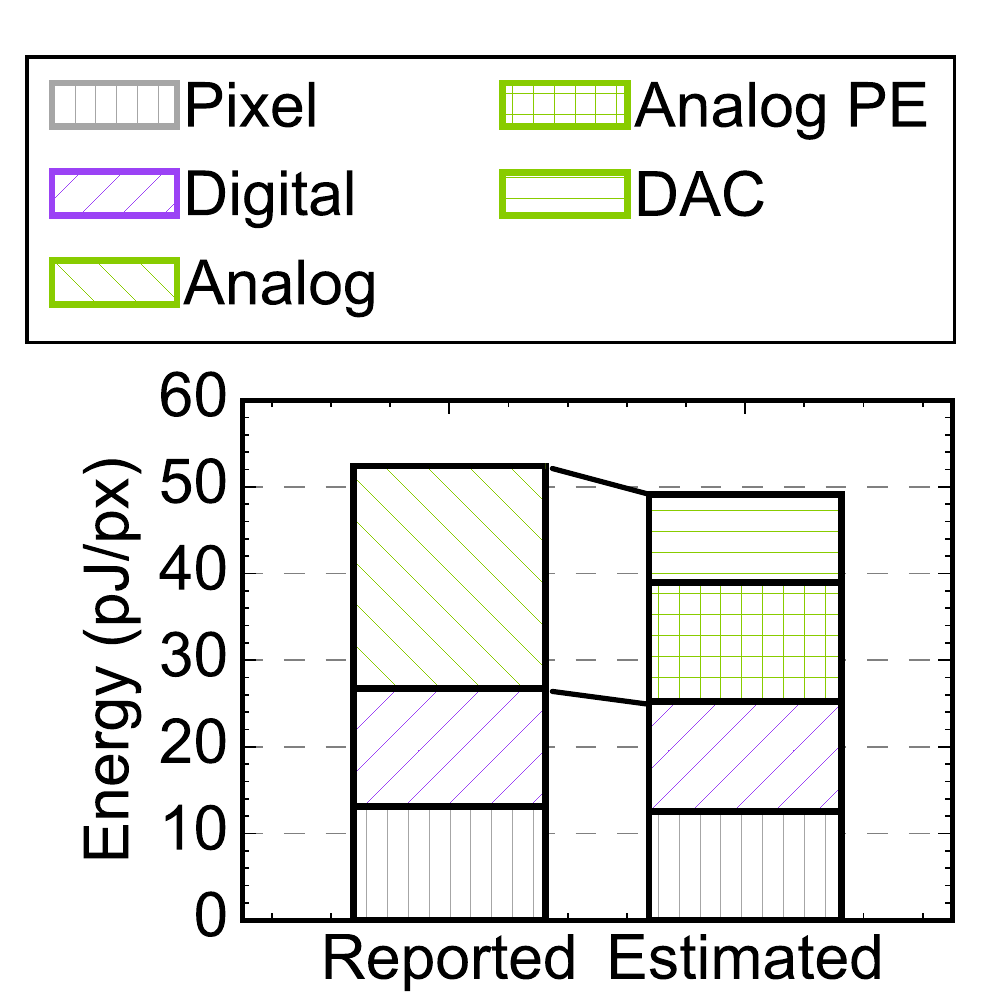} } 
\subfloat[TCAS-I'22~\cite{xu2021senputing}]{
	\label{fig:tcas_i_22}	
	\includegraphics[width=0.38\columnwidth]{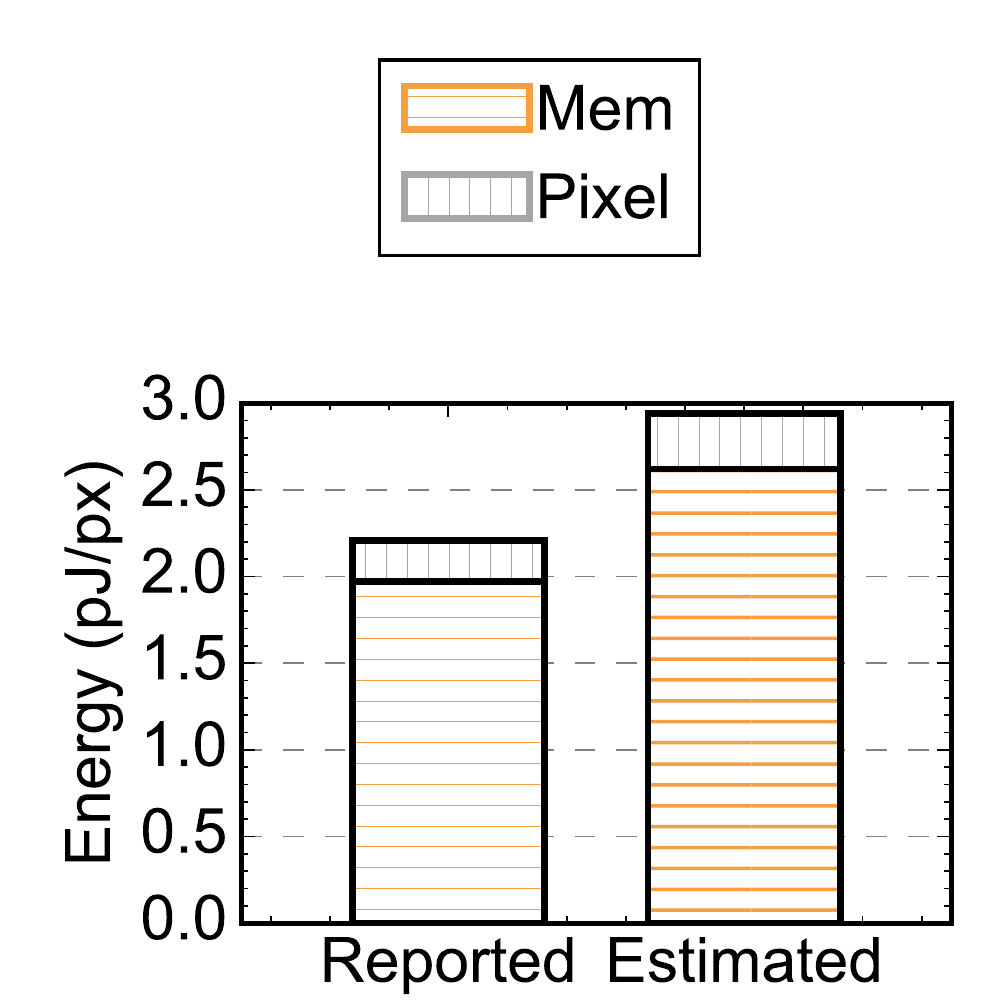} }
\caption{Validation results. \proj achieves a Pearson correlation coefficient of 0.9999. Several papers lump different components into the coarse-grained ``Analog'', ``Digital'', or ``Others'' categories. We show detailed breakdown and indicate when the sum of several fine-grained categories in our estimation corresponds to a coarse-grained category in the original papers.}
\vspace{-10pt}
\label{fig:validation}
\end{figure*}

%% file: eval.tex
\section{\proj Validation}
\label{sec:val}

In this section, we validate \proj against real measurement data from nine recent CIS chips~\cite{bong2017low, young2019data, choi2020design, eki20219, hsu202005, park202151, seo2021, hsu202208, xu2021senputing} shown in Table~\ref{tab:validation_summary}.
These designs span a range of design dimensions including 2D and 3D designs, different process nodes, pixel types, as well as PE designs and memory sizes in the analog and digital domains.


\Fig{fig:validation} compares the estimated and actual energy per pixel reported in the original papers.
Our estimations closely match the measured results, which span several orders of magnitude, showing both the diversity of the CIS design styles and the wide system power/energy scale that \proj can flexibly support and accurately model.
Across all designs, \proj achieves a Mean Absolute Percentage Error of 7.5\% and a Pearson Correlation Coefficient of 0.9999.

\Fig{fig:isscc_17} -- \Fig{fig:tcas_i_22} compare the detailed energy breakdown across the nine designs.
Whenever possible, we use the circuit parameters reported in the papers.
For SRAMs, we use DESTINY~\cite{poremba2015destiny} to obtain per-access energy.
The three papers that perform digital computation all execute DNNs, where a PE is a MAC unit;
we use the synthesis result of a 65~nm MAC unit design for per-MAC energy~\cite{bong2017low}, and scale it to other process nodes based on classic CMOS scaling~\cite{stillmaker2017scaling, sarangi2021deepscaletool}.

While overall \proj provides an accurate  component-level and full-system energy estimation, we find two key reasons behind result mismatches.
First, the results are less accurate when \proj does not have access to detailed design parameters.
For example, the pixel estimation in \Fig{fig:jssc_21_1}, \Fig{fig:jssc_21_2}, and \Fig{fig:tcas_i_22} shows an absolute error of 12.4\%, 38.9\%, and 33.3\%, respectively, due to insufficient circuit parameters on pixel ramp-generator (\Fig{fig:jssc_21_1}), pixel parasitic capacitance (\Fig{fig:jssc_21_2}), and photodiode voltage swing (\Fig{fig:tcas_i_22}). 
Similarly, the analog PE in \Fig{fig:jssc_21_1} and \Fig{fig:isscc_17} shows an absolute error of 9.3\% and 23.7\% due to insufficient circuit parameters on sampling capacitance (\Fig{fig:jssc_21_1}) and sense amplifier conversion energy (\Fig{fig:isscc_17}), respectively.
In contrast for \Fig{fig:jssc_19}, where the detailed design parameters are provided for the analog PE, the estimation error is only 0.4\%.



The other source of inaccuracy comes from the mismatch between the actual circuit design and \proj's default circuit template.
For example, the ADCs in \Fig{fig:jssc_21_2} and \Fig{fig:vlsi_21} show an absolute difference of 31.7\% and 16\%, respectively
\footnote{Both papers consider ADC as a digital unit, which is what we use here.};
the original designs use low-power dynamic technique (\Fig{fig:jssc_21_2}) whereas \proj estimates the energy of ADC based on the FOM survey~\cite{adc_fom}.
The memory in \Fig{fig:tcas_i_22} shows an estimation error of 33.0\% because the original design uses customized 8T SRAMs while \proj uses standard 6T SRAMs from DESTINY~\cite{poremba2015destiny}, resulting in higher leakage power.

%% file: use_case.tex
\section{Architectural Exploration}
\label{sec:usecases}

We now demonstrate three complementary examples of using \proj to explore architectural trade-offs.
The three use-cases complement each other. First, we explore when moving computation inside the CIS brings energy benefits (\Sect{sec:usecase:insensor}). We show that the energy reduction is small in a conventional 2D design.
We then show that introducing 3D stacking, which allows for hybrid integration, further improve the energy efficiency at a cost of higher power density (\Sect{sec:usecase:3d}).
Finally, we discuss the trade-off of moving computation into the analog domain (\Sect{sec:usecase:analog}).

\subsection{Computing Inside vs. Off Sensor}
\label{sec:usecase:insensor}

\begin{figure}
    \centering
    \subfloat[Rhythmic.]{
	\label{fig:rhythmic_pipeline}	
	\includegraphics[height=1.9in]{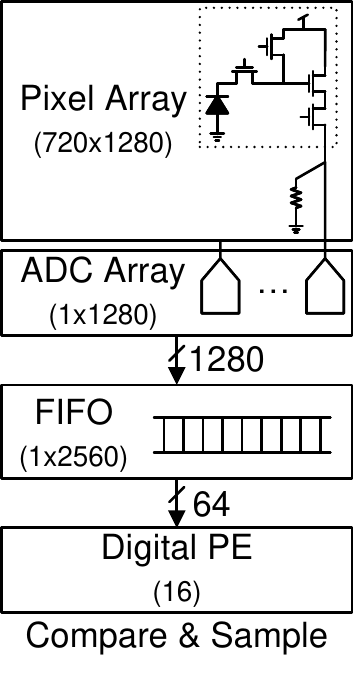} }
\hfill
    \subfloat[Edgaze.]{
	\label{fig:edgaze_pipeline}
	\includegraphics[height=1.9in]{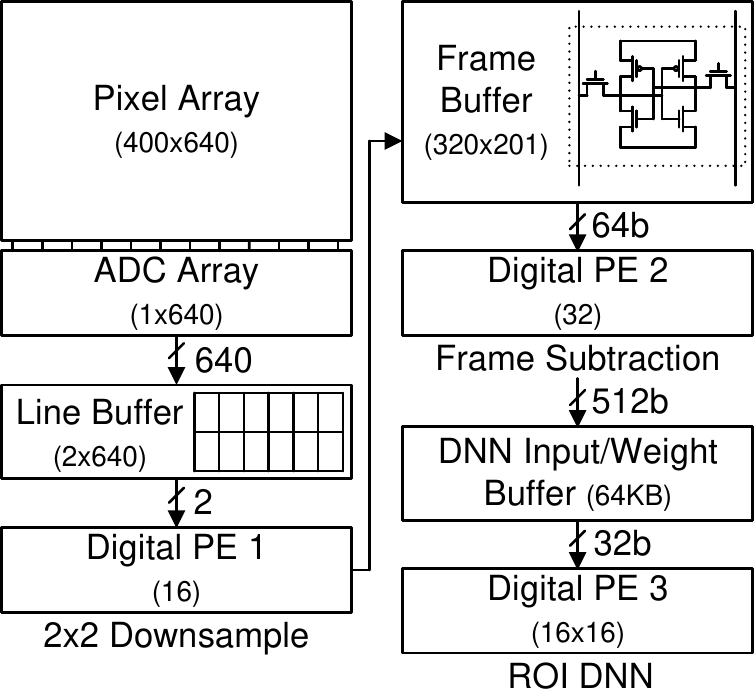} } 
    \caption{Hardware design for \protect\subref{fig:rhythmic_pipeline} Rhythmic Pixel Regions~\cite{kodukula2021rhythmic}, which is a ROI-based image encoder and \protect\subref{fig:edgaze_pipeline} Ed-Gaze, which is a gaze tracking algorithm. In their original designs, everything after ADC takes place off CIS; we evaluate the benefits of bringing the entire execution into the CIS.}
    \label{fig:compute_diagram}
\end{figure}

\begin{figure}[t]
\centering
\subfloat[Results on Rhythmic Pixel Regions.]{
	\label{fig:rhythmic_pixel_comb}	
	\includegraphics[width=\columnwidth]{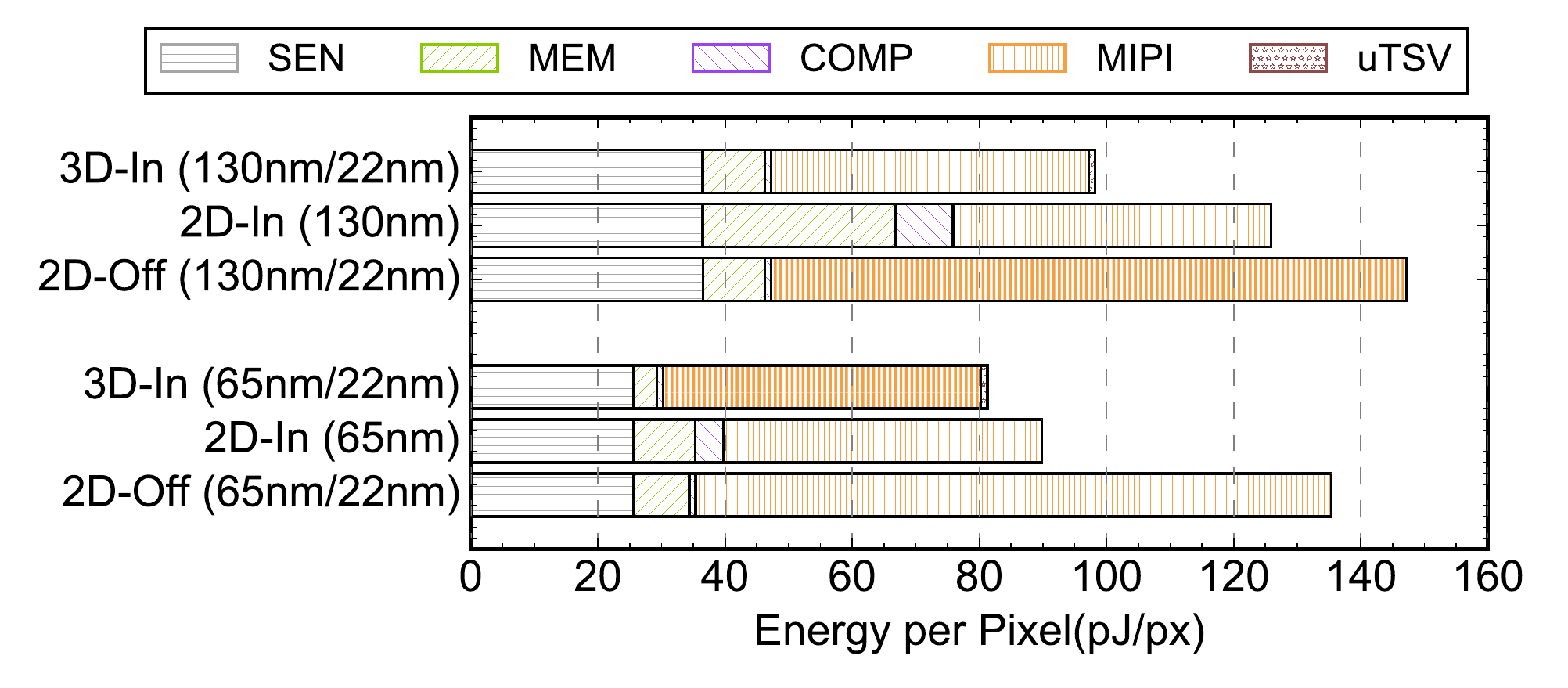} }
\\
\vspace{-5pt}
\subfloat[Results on Ed-Gaze.]{
	\label{fig:ieee_vr_comb}
	\includegraphics[width=\columnwidth]{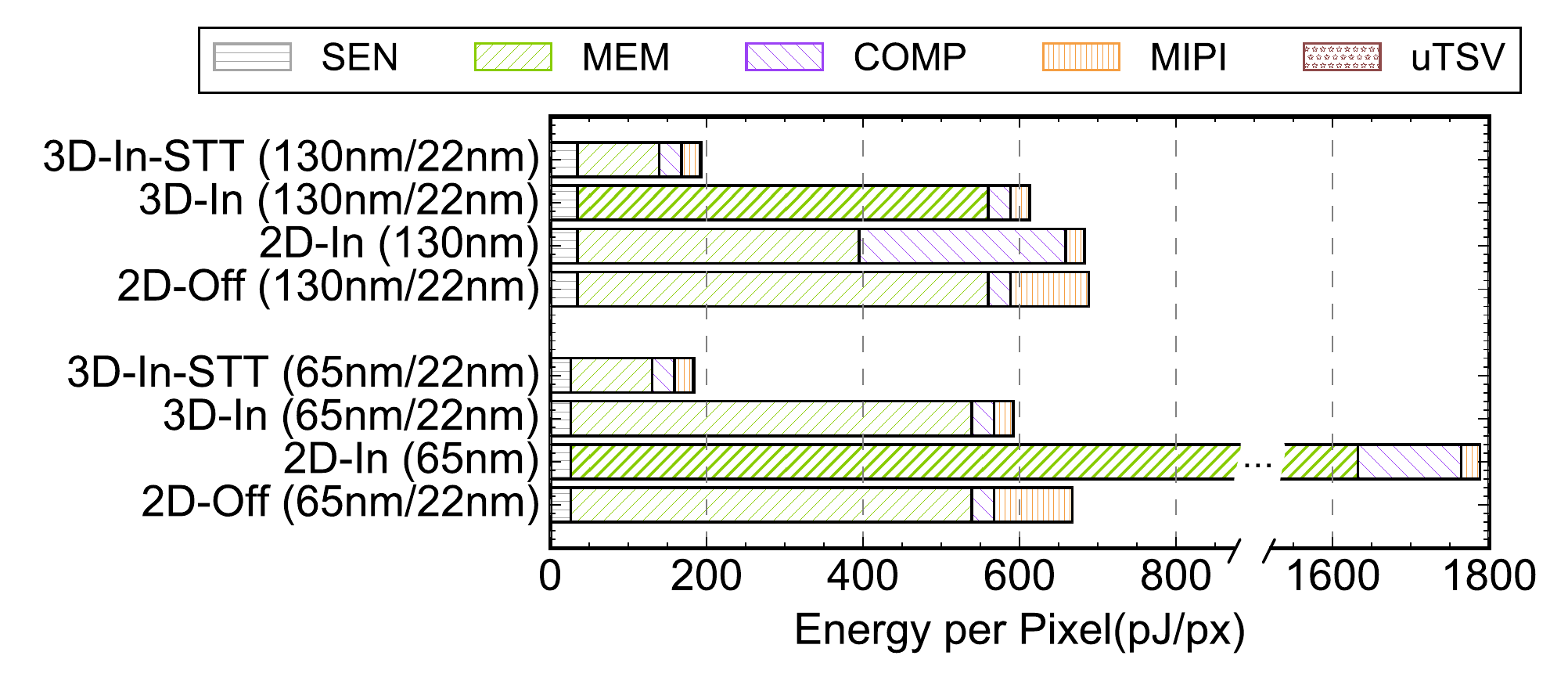} } 
\caption{Energy consumption comparison between 2D in-sensor (2D-In), 2D off-sensor (2D-Off), 3D in-sensor (3D-In), and 3D in-sensor with STT-RAM (3D-In-STT) processing. SEN: everything up to and including ADCs; MEM: memory; COMP: computation; MIPI: MIPI CSI-2 communication; uTSV: $\mu$TSV communication between layers.}
\label{fig:comp_insensor_vs_out}
\end{figure}

Computing inside CIS reduces the data transmission cost by consuming pixel data inside the sensor.
To explore the benefits, we evaluate two recent papers, Rhythmic Pixel Regions~\cite{kodukula2021rhythmic} and Ed-Gaze~\cite{feng2022}, both of which generate a small Region of Interest (ROI) from the original full-resolution image and, thus, can potentially benefit from in-sensor computing by moving the ROI generation inside the sensor.

\Fig{fig:rhythmic_pipeline} and \Fig{fig:edgaze_pipeline} illustrate the execution flow of the two algorithms, respectively. Everything after the ADC is executed outside the sensor in the original papers, and we use \proj to evaluate the energy consumption after moving the entire execution inside the sensor. Specifically:
\begin{itemize}
    \item \textbf{Rhythmic Pixel Regions}: a $1280\times720$ pixel array is  processed by a dedicated accelerator (Compare \& Sample) to generate the ROI, which on average reduces the image size by 50\%. The ROI generation performs roughly $7.4\times10^6$ arithmetic operations per frame.
    \item \textbf{Ed-Gaze}: a $640\times400$ pixel array is first downsampled by $2\times2$, and then processed by an pixel-wise subtraction operation with respect to the previous frame to generate an event map, which is then processed by a DNN to generate the ROI. The ROI, on average, reduces the image size by 25\%. The DNN dominates the computation and performs about $5.76\times10^7$ MAC operations per frame.
\end{itemize}

We use \proj to evaluate two hardware configurations:
\begin{itemize}
    \item \textbf{2D-In (H)}: a 2D CIS fabricated in the \textbf{H} process node; the entire execution is performed inside the CIS.
    \item \textbf{2D-Off (H/L)}: a 2D CIS fabricated in the \textbf{H} process node; everything after the ADC takes place on an SoC, which is fabricated in the \textbf{L} process node.
    We evaluate two CIS process nodes, 130 nm and 65 nm, both common in CIS designs (\Tbl{tab:validation_summary}). We set the SoC process node to 22 nm.
\end{itemize}

\Fig{fig:rhythmic_pixel_comb} shows the energy of Rhythmic Pixel Region under different designs.
Overall, \textbf{2D-In} reduces the energy compared to \textbf{2D-Off}. When CIS process node is 130nm, \textbf{2D-In} saves 14.5\% compared to \textbf{2D-Off}. This saving improves to 33.4\% when the CIS process node is 65nm.
The energy reduction comes from reducing the amount of data (full-resolution image vs. ROI) that has to be communicated through the MIPI CIS-2 interface (MIPI in the figure).
This data communication cost reduction comes at the expense of increasing the computation (COMP) and memory access (MEM) cost, both of which increase because of the older process node in the CIS.

Ed-Gaze's results are shown in \Fig{fig:ieee_vr_comb}, which tells a different story.
Computing inside the CIS ends up consuming much more energy than computing on the SoC.
This is because communication cost is light for Ed-Gaze:
the communication cost contributes to only 15.0\% of the total energy in the off-sensor system to begin with.
Thus, the additional energy costs of computation and memory accesses inside the sensor far outweigh the reduction in communication cost.
Interestingly, 65nm \textbf{2D-In} consumes more energy than its 130nm counterpart, because the 65nm node is known to have high leakage power~\cite{gielen2005analog}.
In Ed-Gaze, the frame buffer must be on during the entire frame time without power gating, since a frame must always be retained in order for frame subtraction.

\vspace{3pt}
\PBox{\textbf{Finding 1:} \textit{\hl{Computation inside CIS is less efficient than that off CIS. In-CIS computing saves energy only when the process node of the CIS is not too far behind that of the SoC and the application is communication-dominant.}}}

\subsection{Comparison of 2D CIS and 3D-stacked CIS}
\label{sec:usecase:3d}


We use \proj to explore the benefits of 3D stacked CIS.
We use the same algorithms in Rhythmic Pixel Regions and Ed-Gaze and consider two additional 3D configurations:

\begin{itemize}
    \item \textbf{3D-In (H/L)}: a two-layer stacked CIS, where the pixel layer is fabricated in the \textbf{H} process node and the compute layer is fabricated in an advanced process node \textbf{L}. All post-ADC operations take place in the compute layer.
    \item \textbf{3D-In-STT (H/L)}: 
    similar to \textbf{3D-In (H/L)} 
    except the SRAM in the compute layer is replaced with a STT-RAM, which we model using NVMExplore~\cite{pentecost2022nvmexplorer}.
    Rhythmic Pixel Regions lacks STT-RAM results, because it requires only a 2K memory, which NVMExplore does not support.
\end{itemize}

Comparing \textbf{3D-In} and \textbf{2D-In} in \Fig{fig:rhythmic_pixel_comb} shows that 3D integration reduces the energy by 15.8\% on average for Rhythmic Pixel Regions.
This is because the digital PEs and the SRAMs now use the same process node as that of the SoC; moving computation inside the sensor no longer increases the computation and memory energy but still enjoys the significant data volume reduction.
The additional cost of moving data via $\mu$TSV is insignificant, due to the low energy cost of $\mu$TSV.

The energy reduction from 3D integration is even higher for Ed-Gaze.
Comparing \textbf{3D-In} and \textbf{2D-In} in \Fig{fig:ieee_vr_comb}, 3D stacking reduces the energy by 38.5\% on average, because memory energy contributes to 71.3\% of the total energy in \textbf{2D-In},
which is significantly reduced when using a 22nm node.

That said, the memory energy still dominates in \textbf{3D-In}, because the frame buffer cannot be power-gated as explained before, consuming non-trivial leakage power.
To further reduce the memory energy, we explore STT-RAM, which is known to have low leakage. By replacing SRAM with STT-RAM, \textbf{3D-In-STT} further reduces the overall energy by 69.1\% and 68.5\% compared to \textbf{3D-In} under the 65nm/22nm and the 130nm/22nm combination, respectively.

\input{table/power_density}

\paragraph{Power Density.}
\hl{While 3D stacking is known to increase power density, we use \mbox{\proj} to show that the impact on power density is algorithm-specific.}
We use a conservative area estimation to obtain a power density upper bound while leaving a comprehensive area modeling to future work.
Specifically, we use the pixel array to approximate the analog area and use SRAM area to approximate the digital area.

\Tbl{tab:power_density} compares the power density across the three sensor variants for the two algorithms.
The power density of Rhythmic Pixel Regions shows no significant difference among the three variants, because the power of Rhythmic Pixel Regions is dominated by communication, which is not greatly affected by stacking.
In contrast, the power density of Ed-Gaze more than doubles by 3D stacking under the 130nm/22nm combination. This is because \textbf{3D-In} is 3$\times$ smaller in area than \textbf{2D-In}. 
Under the 65nm/22nm combination, \textbf{2D-In} has higher power density due to the high leakage power of 65nm process, which is avoided in the stacked design.

\hl{It is worth noting that the \textit{absolute} power density of 3D-stacked sensors is still very low, in fact three to four orders of magnitude lower than the power density of typical CPUs (up to 1$\text{W/mm}^\text{2}$~\mbox{\cite{danowitz2012cpu}}) and GPUs (up to 0.3$\text{W/mm}^\text{2}$~\mbox{\cite{chen2009gpu}}). Such a low power density will unlikely lead to thermal hotspots and create a cooling challenge~\mbox{\cite{yu2018designing}}. However, higher power density increases the thermal-induced noise and worsens the imaging and computing quality~\mbox{\cite{kodukula2021dynamic}}. End-to-end application optimization must take into account the noise impact, an exploration that \mbox{\proj} enables and that we leave to future work.}

\vspace{5pt}
\PBox{\textbf{Finding 2:} \textit{\hl{3D stacking saves energy but increases power density --- for compute-dominant applications. The absolute power density is not high enough to create thermal hotspots but could increase noise, warranting further studies.}}}





\subsection{Comparison of Analog and Digital Computing}
\label{sec:usecase:analog}

\begin{figure}[t]
    \centering
    \includegraphics[width=\columnwidth]{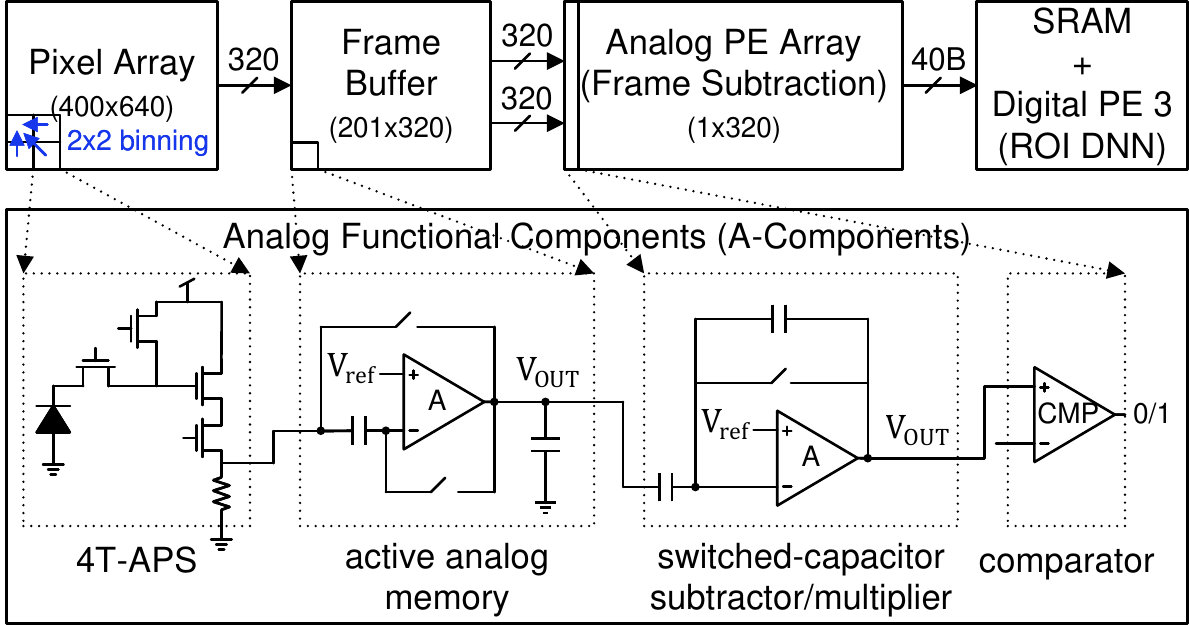}
    \caption{Mixed-signal CIS design for Ed-Gaze. We move the first digital stages in \Fig{fig:edgaze_pipeline} to the analog domain.}
    \label{fig:analog_edgaze}
\end{figure}

\begin{figure}[t]
    \centering
    \includegraphics[width=\columnwidth]{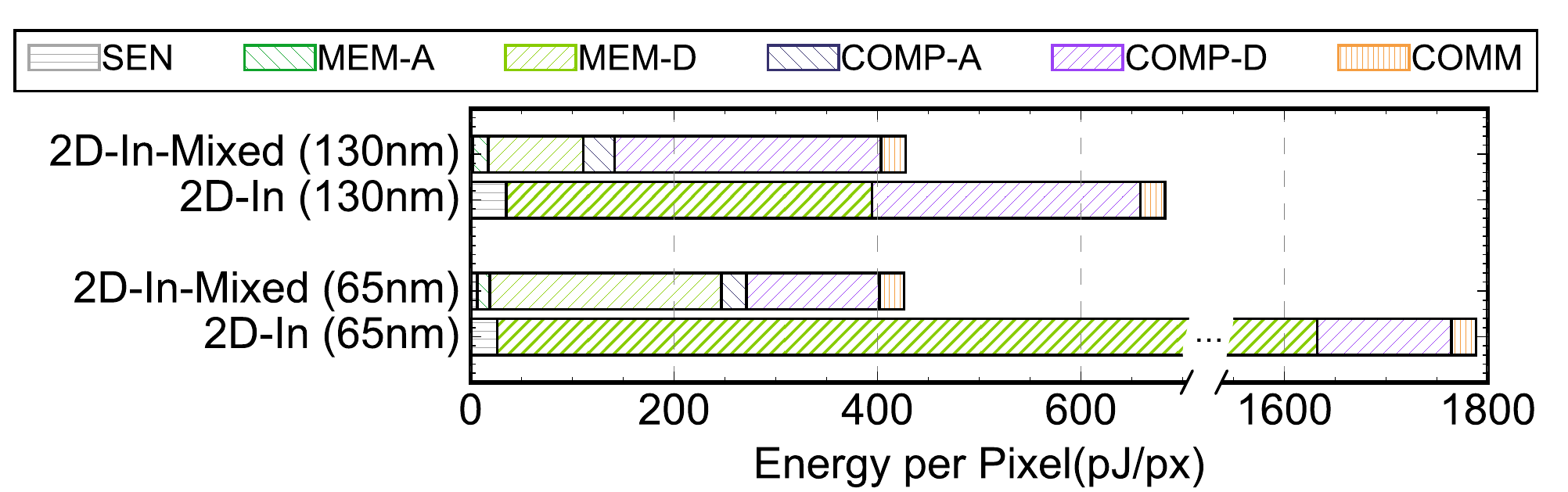}
    \caption{Energy comparison between mixed-signal in-sensor computation and fully-digital in-sensor computation on Ed-Gaze. COMP/MEM-D: digital compute and memory; COMP/MEM-A: analog compute and memory.}
    \label{fig:ieee_vr_analog_vs_digital}
\end{figure}

We use \proj to explore the benefits of in-sensor analog computing.
In particular, we use Ed-Gaze for the case study and consider a mixed-signal configuration:
\begin{itemize}
    \item \textbf{2D-In-Mixed}: a 2D CIS, where the first two stages in the algorithm (\Fig{fig:edgaze_pipeline}), $2\times2$ downsampling and frame subtraction, are implemented in analog while last stage (ROI DNN) is implemented in the digital domain.
\end{itemize}


\Fig{fig:analog_edgaze} shows how the Ed-Gaze is mapped to a mixed-signal CIS.
Inside the pixel array, the $2\times2$ downsampling is done through pixel binning (similar to that in \Fig{fig:code_example}).
The analog frame buffer stores the downsampled analog pixel values,
which are read by an analog PE array for frame subtraction.
Each analog PE consists of a switched-capacitor subtractor/ multiplier for absolute subtraction and a comparator for frame delta digitization.
The output of the Analog PE array enters the SRAM array, at which point the hardware is the same as that in \textbf{2D-In}.
For a fair comparison and to ensure area overhead is well accounted for, we conservatively set all the capacitors to 100fF.
Despite the over-sizing, analog design still yields at least 27\% less area than the digital counterpart.

\begin{figure}[t]
\centering
\begin{minipage}[t]{0.48\columnwidth}
  \centering
  \includegraphics[width=\columnwidth]{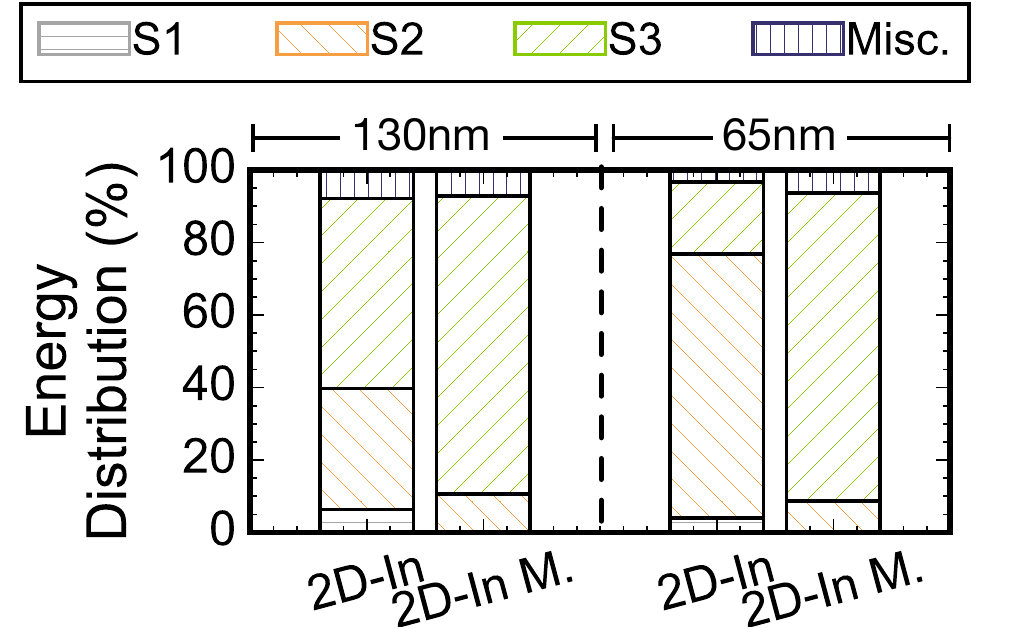}
  \caption{Normalized energy breakdown among the three stages (S1, S2, S3).}
  \label{fig:energy_pct}
\end{minipage}
\hspace{2pt}
\begin{minipage}[t]{0.48\columnwidth}
  \centering
  \includegraphics[width=\columnwidth]{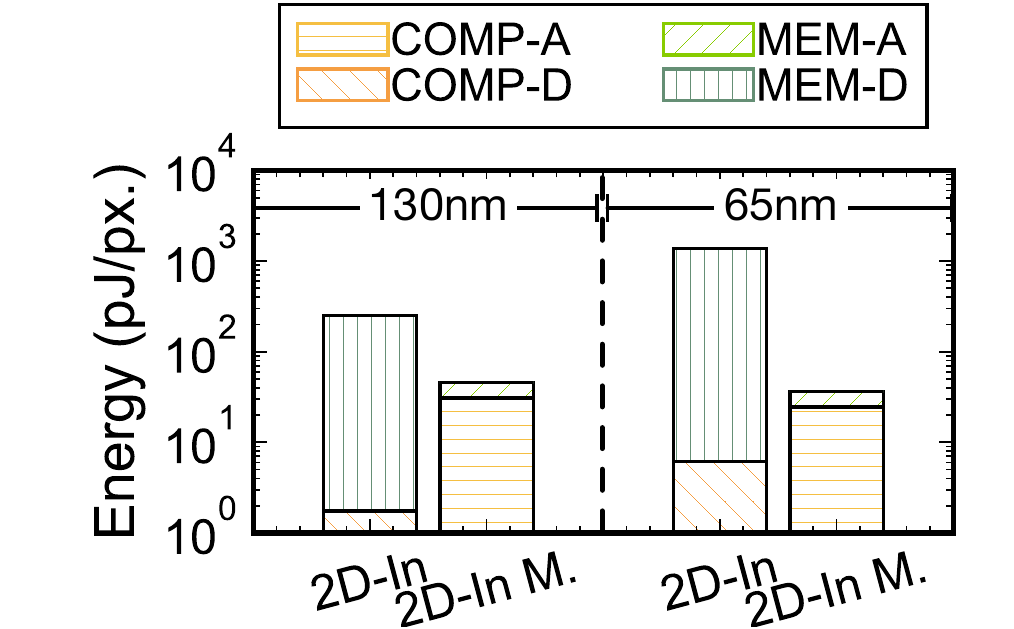}
  \caption{Energy breakdown of first two stages.}
  \label{fig:energy_details}
\end{minipage}
\end{figure}

\Fig{fig:ieee_vr_analog_vs_digital} compares \textbf{2D-In-Mixed} and \textbf{2D-In}.
Moving the first stages of the Ed-Gaze algorithm to the analog domain reduces the energy by 38.8\% and 77.1\%.
The energy reduction comes from two sources: removing the ADCs (indicated by lower SEN) and replacing SRAMs in the first two stages with analog buffers (indicated by lower MEM-D).
The reduction in MEM-D is particularly significant for the 65nm node, where the SRAM leakage power is high.
To corroborate the results, 
 \Fig{fig:energy_pct} shows the normalized energy breakdown among the three stages (S1, S2, and S3).
S3 (DNN) becomes the dominant stage after moving first two stages into analog domain, showing the effectiveness of analog processing.

Interestingly, the energy reduction is obtained when the compute energy of the first two stages slightly increase.
\Fig{fig:energy_details} shows the energy breakdown of the first two stages.
While the memory energy reduces, the compute energy increases in the mixed-signal mode.
This is because to maintain an 8-bit precision the OpAmp consumes too much energy (\Equ{eq:thermal}).
A caveat is that the analog design 
presented here, which uses active switched-capacitor circuits, is based on our specific implementation choice.
It is conceivable that different designs would yield different efficiency results.

\vspace{5pt}
\PBox{\textbf{Finding 3:} \textit{\hl{While analog computing is known for reducing ADC and computing energy, the energy saving is also attributed to lower analog memory energy, especially for memory-intensive applications, which many in-CIS use-cases are.}
}}

%% file: table/power_density.tex
\begin{table} 
\caption{Power Density. Unit is $\text{mW/mm}^\text{2}$.}
\resizebox{\columnwidth}{!}{
\renewcommand*{\arraystretch}{1}
\renewcommand*{\tabcolsep}{5pt}
\begin{tabular}{ c|ccc|ccc } 
\toprule[0.15em]
  \textbf{Process Node} & \multicolumn{3}{c|}{\textbf{Rhythmic Pixel Regions}}  & \multicolumn{3}{c}{\textbf{Ed-Gaze}} \\ 
  Pixel/Compute & 2D-Off & 2D-In & 3D-In & 2D-Off & 2D-In & 3D-In \\
\midrule[0.05em]
  130nm/22nm & 0.05 & 0.09 & 0.06 & 0.19 & 0.30 & 0.78 \\
  65nm/22nm &  0.03 & 0.05 & 0.04 & 0.11 & 2.24 & 0.70 \\
\bottomrule[0.15em]
\end{tabular}
}
\label{tab:power_density}
\end{table}

%% file: related.tex
\section{Related Work}
\label{sec:related}

Emerging CIS designs are discussed in \Sect{sec:bck} and \Sect{sec:val}.


\paragraph{Power Modeling.}
Power/energy modeling is a cornerstone of architectural exploration.
Prior power models of CPUs~\cite{brooks2000wattch, li2009mcpat, shao2013energy}, GPUs~\cite{leng2013gpuwattch, kandiah2021accelwattch, hong2010integrated}, and memory~\cite{balasubramonian2017cacti, poremba2015destiny, guthaus2016openram, pentecost2022nvmexplorer} have enabled a plethora of power/energy optimizations.
Fundamentally, \proj shares the same, bottom-up modeling methodology,
where energy is estimated from access counts and per-access energy.
Additionally, \proj provides a clean programming interface to integrate other architectural simulators~\cite{shao2014aladdin, samajdar2018scale, feng2019asv, gao2017tetris} and memory modeling tools~\cite{poremba2015destiny, pentecost2022nvmexplorer} to model bespoke accelerators and memories.

Prior analog power modeling requires either detailed transistor-level parameters~\cite{svensson2010power} or is based on the statistic models of particular analog circuits~\cite{Lauwers20002}.
Lim et al.~\cite{lim2019} decomposes a mixed-signal circuit into basic cells and accelerate the mixed-signal simulation by approximating the transfer function of each cell.
\proj uses a similar decomposition methodology but specifically targets CIS.

\paragraph{CIS Modeling.}
No comprehensive CIS modeling framework exists.
Two recent papers from Meta use first-order analytical model to estimate the energy of their custom CIS design, i.e., 3D stacking with DPS~\cite{gomez2022distributed, liu2019intelligent}.
It does not provide the level of flexibility to accommodate general CIS design and architecture  exploration as supported by \proj.

LiKamWa et al.~\cite{likamwa2013energy} provide a coarse-grained CIS power model using the idle and active period/power without considering the hardware implementation details.
\proj, instead, models the hardware with finer granularity to achieve finer-grained architectural exploration.
Kodukula et al.~\cite{kodukula2021dynamic} cite coarse-grained component energy of typical CIS designs and builds a thermal model.
\proj, can provide more accurate power/energy modeling that feeds into such thermal model.

\paragraph{Visual Computing Optimizations.}
Recent work discusses the possibility of processing inside an CIS to reduce the data transmission cost, e.g., Ed-Gaze~\cite{feng2022}, Rhythmic Pixel Regions~\cite{kodukula2021rhythmic}, Reid et al.~\cite{pinkham2020algorithm}, and SplitNets~\cite{dong2022splitnets}.
All, however, rely on first-order energy models.
Using \proj, we study the algorithms in Ed-Gaze and Rhythmic Pixel Regions and quantify their benefits (\Sect{sec:usecases}).
Many recent visual computing optimizations use motion vectors that  can be naturally generated during imaging to simplify downstream vision processing~\cite{zhu2018euphrates, feng2019asv}.
It is interesting to explore how motion estimation can be integrated into the CIS using \proj.
\proj can also be integrated with visual computing benchmarks~\cite{kwon2022xrbench, huzaifa2021illixr} to study in-CIS computing for different workloads.

%% file: conc.tex
\section{Conclusion}
\label{sec:conc}

\proj is a silicon-validated, component-level energy modeling and architecture exploration framework for computational CIS.
It accepts high-level descriptions of image processing algorithms and hardware designs, and enables exploration of architectural trade-offs, e.g., in vs. off CIS, 2D vs. 3D design, and analog vs. digital processing.
\proj is the first step in the promising field of in-sensor visual computing.

%% file: main.bbl

\begin{thebibliography}{74}


\ifx \showCODEN    \undefined \def \showCODEN     #1{\unskip}     \fi
\ifx \showDOI      \undefined \def \showDOI       #1{#1}\fi
\ifx \showISBNx    \undefined \def \showISBNx     #1{\unskip}     \fi
\ifx \showISBNxiii \undefined \def \showISBNxiii  #1{\unskip}     \fi
\ifx \showISSN     \undefined \def \showISSN      #1{\unskip}     \fi
\ifx \showLCCN     \undefined \def \showLCCN      #1{\unskip}     \fi
\ifx \shownote     \undefined \def \shownote      #1{#1}          \fi
\ifx \showarticletitle \undefined \def \showarticletitle #1{#1}   \fi
\ifx \showURL      \undefined \def \showURL       {\relax}        \fi
\providecommand\bibfield[2]{#2}
\providecommand\bibinfo[2]{#2}
\providecommand\natexlab[1]{#1}
\providecommand\showeprint[2][]{arXiv:#2}

\bibitem[ird({[n.\,d.]})]%
        {irds}
 \bibinfo{year}{[n.\,d.]}\natexlab{}.
\newblock \bibinfo{title}{International Roadmap for Devices and Systems}.
\newblock \bibinfo{howpublished}{\url{https://irds.ieee.org/}}.
\newblock


\bibitem[Akhlaghi et~al\mbox{.}(2018)]%
        {akhlaghi2018snapea}
\bibfield{author}{\bibinfo{person}{Vahideh Akhlaghi}, \bibinfo{person}{Amir
  Yazdanbakhsh}, \bibinfo{person}{Kambiz Samadi}, \bibinfo{person}{Rajesh~K
  Gupta}, {and} \bibinfo{person}{Hadi Esmaeilzadeh}.}
  \bibinfo{year}{2018}\natexlab{}.
\newblock \showarticletitle{Snapea: Predictive early activation for reducing
  computation in deep convolutional neural networks}. In
  \bibinfo{booktitle}{\emph{2018 ACM/IEEE 45th Annual International Symposium
  on Computer Architecture (ISCA)}}. IEEE, \bibinfo{pages}{662--673}.
\newblock


\bibitem[Balasubramonian et~al\mbox{.}(2017)]%
        {balasubramonian2017cacti}
\bibfield{author}{\bibinfo{person}{Rajeev Balasubramonian},
  \bibinfo{person}{Andrew~B Kahng}, \bibinfo{person}{Naveen Muralimanohar},
  \bibinfo{person}{Ali Shafiee}, {and} \bibinfo{person}{Vaishnav Srinivas}.}
  \bibinfo{year}{2017}\natexlab{}.
\newblock \showarticletitle{CACTI 7: New tools for interconnect exploration in
  innovative off-chip memories}.
\newblock \bibinfo{journal}{\emph{ACM Transactions on Architecture and Code
  Optimization (TACO)}} \bibinfo{volume}{14}, \bibinfo{number}{2}
  (\bibinfo{year}{2017}), \bibinfo{pages}{1--25}.
\newblock


\bibitem[Bigas et~al\mbox{.}(2006)]%
        {bigas2006review}
\bibfield{author}{\bibinfo{person}{M Bigas}, \bibinfo{person}{Enric Cabruja},
  \bibinfo{person}{Josep Forest}, {and} \bibinfo{person}{Joaquim Salvi}.}
  \bibinfo{year}{2006}\natexlab{}.
\newblock \showarticletitle{Review of CMOS image sensors}.
\newblock \bibinfo{journal}{\emph{Microelectronics journal}}
  \bibinfo{volume}{37}, \bibinfo{number}{5} (\bibinfo{year}{2006}),
  \bibinfo{pages}{433--451}.
\newblock


\bibitem[Bong et~al\mbox{.}(2017a)]%
        {bong2017low}
\bibfield{author}{\bibinfo{person}{Kyeongryeol Bong}, \bibinfo{person}{Sungpill
  Choi}, \bibinfo{person}{Changhyeon Kim}, \bibinfo{person}{Donghyeon Han},
  {and} \bibinfo{person}{Hoi-Jun Yoo}.} \bibinfo{year}{2017}\natexlab{a}.
\newblock \showarticletitle{A low-power convolutional neural network face
  recognition processor and a CIS integrated with always-on face detector}.
\newblock \bibinfo{journal}{\emph{IEEE Journal of Solid-State Circuits}}
  \bibinfo{volume}{53}, \bibinfo{number}{1} (\bibinfo{year}{2017}),
  \bibinfo{pages}{115--123}.
\newblock


\bibitem[Bong et~al\mbox{.}(2017b)]%
        {bong201714}
\bibfield{author}{\bibinfo{person}{Kyeongryeol Bong}, \bibinfo{person}{Sungpill
  Choi}, \bibinfo{person}{Changhyeon Kim}, \bibinfo{person}{Sanghoon Kang},
  \bibinfo{person}{Youchang Kim}, {and} \bibinfo{person}{Hoi-Jun Yoo}.}
  \bibinfo{year}{2017}\natexlab{b}.
\newblock \showarticletitle{14.6 A 0.62 mW ultra-low-power
  convolutional-neural-network face-recognition processor and a CIS integrated
  with always-on haar-like face detector}. In \bibinfo{booktitle}{\emph{2017
  IEEE International Solid-State Circuits Conference (ISSCC)}}. IEEE,
  \bibinfo{pages}{248--249}.
\newblock


\bibitem[Brooks et~al\mbox{.}(2000)]%
        {brooks2000wattch}
\bibfield{author}{\bibinfo{person}{David Brooks}, \bibinfo{person}{Vivek
  Tiwari}, {and} \bibinfo{person}{Margaret Martonosi}.}
  \bibinfo{year}{2000}\natexlab{}.
\newblock \showarticletitle{Wattch: A framework for architectural-level power
  analysis and optimizations}.
\newblock \bibinfo{journal}{\emph{ACM SIGARCH Computer Architecture News}}
  \bibinfo{volume}{28}, \bibinfo{number}{2} (\bibinfo{year}{2000}),
  \bibinfo{pages}{83--94}.
\newblock


\bibitem[Cao et~al\mbox{.}(2022)]%
        {cao2022}
\bibfield{author}{\bibinfo{person}{Weidong Cao}, \bibinfo{person}{Yilong Zhao},
  \bibinfo{person}{Adith Boloor}, \bibinfo{person}{Yinhe Han},
  \bibinfo{person}{Xuan Zhang}, {and} \bibinfo{person}{Li Jiang}.}
  \bibinfo{year}{2022}\natexlab{}.
\newblock \showarticletitle{Neural-PIM: Efficient Processing-In-Memory With
  Neural Approximation of Peripherals}.
\newblock \bibinfo{journal}{\emph{IEEE Trans. Comput.}} \bibinfo{volume}{71},
  \bibinfo{number}{9} (\bibinfo{year}{2022}), \bibinfo{pages}{2142--2155}.
\newblock


\bibitem[Capoccia et~al\mbox{.}(2020)]%
        {cds}
\bibfield{author}{\bibinfo{person}{Raffaele Capoccia}, \bibinfo{person}{Assim
  Boukhayma}, {and} \bibinfo{person}{Christian Enz}.}
  \bibinfo{year}{2020}\natexlab{}.
\newblock \showarticletitle{Experimental Verification of the Impact of Analog
  CMS on CIS Readout Noise}.
\newblock \bibinfo{journal}{\emph{IEEE Transactions on Circuits and Systems I:
  Regular Papers}} \bibinfo{volume}{67}, \bibinfo{number}{3}
  (\bibinfo{year}{2020}), \bibinfo{pages}{774--784}.
\newblock


\bibitem[Chandrasekar et~al\mbox{.}(2012)]%
        {chandrasekar2012drampower}
\bibfield{author}{\bibinfo{person}{Karthik Chandrasekar},
  \bibinfo{person}{Christian Weis}, \bibinfo{person}{Yonghui Li},
  \bibinfo{person}{Benny Akesson}, \bibinfo{person}{Norbert Wehn}, {and}
  \bibinfo{person}{Kees Goossens}.} \bibinfo{year}{2012}\natexlab{}.
\newblock \showarticletitle{DRAMPower: Open-source DRAM power \& energy
  estimation tool}.
\newblock \bibinfo{journal}{\emph{URL: http://www. drampower. info}}
  \bibinfo{volume}{22} (\bibinfo{year}{2012}).
\newblock


\bibitem[Chen(2009)]%
        {chen2009gpu}
\bibfield{author}{\bibinfo{person}{John~Y Chen}.}
  \bibinfo{year}{2009}\natexlab{}.
\newblock \showarticletitle{GPU technology trends and future requirements}. In
  \bibinfo{booktitle}{\emph{2009 IEEE International Electron Devices Meeting
  (IEDM)}}. IEEE, \bibinfo{pages}{1--6}.
\newblock


\bibitem[Cheng et~al\mbox{.}(2008)]%
        {cheng2008ivisual}
\bibfield{author}{\bibinfo{person}{Chih-Chi Cheng}, \bibinfo{person}{Chia-Hua
  Lin}, \bibinfo{person}{Chung-Te Li}, {and} \bibinfo{person}{Liang-Gee Chen}.}
  \bibinfo{year}{2008}\natexlab{}.
\newblock \showarticletitle{iVisual: An intelligent visual sensor SoC with 2790
  fps CMOS image sensor and 205 GOPS/W vision processor}.
\newblock \bibinfo{journal}{\emph{IEEE Journal of Solid-State Circuits}}
  \bibinfo{volume}{44}, \bibinfo{number}{1} (\bibinfo{year}{2008}),
  \bibinfo{pages}{127--135}.
\newblock


\bibitem[Choi et~al\mbox{.}(2020)]%
        {choi2020design}
\bibfield{author}{\bibinfo{person}{Jaihyuk Choi}, \bibinfo{person}{Sungjae
  Lee}, \bibinfo{person}{Youngdoo Son}, {and} \bibinfo{person}{Soo~Youn Kim}.}
  \bibinfo{year}{2020}\natexlab{}.
\newblock \showarticletitle{Design of an always-on image sensor using an analog
  lightweight convolutional neural network}.
\newblock \bibinfo{journal}{\emph{Sensors}} \bibinfo{volume}{20},
  \bibinfo{number}{11} (\bibinfo{year}{2020}), \bibinfo{pages}{3101}.
\newblock


\bibitem[Danowitz et~al\mbox{.}(2012)]%
        {danowitz2012cpu}
\bibfield{author}{\bibinfo{person}{Andrew Danowitz}, \bibinfo{person}{Kyle
  Kelley}, \bibinfo{person}{James Mao}, \bibinfo{person}{John~P Stevenson},
  {and} \bibinfo{person}{Mark Horowitz}.} \bibinfo{year}{2012}\natexlab{}.
\newblock \showarticletitle{CPU DB: recording microprocessor history}.
\newblock \bibinfo{journal}{\emph{Commun. ACM}} \bibinfo{volume}{55},
  \bibinfo{number}{4} (\bibinfo{year}{2012}), \bibinfo{pages}{55--63}.
\newblock


\bibitem[Dong et~al\mbox{.}(2022)]%
        {dong2022splitnets}
\bibfield{author}{\bibinfo{person}{Xin Dong}, \bibinfo{person}{Barbara
  De~Salvo}, \bibinfo{person}{Meng Li}, \bibinfo{person}{Chiao Liu},
  \bibinfo{person}{Zhongnan Qu}, \bibinfo{person}{HT Kung}, {and}
  \bibinfo{person}{Ziyun Li}.} \bibinfo{year}{2022}\natexlab{}.
\newblock \showarticletitle{SplitNets: Designing Neural Architectures for
  Efficient Distributed Computing on Head-Mounted Systems}. In
  \bibinfo{booktitle}{\emph{Proceedings of the IEEE/CVF Conference on Computer
  Vision and Pattern Recognition}}. \bibinfo{pages}{12559--12569}.
\newblock


\bibitem[Eki et~al\mbox{.}(2021)]%
        {eki20219}
\bibfield{author}{\bibinfo{person}{Ryoji Eki}, \bibinfo{person}{Satoshi
  Yamada}, \bibinfo{person}{Hiroyuki Ozawa}, \bibinfo{person}{Hitoshi Kai},
  \bibinfo{person}{Kazuyuki Okuike}, \bibinfo{person}{Hareesh Gowtham},
  \bibinfo{person}{Hidetomo Nakanishi}, \bibinfo{person}{Edan Almog},
  \bibinfo{person}{Yoel Livne}, \bibinfo{person}{Gadi Yuval}, {et~al\mbox{.}}}
  \bibinfo{year}{2021}\natexlab{}.
\newblock \showarticletitle{9.6 A 1/2.3 inch 12.3 Mpixel with on-chip 4.97
  TOPS/W CNN processor back-illuminated stacked CMOS image sensor}. In
  \bibinfo{booktitle}{\emph{2021 IEEE International Solid-State Circuits
  Conference (ISSCC)}}, Vol.~\bibinfo{volume}{64}. IEEE,
  \bibinfo{pages}{154--156}.
\newblock


\bibitem[Feng et~al\mbox{.}(2022)]%
        {feng2022}
\bibfield{author}{\bibinfo{person}{Yu Feng}, \bibinfo{person}{Nathan
  Goulding-Hotta}, \bibinfo{person}{Asif Khan}, \bibinfo{person}{Hans
  Reyserhove}, {and} \bibinfo{person}{Yuhao Zhu}.}
  \bibinfo{year}{2022}\natexlab{}.
\newblock \showarticletitle{Real-Time Gaze Tracking with Event-Driven Eye
  Segmentation}. In \bibinfo{booktitle}{\emph{2022 IEEE Conference on Virtual
  Reality and 3D User Interfaces (VR)}}. IEEE, \bibinfo{pages}{399--408}.
\newblock


\bibitem[Feng et~al\mbox{.}(2019)]%
        {feng2019asv}
\bibfield{author}{\bibinfo{person}{Yu Feng}, \bibinfo{person}{Paul Whatmough},
  {and} \bibinfo{person}{Yuhao Zhu}.} \bibinfo{year}{2019}\natexlab{}.
\newblock \showarticletitle{Asv: Accelerated stereo vision system}. In
  \bibinfo{booktitle}{\emph{Proceedings of the 52nd Annual IEEE/ACM
  International Symposium on Microarchitecture}}. \bibinfo{pages}{643--656}.
\newblock


\bibitem[Gao et~al\mbox{.}(2017)]%
        {gao2017tetris}
\bibfield{author}{\bibinfo{person}{Mingyu Gao}, \bibinfo{person}{Jing Pu},
  \bibinfo{person}{Xuan Yang}, \bibinfo{person}{Mark Horowitz}, {and}
  \bibinfo{person}{Christos Kozyrakis}.} \bibinfo{year}{2017}\natexlab{}.
\newblock \showarticletitle{Tetris: Scalable and efficient neural network
  acceleration with 3d memory}. In \bibinfo{booktitle}{\emph{Proceedings of the
  Twenty-Second International Conference on Architectural Support for
  Programming Languages and Operating Systems}}. \bibinfo{pages}{751--764}.
\newblock


\bibitem[Gielen and Dehaene(2005)]%
        {gielen2005analog}
\bibfield{author}{\bibinfo{person}{Georges Gielen} {and} \bibinfo{person}{Wim
  Dehaene}.} \bibinfo{year}{2005}\natexlab{}.
\newblock \showarticletitle{Analog and digital circuit design in 65 nm CMOS:
  End of the road?}. In \bibinfo{booktitle}{\emph{Design, Automation and Test
  in Europe}}. IEEE, \bibinfo{pages}{37--42}.
\newblock


\bibitem[Gomez et~al\mbox{.}(2022)]%
        {gomez2022distributed}
\bibfield{author}{\bibinfo{person}{Jorge Gomez}, \bibinfo{person}{Saavan
  Patel}, \bibinfo{person}{Syed~Shakib Sarwar}, \bibinfo{person}{Ziyun Li},
  \bibinfo{person}{Raffaele Capoccia}, \bibinfo{person}{Zhao Wang},
  \bibinfo{person}{Reid Pinkham}, \bibinfo{person}{Andrew Berkovich},
  \bibinfo{person}{Tsung-Hsun Tsai}, \bibinfo{person}{Barbara De~Salvo},
  {et~al\mbox{.}}} \bibinfo{year}{2022}\natexlab{}.
\newblock \showarticletitle{Distributed On-Sensor Compute System for AR/VR
  Devices: A Semi-Analytical Simulation Framework for Power Estimation}.
\newblock \bibinfo{journal}{\emph{arXiv preprint arXiv:2203.07474}}
  (\bibinfo{year}{2022}).
\newblock


\bibitem[Group(2021)]%
        {mipi_csi}
\bibfield{author}{\bibinfo{person}{Camare~Working Group}.}
  \bibinfo{year}{2021}\natexlab{}.
\newblock \bibinfo{title}{MIPI White Paper: An Introductory Guide to MIPI
  Automotive SerDes Solutions (MASS)}.
\newblock
\newblock
\urldef\tempurl%
\url{https://www.mipi.org/introductory-guide-to-mass}
\showURL{%
\tempurl}


\bibitem[Guthaus et~al\mbox{.}(2016)]%
        {guthaus2016openram}
\bibfield{author}{\bibinfo{person}{Matthew~R Guthaus}, \bibinfo{person}{James~E
  Stine}, \bibinfo{person}{Samira Ataei}, \bibinfo{person}{Brian Chen},
  \bibinfo{person}{Bin Wu}, {and} \bibinfo{person}{Mehedi Sarwar}.}
  \bibinfo{year}{2016}\natexlab{}.
\newblock \showarticletitle{OpenRAM: An open-source memory compiler}. In
  \bibinfo{booktitle}{\emph{2016 IEEE/ACM International Conference on
  Computer-Aided Design (ICCAD)}}. IEEE, \bibinfo{pages}{1--6}.
\newblock


\bibitem[Han et~al\mbox{.}(2016)]%
        {han2016eie}
\bibfield{author}{\bibinfo{person}{Song Han}, \bibinfo{person}{Xingyu Liu},
  \bibinfo{person}{Huizi Mao}, \bibinfo{person}{Jing Pu},
  \bibinfo{person}{Ardavan Pedram}, \bibinfo{person}{Mark~A Horowitz}, {and}
  \bibinfo{person}{William~J Dally}.} \bibinfo{year}{2016}\natexlab{}.
\newblock \showarticletitle{EIE: Efficient inference engine on compressed deep
  neural network}.
\newblock \bibinfo{journal}{\emph{ACM SIGARCH Computer Architecture News}}
  \bibinfo{volume}{44}, \bibinfo{number}{3} (\bibinfo{year}{2016}),
  \bibinfo{pages}{243--254}.
\newblock


\bibitem[Haruta et~al\mbox{.}(2017)]%
        {haruta20174}
\bibfield{author}{\bibinfo{person}{Tsutomu Haruta}, \bibinfo{person}{Tsutomu
  Nakajima}, \bibinfo{person}{Jun Hashizume}, \bibinfo{person}{Taku
  Umebayashi}, \bibinfo{person}{Hiroshi Takahashi}, \bibinfo{person}{Kazuo
  Taniguchi}, \bibinfo{person}{Masami Kuroda}, \bibinfo{person}{Hiroshi
  Sumihiro}, \bibinfo{person}{Koji Enoki}, \bibinfo{person}{Takatsugu
  Yamasaki}, {et~al\mbox{.}}} \bibinfo{year}{2017}\natexlab{}.
\newblock \showarticletitle{4.6 A 1/2.3 inch 20Mpixel 3-layer stacked CMOS
  Image Sensor with DRAM}. In \bibinfo{booktitle}{\emph{2017 IEEE International
  Solid-State Circuits Conference (ISSCC)}}. IEEE, \bibinfo{pages}{76--77}.
\newblock


\bibitem[Hegarty et~al\mbox{.}(2014)]%
        {hegarty2014darkroom}
\bibfield{author}{\bibinfo{person}{James Hegarty}, \bibinfo{person}{John
  Brunhaver}, \bibinfo{person}{Zachary DeVito}, \bibinfo{person}{Jonathan
  Ragan-Kelley}, \bibinfo{person}{Noy Cohen}, \bibinfo{person}{Steven Bell},
  \bibinfo{person}{Artem Vasilyev}, \bibinfo{person}{Mark Horowitz}, {and}
  \bibinfo{person}{Pat Hanrahan}.} \bibinfo{year}{2014}\natexlab{}.
\newblock \showarticletitle{Darkroom: compiling high-level image processing
  code into hardware pipelines.}
\newblock \bibinfo{journal}{\emph{ACM Trans. Graph.}} \bibinfo{volume}{33},
  \bibinfo{number}{4} (\bibinfo{year}{2014}), \bibinfo{pages}{144--1}.
\newblock


\bibitem[Hirata et~al\mbox{.}(2021)]%
        {hirata20217}
\bibfield{author}{\bibinfo{person}{Tomoki Hirata}, \bibinfo{person}{Hironobu
  Murata}, \bibinfo{person}{Hideaki Matsuda}, \bibinfo{person}{Yojiro Tezuka},
  {and} \bibinfo{person}{Shiro Tsunai}.} \bibinfo{year}{2021}\natexlab{}.
\newblock \showarticletitle{7.8 A 1-inch 17Mpixel 1000fps Block-Controlled
  Coded-Exposure Back-Illuminated Stacked CMOS Image Sensor for Computational
  Imaging and Adaptive Dynamic Range Control}. In
  \bibinfo{booktitle}{\emph{2021 IEEE International Solid-State Circuits
  Conference (ISSCC)}}, Vol.~\bibinfo{volume}{64}. IEEE,
  \bibinfo{pages}{120--122}.
\newblock


\bibitem[Hong and Kim(2010)]%
        {hong2010integrated}
\bibfield{author}{\bibinfo{person}{Sunpyo Hong} {and} \bibinfo{person}{Hyesoon
  Kim}.} \bibinfo{year}{2010}\natexlab{}.
\newblock \showarticletitle{An integrated GPU power and performance model}. In
  \bibinfo{booktitle}{\emph{Proceedings of the 37th annual international
  symposium on Computer architecture}}. \bibinfo{pages}{280--289}.
\newblock


\bibitem[Hsu et~al\mbox{.}(2022)]%
        {hsu202208}
\bibfield{author}{\bibinfo{person}{Tzu-Hsiang Hsu}, \bibinfo{person}{Guan-Cheng
  Chen}, \bibinfo{person}{Yi-Ren Chen}, \bibinfo{person}{Chung-Chuan Lo},
  \bibinfo{person}{Ren-Shuo Liu}, \bibinfo{person}{Meng-Fan Chang},
  \bibinfo{person}{Kea-Tiong Tang}, {and} \bibinfo{person}{Chih-Cheng Hsieh}.}
  \bibinfo{year}{2022}\natexlab{}.
\newblock \showarticletitle{A 0.8 V Intelligent Vision Sensor with Tiny
  Convolutional Neural Network and Programmable Weights Using Mixed-Mode
  Processing-in-Sensor Technique for Image Classification}. In
  \bibinfo{booktitle}{\emph{2022 IEEE International Solid-State Circuits
  Conference (ISSCC)}}, Vol.~\bibinfo{volume}{65}. IEEE, \bibinfo{pages}{1--3}.
\newblock


\bibitem[Hsu et~al\mbox{.}(2020)]%
        {hsu202005}
\bibfield{author}{\bibinfo{person}{Tzu-Hsiang Hsu}, \bibinfo{person}{Yi-Ren
  Chen}, \bibinfo{person}{Ren-Shuo Liu}, \bibinfo{person}{Chung-Chuan Lo},
  \bibinfo{person}{Kea-Tiong Tang}, \bibinfo{person}{Meng-Fan Chang}, {and}
  \bibinfo{person}{Chih-Cheng Hsieh}.} \bibinfo{year}{2020}\natexlab{}.
\newblock \showarticletitle{A 0.5-V real-time computational CMOS image sensor
  with programmable kernel for feature extraction}.
\newblock \bibinfo{journal}{\emph{IEEE Journal of Solid-State Circuits}}
  \bibinfo{volume}{56}, \bibinfo{number}{5} (\bibinfo{year}{2020}),
  \bibinfo{pages}{1588--1596}.
\newblock


\bibitem[Huzaifa et~al\mbox{.}(2021)]%
        {huzaifa2021illixr}
\bibfield{author}{\bibinfo{person}{Muhammad Huzaifa}, \bibinfo{person}{Rishi
  Desai}, \bibinfo{person}{Samuel Grayson}, \bibinfo{person}{Xutao Jiang},
  \bibinfo{person}{Ying Jing}, \bibinfo{person}{Jae Lee}, \bibinfo{person}{Fang
  Lu}, \bibinfo{person}{Yihan Pang}, \bibinfo{person}{Joseph Ravichandran},
  \bibinfo{person}{Finn Sinclair}, {et~al\mbox{.}}}
  \bibinfo{year}{2021}\natexlab{}.
\newblock \showarticletitle{ILLIXR: Enabling End-to-End Extended Reality
  Research}. In \bibinfo{booktitle}{\emph{2021 IEEE International Symposium on
  Workload Characterization (IISWC)}}. IEEE, \bibinfo{pages}{24--38}.
\newblock


\bibitem[Jespers(2010)]%
        {gm_id}
\bibfield{author}{\bibinfo{person}{Paul~G Jespers}.}
  \bibinfo{year}{2010}\natexlab{}.
\newblock \bibinfo{booktitle}{\emph{The gm/ID Methodology, a sizing tool for
  low-voltage analog CMOS Circuits}}.
\newblock \bibinfo{publisher}{Springer}.
\newblock


\bibitem[Kandiah et~al\mbox{.}(2021)]%
        {kandiah2021accelwattch}
\bibfield{author}{\bibinfo{person}{Vijay Kandiah}, \bibinfo{person}{Scott
  Peverelle}, \bibinfo{person}{Mahmoud Khairy}, \bibinfo{person}{Junrui Pan},
  \bibinfo{person}{Amogh Manjunath}, \bibinfo{person}{Timothy~G Rogers},
  \bibinfo{person}{Tor~M Aamodt}, {and} \bibinfo{person}{Nikos Hardavellas}.}
  \bibinfo{year}{2021}\natexlab{}.
\newblock \showarticletitle{AccelWattch: A Power Modeling Framework for Modern
  GPUs}. In \bibinfo{booktitle}{\emph{MICRO-54: 54th Annual IEEE/ACM
  International Symposium on Microarchitecture}}. \bibinfo{pages}{738--753}.
\newblock


\bibitem[Kaur et~al\mbox{.}(2020)]%
        {kaur2020array}
\bibfield{author}{\bibinfo{person}{Amandeep Kaur}, \bibinfo{person}{Deepak
  Mishra}, \bibinfo{person}{KM Amogh}, {and} \bibinfo{person}{Mukul Sarkar}.}
  \bibinfo{year}{2020}\natexlab{}.
\newblock \showarticletitle{On-array compressive acquisition in cmos image
  sensors using accumulated spatial gradients}.
\newblock \bibinfo{journal}{\emph{IEEE Transactions on Circuits and Systems for
  Video Technology}} \bibinfo{volume}{31}, \bibinfo{number}{2}
  (\bibinfo{year}{2020}), \bibinfo{pages}{523--532}.
\newblock


\bibitem[Kim et~al\mbox{.}(2005)]%
        {kim2005200}
\bibfield{author}{\bibinfo{person}{Seong-Jin Kim}, \bibinfo{person}{Kwang-Hyun
  Lee}, \bibinfo{person}{Sang-Wook Han}, {and} \bibinfo{person}{Euisik Yoon}.}
  \bibinfo{year}{2005}\natexlab{}.
\newblock \showarticletitle{A 200/spl times/160 pixel CMOS fingerprint
  recognition SoC with adaptable column-parallel processors}. In
  \bibinfo{booktitle}{\emph{ISSCC. 2005 IEEE International Digest of Technical
  Papers. Solid-State Circuits Conference, 2005.}} IEEE,
  \bibinfo{pages}{250--596}.
\newblock


\bibitem[Kodukula et~al\mbox{.}(2021a)]%
        {kodukula2021dynamic}
\bibfield{author}{\bibinfo{person}{Venkatesh Kodukula}, \bibinfo{person}{Saad
  Katrawala}, \bibinfo{person}{Britton Jones}, \bibinfo{person}{Carole-Jean
  Wu}, {and} \bibinfo{person}{Robert LiKamWa}.}
  \bibinfo{year}{2021}\natexlab{a}.
\newblock \showarticletitle{Dynamic temperature management of near-sensor
  processing for energy-efficient high-fidelity imaging}.
\newblock \bibinfo{journal}{\emph{Sensors}} \bibinfo{volume}{21},
  \bibinfo{number}{3} (\bibinfo{year}{2021}), \bibinfo{pages}{926}.
\newblock


\bibitem[Kodukula et~al\mbox{.}(2021b)]%
        {kodukula2021rhythmic}
\bibfield{author}{\bibinfo{person}{Venkatesh Kodukula},
  \bibinfo{person}{Alexander Shearer}, \bibinfo{person}{Van Nguyen},
  \bibinfo{person}{Srinivas Lingutla}, \bibinfo{person}{Yifei Liu}, {and}
  \bibinfo{person}{Robert LiKamWa}.} \bibinfo{year}{2021}\natexlab{b}.
\newblock \showarticletitle{Rhythmic pixel regions: multi-resolution visual
  sensing system towards high-precision visual computing at low power}. In
  \bibinfo{booktitle}{\emph{Proceedings of the 26th ACM International
  Conference on Architectural Support for Programming Languages and Operating
  Systems}}. \bibinfo{pages}{573--586}.
\newblock


\bibitem[Kumagai et~al\mbox{.}(2018)]%
        {kumagai20181}
\bibfield{author}{\bibinfo{person}{Oichi Kumagai}, \bibinfo{person}{Atsumi
  Niwa}, \bibinfo{person}{Katsuhiko Hanzawa}, \bibinfo{person}{Hidetaka Kato},
  \bibinfo{person}{Shinichiro Futami}, \bibinfo{person}{Toshio Ohyama},
  \bibinfo{person}{Tsutomu Imoto}, \bibinfo{person}{Masahiko Nakamizo},
  \bibinfo{person}{Hirotaka Murakami}, \bibinfo{person}{Tatsuki Nishino},
  {et~al\mbox{.}}} \bibinfo{year}{2018}\natexlab{}.
\newblock \showarticletitle{A 1/4-inch 3.9 Mpixel low-power event-driven
  back-illuminated stacked CMOS image sensor}. In
  \bibinfo{booktitle}{\emph{2018 IEEE International Solid-State Circuits
  Conference-(ISSCC)}}. IEEE, \bibinfo{pages}{86--88}.
\newblock


\bibitem[Kwon et~al\mbox{.}(2022)]%
        {kwon2022xrbench}
\bibfield{author}{\bibinfo{person}{Hyoukjun Kwon},
  \bibinfo{person}{Krishnakumar Nair}, \bibinfo{person}{Jamin Seo},
  \bibinfo{person}{Jason Yik}, \bibinfo{person}{Debabrata Mohapatra},
  \bibinfo{person}{Dongyuan Zhan}, \bibinfo{person}{Jinook Song},
  \bibinfo{person}{Peter Capak}, \bibinfo{person}{Peizhao Zhang},
  \bibinfo{person}{Peter Vajda}, {et~al\mbox{.}}}
  \bibinfo{year}{2022}\natexlab{}.
\newblock \showarticletitle{XRBench: An Extended Reality (XR) Machine Learning
  Benchmark Suite for the Metaverse}.
\newblock \bibinfo{journal}{\emph{arXiv preprint arXiv:2211.08675}}
  (\bibinfo{year}{2022}).
\newblock


\bibitem[Kwon et~al\mbox{.}(2020)]%
        {kwon2020low}
\bibfield{author}{\bibinfo{person}{Minho Kwon}, \bibinfo{person}{Seunghyun
  Lim}, \bibinfo{person}{Hyeokjong Lee}, \bibinfo{person}{Il-Seon Ha},
  \bibinfo{person}{Moo-Young Kim}, \bibinfo{person}{Il-Jin Seo},
  \bibinfo{person}{Suho Lee}, \bibinfo{person}{Yongsuk Choi},
  \bibinfo{person}{Kyunghoon Kim}, \bibinfo{person}{Hansoo Lee},
  {et~al\mbox{.}}} \bibinfo{year}{2020}\natexlab{}.
\newblock \showarticletitle{A Low-Power 65/14nm Stacked CMOS Image Sensor}. In
  \bibinfo{booktitle}{\emph{2020 IEEE International Symposium on Circuits and
  Systems (ISCAS)}}. IEEE, \bibinfo{pages}{1--4}.
\newblock


\bibitem[Lauwers and Gielen(2002)]%
        {Lauwers20002}
\bibfield{author}{\bibinfo{person}{E. Lauwers} {and} \bibinfo{person}{G.
  Gielen}.} \bibinfo{year}{2002}\natexlab{}.
\newblock \showarticletitle{Power estimation methods for analog circuits for
  architectural exploration of integrated systems}.
\newblock \bibinfo{journal}{\emph{IEEE Transactions on Very Large Scale
  Integration (VLSI) Systems}} \bibinfo{volume}{10}, \bibinfo{number}{2}
  (\bibinfo{year}{2002}), \bibinfo{pages}{155--162}.
\newblock


\bibitem[Lee and Wong(2017)]%
        {lee2017}
\bibfield{author}{\bibinfo{person}{Edward~H. Lee} {and}
  \bibinfo{person}{S.~Simon Wong}.} \bibinfo{year}{2017}\natexlab{}.
\newblock \showarticletitle{Analysis and Design of a Passive Switched-Capacitor
  Matrix Multiplier for Approximate Computing}.
\newblock \bibinfo{journal}{\emph{IEEE Journal of Solid-State Circuits}}
  \bibinfo{volume}{52}, \bibinfo{number}{1} (\bibinfo{year}{2017}),
  \bibinfo{pages}{261--271}.
\newblock


\bibitem[Leng et~al\mbox{.}(2013)]%
        {leng2013gpuwattch}
\bibfield{author}{\bibinfo{person}{Jingwen Leng}, \bibinfo{person}{Tayler
  Hetherington}, \bibinfo{person}{Ahmed ElTantawy}, \bibinfo{person}{Syed
  Gilani}, \bibinfo{person}{Nam~Sung Kim}, \bibinfo{person}{Tor~M Aamodt},
  {and} \bibinfo{person}{Vijay~Janapa Reddi}.} \bibinfo{year}{2013}\natexlab{}.
\newblock \showarticletitle{GPUWattch: Enabling energy optimizations in
  GPGPUs}.
\newblock \bibinfo{journal}{\emph{ACM SIGARCH Computer Architecture News}}
  \bibinfo{volume}{41}, \bibinfo{number}{3} (\bibinfo{year}{2013}),
  \bibinfo{pages}{487--498}.
\newblock


\bibitem[Li et~al\mbox{.}(2009)]%
        {li2009mcpat}
\bibfield{author}{\bibinfo{person}{Sheng Li}, \bibinfo{person}{Jung~Ho Ahn},
  \bibinfo{person}{Richard~D Strong}, \bibinfo{person}{Jay~B Brockman},
  \bibinfo{person}{Dean~M Tullsen}, {and} \bibinfo{person}{Norman~P Jouppi}.}
  \bibinfo{year}{2009}\natexlab{}.
\newblock \showarticletitle{McPAT: An integrated power, area, and timing
  modeling framework for multicore and manycore architectures}. In
  \bibinfo{booktitle}{\emph{Proceedings of the 42nd annual ieee/acm
  international symposium on microarchitecture}}. \bibinfo{pages}{469--480}.
\newblock


\bibitem[LiKamWa et~al\mbox{.}(2016)]%
        {likamwa2016redeye}
\bibfield{author}{\bibinfo{person}{Robert LiKamWa}, \bibinfo{person}{Yunhui
  Hou}, \bibinfo{person}{Julian Gao}, \bibinfo{person}{Mia Polansky}, {and}
  \bibinfo{person}{Lin Zhong}.} \bibinfo{year}{2016}\natexlab{}.
\newblock \showarticletitle{Redeye: analog convnet image sensor architecture
  for continuous mobile vision}.
\newblock \bibinfo{journal}{\emph{ACM SIGARCH Computer Architecture News}}
  \bibinfo{volume}{44}, \bibinfo{number}{3} (\bibinfo{year}{2016}),
  \bibinfo{pages}{255--266}.
\newblock


\bibitem[LiKamWa et~al\mbox{.}(2013)]%
        {likamwa2013energy}
\bibfield{author}{\bibinfo{person}{Robert LiKamWa}, \bibinfo{person}{Bodhi
  Priyantha}, \bibinfo{person}{Matthai Philipose}, \bibinfo{person}{Lin Zhong},
  {and} \bibinfo{person}{Paramvir Bahl}.} \bibinfo{year}{2013}\natexlab{}.
\newblock \showarticletitle{Energy characterization and optimization of image
  sensing toward continuous mobile vision}. In
  \bibinfo{booktitle}{\emph{Proceeding of the 11th annual international
  conference on Mobile systems, applications, and services}}.
  \bibinfo{pages}{69--82}.
\newblock


\bibitem[Lim and Horowitz(2019)]%
        {lim2019}
\bibfield{author}{\bibinfo{person}{Byong~Chan Lim} {and} \bibinfo{person}{Mark
  Horowitz}.} \bibinfo{year}{2019}\natexlab{}.
\newblock \showarticletitle{An Analog Model Template Library: Simplifying
  Chip-Level, Mixed-Signal Design Verification}.
\newblock \bibinfo{journal}{\emph{IEEE Transactions on Very Large Scale
  Integration (VLSI) Systems}} \bibinfo{volume}{27}, \bibinfo{number}{1}
  (\bibinfo{year}{2019}), \bibinfo{pages}{193--204}.
\newblock


\bibitem[Liu et~al\mbox{.}(2019)]%
        {liu2019intelligent}
\bibfield{author}{\bibinfo{person}{Chiao Liu}, \bibinfo{person}{Andrew
  Berkovich}, \bibinfo{person}{Song Chen}, \bibinfo{person}{Hans Reyserhove},
  \bibinfo{person}{Syed~Shakib Sarwar}, {and} \bibinfo{person}{Tsung-Hsun
  Tsai}.} \bibinfo{year}{2019}\natexlab{}.
\newblock \showarticletitle{Intelligent vision systems--bringing human-machine
  interface to AR/VR}. In \bibinfo{booktitle}{\emph{2019 IEEE International
  Electron Devices Meeting (IEDM)}}. IEEE, \bibinfo{pages}{10--5}.
\newblock


\bibitem[Liu et~al\mbox{.}(2022)]%
        {liu2022augmented}
\bibfield{author}{\bibinfo{person}{Chiao Liu}, \bibinfo{person}{Song Chen},
  \bibinfo{person}{Tsung-Hsun Tsai}, \bibinfo{person}{Barbara De~Salvo}, {and}
  \bibinfo{person}{Jorge Gomez}.} \bibinfo{year}{2022}\natexlab{}.
\newblock \showarticletitle{Augmented Reality-The Next Frontier of Image
  Sensors and Compute Systems}. In \bibinfo{booktitle}{\emph{2022 IEEE
  International Solid-State Circuits Conference (ISSCC)}},
  Vol.~\bibinfo{volume}{65}. IEEE, \bibinfo{pages}{426--428}.
\newblock


\bibitem[Ma et~al\mbox{.}(2022)]%
        {ma2022hogeye}
\bibfield{author}{\bibinfo{person}{Tianrui Ma}, \bibinfo{person}{Weidong Cao},
  \bibinfo{person}{Fei Qiao}, \bibinfo{person}{Ayan Chakrabarti}, {and}
  \bibinfo{person}{Xuan Zhang}.} \bibinfo{year}{2022}\natexlab{}.
\newblock \showarticletitle{HOGEye: neural approximation of hog feature
  extraction in rram-based 3d-stacked image sensors}. In
  \bibinfo{booktitle}{\emph{Proceedings of the ACM/IEEE International Symposium
  on Low Power Electronics and Design}}. \bibinfo{pages}{1--6}.
\newblock


\bibitem[Mittal et~al\mbox{.}(2017)]%
        {mittal2017destiny}
\bibfield{author}{\bibinfo{person}{Sparsh Mittal}, \bibinfo{person}{Rujia
  Wang}, {and} \bibinfo{person}{Jeffrey Vetter}.}
  \bibinfo{year}{2017}\natexlab{}.
\newblock \showarticletitle{DESTINY: A comprehensive tool with 3D and
  multi-level cell memory modeling capability}.
\newblock \bibinfo{journal}{\emph{Journal of Low Power Electronics and
  Applications}} \bibinfo{volume}{7}, \bibinfo{number}{3}
  (\bibinfo{year}{2017}), \bibinfo{pages}{23}.
\newblock


\bibitem[Murakami et~al\mbox{.}(2022)]%
        {murakami20224}
\bibfield{author}{\bibinfo{person}{Hirotaka Murakami}, \bibinfo{person}{Eric
  Bohannon}, \bibinfo{person}{John Childs}, \bibinfo{person}{Grace Gui},
  \bibinfo{person}{Eric Moule}, \bibinfo{person}{Katsuhiko Hanzawa},
  \bibinfo{person}{Tomofumi Koda}, \bibinfo{person}{Chiaki Takano},
  \bibinfo{person}{Toshimasa Shimizu}, \bibinfo{person}{Yuki Takizawa},
  {et~al\mbox{.}}} \bibinfo{year}{2022}\natexlab{}.
\newblock \showarticletitle{A 4.9 Mpixel Programmable-Resolution Multi-Purpose
  CMOS Image Sensor for Computer Vision}. In \bibinfo{booktitle}{\emph{2022
  IEEE International Solid-State Circuits Conference (ISSCC)}},
  Vol.~\bibinfo{volume}{65}. IEEE, \bibinfo{pages}{104--106}.
\newblock


\bibitem[Murmann({[n.\,d.]})]%
        {adc_fom}
\bibfield{author}{\bibinfo{person}{B. Murmann}.}
  \bibinfo{year}{[n.\,d.]}\natexlab{}.
\newblock \bibinfo{title}{ADC Performance Survey 1997-2022}.
\newblock
  \bibinfo{howpublished}{\url{http://web.stanford.edu/~murmann/adcsurvey.html}}.
\newblock


\bibitem[Park et~al\mbox{.}(2021)]%
        {park202151}
\bibfield{author}{\bibinfo{person}{Chanmin Park}, \bibinfo{person}{Wenda Zhao},
  \bibinfo{person}{Injun Park}, \bibinfo{person}{Nan Sun}, {and}
  \bibinfo{person}{Youngcheol Chae}.} \bibinfo{year}{2021}\natexlab{}.
\newblock \showarticletitle{A 51-pJ/pixel 33.7-dB PSNR 4$\times$ compressive
  CMOS image sensor with column-parallel single-shot compressive sensing}.
\newblock \bibinfo{journal}{\emph{IEEE Journal of Solid-State Circuits}}
  \bibinfo{volume}{56}, \bibinfo{number}{8} (\bibinfo{year}{2021}),
  \bibinfo{pages}{2503--2515}.
\newblock


\bibitem[Pentecost et~al\mbox{.}(2022)]%
        {pentecost2022nvmexplorer}
\bibfield{author}{\bibinfo{person}{Lillian Pentecost},
  \bibinfo{person}{Alexander Hankin}, \bibinfo{person}{Marco Donato},
  \bibinfo{person}{Mark Hempstead}, \bibinfo{person}{Gu-Yeon Wei}, {and}
  \bibinfo{person}{David Brooks}.} \bibinfo{year}{2022}\natexlab{}.
\newblock \showarticletitle{NVMExplorer: A Framework for Cross-Stack
  Comparisons of Embedded Non-Volatile Memories}. In
  \bibinfo{booktitle}{\emph{2022 IEEE International Symposium on
  High-Performance Computer Architecture (HPCA)}}. IEEE,
  \bibinfo{pages}{938--956}.
\newblock


\bibitem[Pinkham et~al\mbox{.}(2020)]%
        {pinkham2020algorithm}
\bibfield{author}{\bibinfo{person}{Reid Pinkham}, \bibinfo{person}{Tanner
  Schmidt}, {and} \bibinfo{person}{Andrew Berkovich}.}
  \bibinfo{year}{2020}\natexlab{}.
\newblock \showarticletitle{Algorithm-aware neural network based image
  compression for high-speed imaging}. In \bibinfo{booktitle}{\emph{2020 IEEE
  International Conference on Artificial Intelligence and Virtual Reality
  (AIVR)}}. IEEE, \bibinfo{pages}{196--199}.
\newblock


\bibitem[Poremba et~al\mbox{.}(2015)]%
        {poremba2015destiny}
\bibfield{author}{\bibinfo{person}{Matt Poremba}, \bibinfo{person}{Sparsh
  Mittal}, \bibinfo{person}{Dong Li}, \bibinfo{person}{Jeffrey~S Vetter}, {and}
  \bibinfo{person}{Yuan Xie}.} \bibinfo{year}{2015}\natexlab{}.
\newblock \showarticletitle{Destiny: A tool for modeling emerging 3d nvm and
  edram caches}. In \bibinfo{booktitle}{\emph{2015 Design, Automation \& Test
  in Europe Conference \& Exhibition (DATE)}}. IEEE,
  \bibinfo{pages}{1543--1546}.
\newblock


\bibitem[Qadeer et~al\mbox{.}(2013)]%
        {qadeer2013convolution}
\bibfield{author}{\bibinfo{person}{Wajahat Qadeer}, \bibinfo{person}{Rehan
  Hameed}, \bibinfo{person}{Ofer Shacham}, \bibinfo{person}{Preethi
  Venkatesan}, \bibinfo{person}{Christos Kozyrakis}, {and}
  \bibinfo{person}{Mark~A Horowitz}.} \bibinfo{year}{2013}\natexlab{}.
\newblock \showarticletitle{Convolution engine: balancing efficiency \&
  flexibility in specialized computing}. In
  \bibinfo{booktitle}{\emph{Proceedings of the 40th IEEE Annual International
  Symposium on Computer Architecture}}.
\newblock


\bibitem[Samajdar et~al\mbox{.}(2018)]%
        {samajdar2018scale}
\bibfield{author}{\bibinfo{person}{Ananda Samajdar}, \bibinfo{person}{Yuhao
  Zhu}, \bibinfo{person}{Paul Whatmough}, \bibinfo{person}{Matthew Mattina},
  {and} \bibinfo{person}{Tushar Krishna}.} \bibinfo{year}{2018}\natexlab{}.
\newblock \showarticletitle{Scale-sim: Systolic cnn accelerator simulator}.
\newblock \bibinfo{journal}{\emph{arXiv preprint arXiv:1811.02883}}
  (\bibinfo{year}{2018}).
\newblock


\bibitem[Sarangi and Baas(2021)]%
        {sarangi2021deepscaletool}
\bibfield{author}{\bibinfo{person}{Satyabrata Sarangi} {and}
  \bibinfo{person}{Bevan Baas}.} \bibinfo{year}{2021}\natexlab{}.
\newblock \showarticletitle{DeepScaleTool: A tool for the accurate estimation
  of technology scaling in the deep-submicron era}. In
  \bibinfo{booktitle}{\emph{2021 IEEE International Symposium on Circuits and
  Systems (ISCAS)}}. IEEE, \bibinfo{pages}{1--5}.
\newblock


\bibitem[Seo et~al\mbox{.}(2021)]%
        {seo2021}
\bibfield{author}{\bibinfo{person}{Min-Woong Seo}, \bibinfo{person}{Myunglae
  Chu}, \bibinfo{person}{Hyun-Yong Jung}, \bibinfo{person}{Suksan Kim},
  \bibinfo{person}{Jiyoun Song}, \bibinfo{person}{Junan Lee},
  \bibinfo{person}{Sung-Yong Kim}, \bibinfo{person}{Jongyeon Lee},
  \bibinfo{person}{Sung-Jae Byun}, \bibinfo{person}{Daehee Bae},
  \bibinfo{person}{Minkyung Kim}, \bibinfo{person}{Gwi-Deok Lee},
  \bibinfo{person}{Heesung Shim}, \bibinfo{person}{Changyong Um},
  \bibinfo{person}{Changhwa Kim}, \bibinfo{person}{In-Gyu Baek},
  \bibinfo{person}{Doowon Kwon}, \bibinfo{person}{Hongki Kim},
  \bibinfo{person}{Hyuksoon Choi}, \bibinfo{person}{Jonghyun Go},
  \bibinfo{person}{JungChak Ahn}, \bibinfo{person}{Jaekyu Lee},
  \bibinfo{person}{Changrok Moon}, \bibinfo{person}{Kyupil Lee}, {and}
  \bibinfo{person}{Hyoung-Sub Kim}.} \bibinfo{year}{2021}\natexlab{}.
\newblock \showarticletitle{A 2.6 e-rms Low-Random-Noise, 116.2 mW Low-Power
  2-Mp Global Shutter CMOS Image Sensor with Pixel-Level ADC and In-Pixel
  Memory}. In \bibinfo{booktitle}{\emph{2021 Symposium on VLSI Circuits}}.
  \bibinfo{pages}{1--2}.
\newblock


\bibitem[Shao and Brooks(2013)]%
        {shao2013energy}
\bibfield{author}{\bibinfo{person}{Yakun~Sophia Shao} {and}
  \bibinfo{person}{David Brooks}.} \bibinfo{year}{2013}\natexlab{}.
\newblock \showarticletitle{Energy characterization and instruction-level
  energy model of Intel's Xeon Phi processor}. In
  \bibinfo{booktitle}{\emph{International Symposium on Low Power Electronics
  and Design (ISLPED)}}. IEEE, \bibinfo{pages}{389--394}.
\newblock


\bibitem[Shao et~al\mbox{.}(2014)]%
        {shao2014aladdin}
\bibfield{author}{\bibinfo{person}{Yakun~Sophia Shao}, \bibinfo{person}{Brandon
  Reagen}, \bibinfo{person}{Gu-Yeon Wei}, {and} \bibinfo{person}{David
  Brooks}.} \bibinfo{year}{2014}\natexlab{}.
\newblock \showarticletitle{Aladdin: A pre-rtl, power-performance accelerator
  simulator enabling large design space exploration of customized
  architectures}. In \bibinfo{booktitle}{\emph{2014 ACM/IEEE 41st International
  Symposium on Computer Architecture (ISCA)}}. IEEE, \bibinfo{pages}{97--108}.
\newblock


\bibitem[Stillmaker and Baas(2017)]%
        {stillmaker2017scaling}
\bibfield{author}{\bibinfo{person}{Aaron Stillmaker} {and}
  \bibinfo{person}{Bevan Baas}.} \bibinfo{year}{2017}\natexlab{}.
\newblock \showarticletitle{Scaling equations for the accurate prediction of
  CMOS device performance from 180 nm to 7 nm}.
\newblock \bibinfo{journal}{\emph{Integration}}  \bibinfo{volume}{58}
  (\bibinfo{year}{2017}), \bibinfo{pages}{74--81}.
\newblock


\bibitem[Svensson and Wikner(2010)]%
        {svensson2010power}
\bibfield{author}{\bibinfo{person}{Christer Svensson} {and}
  \bibinfo{person}{J~Jacob Wikner}.} \bibinfo{year}{2010}\natexlab{}.
\newblock \showarticletitle{Power consumption of analog circuits: a tutorial}.
\newblock \bibinfo{journal}{\emph{Analog Integrated Circuits and Signal
  Processing}} \bibinfo{volume}{65}, \bibinfo{number}{2}
  (\bibinfo{year}{2010}), \bibinfo{pages}{171--184}.
\newblock


\bibitem[Theuwissen(2021)]%
        {theuwissen20211}
\bibfield{author}{\bibinfo{person}{Albert Theuwissen}.}
  \bibinfo{year}{2021}\natexlab{}.
\newblock \showarticletitle{1.4 There’s More to the Picture Than Meets the
  Eye*, and in the future it will only become more so}. In
  \bibinfo{booktitle}{\emph{2021 IEEE International Solid-State Circuits
  Conference (ISSCC)}}, Vol.~\bibinfo{volume}{64}. IEEE,
  \bibinfo{pages}{30--35}.
\newblock


\bibitem[Tsugawa et~al\mbox{.}(2017)]%
        {Tsugawa2017tsv}
\bibfield{author}{\bibinfo{person}{H. Tsugawa}, \bibinfo{person}{H. Takahashi},
  \bibinfo{person}{R. Nakamura}, \bibinfo{person}{T. Umebayashi},
  \bibinfo{person}{T. Ogita}, \bibinfo{person}{H. Okano}, \bibinfo{person}{K.
  Iwase}, \bibinfo{person}{H. Kawashima}, \bibinfo{person}{T. Yamasaki},
  \bibinfo{person}{D. Yoneyama}, \bibinfo{person}{J. Hashizume},
  \bibinfo{person}{T. Nakajima}, \bibinfo{person}{K. Murata},
  \bibinfo{person}{Y. Kanaishi}, \bibinfo{person}{K. Ikeda},
  \bibinfo{person}{K. Tatani}, \bibinfo{person}{T. Nagano}, \bibinfo{person}{H.
  Nakayama}, \bibinfo{person}{T. Haruta}, {and} \bibinfo{person}{T. Nomoto}.}
  \bibinfo{year}{2017}\natexlab{}.
\newblock \showarticletitle{Pixel/DRAM/logic 3-layer stacked CMOS image sensor
  technology}. In \bibinfo{booktitle}{\emph{2017 IEEE International Electron
  Devices Meeting (IEDM)}}. \bibinfo{pages}{3.2.1--3.2.4}.
\newblock


\bibitem[Whatmough et~al\mbox{.}(2019)]%
        {whatmough2019fixynn}
\bibfield{author}{\bibinfo{person}{Paul~N Whatmough}, \bibinfo{person}{Chuteng
  Zhou}, \bibinfo{person}{Patrick Hansen}, \bibinfo{person}{Shreyas~Kolala
  Venkataramanaiah}, \bibinfo{person}{Jae-sun Seo}, {and}
  \bibinfo{person}{Matthew Mattina}.} \bibinfo{year}{2019}\natexlab{}.
\newblock \showarticletitle{Fixynn: Efficient hardware for mobile computer
  vision via transfer learning}.
\newblock \bibinfo{journal}{\emph{arXiv preprint arXiv:1902.11128}}
  (\bibinfo{year}{2019}).
\newblock


\bibitem[Xie and Zhao(2015)]%
        {xie2015stacking}
\bibfield{author}{\bibinfo{person}{Yuan Xie} {and} \bibinfo{person}{Jishen
  Zhao}.} \bibinfo{year}{2015}\natexlab{}.
\newblock \showarticletitle{Die-stacking architecture}.
\newblock \bibinfo{journal}{\emph{Synthesis Lectures on Computer Architecture}}
  \bibinfo{volume}{10}, \bibinfo{number}{2} (\bibinfo{year}{2015}),
  \bibinfo{pages}{1--127}.
\newblock


\bibitem[Xu et~al\mbox{.}(2021)]%
        {xu2021senputing}
\bibfield{author}{\bibinfo{person}{Han Xu}, \bibinfo{person}{Ningchao Lin},
  \bibinfo{person}{Li Luo}, \bibinfo{person}{Qi Wei}, \bibinfo{person}{Runsheng
  Wang}, \bibinfo{person}{Cheng Zhuo}, \bibinfo{person}{Xunzhao Yin},
  \bibinfo{person}{Fei Qiao}, {and} \bibinfo{person}{Huazhong Yang}.}
  \bibinfo{year}{2021}\natexlab{}.
\newblock \showarticletitle{Senputing: An ultra-low-power always-on vision
  perception chip featuring the deep fusion of sensing and computing}.
\newblock \bibinfo{journal}{\emph{IEEE Transactions on Circuits and Systems I:
  Regular Papers}} \bibinfo{volume}{69}, \bibinfo{number}{1}
  (\bibinfo{year}{2021}), \bibinfo{pages}{232--243}.
\newblock


\bibitem[Yang et~al\mbox{.}(2015)]%
        {yang2015}
\bibfield{author}{\bibinfo{person}{Minhao Yang}, \bibinfo{person}{Shih-Chii
  Liu}, {and} \bibinfo{person}{Tobi Delbruck}.}
  \bibinfo{year}{2015}\natexlab{}.
\newblock \showarticletitle{A Dynamic Vision Sensor With 1\% Temporal Contrast
  Sensitivity and In-Pixel Asynchronous Delta Modulator for Event Encoding}.
\newblock \bibinfo{journal}{\emph{IEEE Journal of Solid-State Circuits}}
  \bibinfo{volume}{50}, \bibinfo{number}{9} (\bibinfo{year}{2015}),
  \bibinfo{pages}{2149--2160}.
\newblock


\bibitem[Young et~al\mbox{.}(2019)]%
        {young2019data}
\bibfield{author}{\bibinfo{person}{Christopher Young}, \bibinfo{person}{Alex
  Omid-Zohoor}, \bibinfo{person}{Pedram Lajevardi}, {and}
  \bibinfo{person}{Boris Murmann}.} \bibinfo{year}{2019}\natexlab{}.
\newblock \showarticletitle{A data-compressive 1.5/2.75-bit log-gradient QVGA
  image sensor with multi-scale readout for always-on object detection}.
\newblock \bibinfo{journal}{\emph{IEEE Journal of Solid-State Circuits}}
  \bibinfo{volume}{54}, \bibinfo{number}{11} (\bibinfo{year}{2019}),
  \bibinfo{pages}{2932--2946}.
\newblock


\bibitem[Yu and Wu(2018)]%
        {yu2018designing}
\bibfield{author}{\bibinfo{person}{Ying-Ju Yu} {and}
  \bibinfo{person}{Carole-Jean Wu}.} \bibinfo{year}{2018}\natexlab{}.
\newblock \showarticletitle{Designing a temperature model to understand the
  thermal challenges of portable computing platforms}. In
  \bibinfo{booktitle}{\emph{2018 17th IEEE Intersociety Conference on Thermal
  and Thermomechanical Phenomena in Electronic Systems (ITherm)}}. IEEE,
  \bibinfo{pages}{992--999}.
\newblock


\bibitem[Zhu et~al\mbox{.}(2018)]%
        {zhu2018euphrates}
\bibfield{author}{\bibinfo{person}{Yuhao Zhu}, \bibinfo{person}{Anand
  Samajdar}, \bibinfo{person}{Matthew Mattina}, {and} \bibinfo{person}{Paul
  Whatmough}.} \bibinfo{year}{2018}\natexlab{}.
\newblock \showarticletitle{Euphrates: Algorithm-SoC Co-Design for Low-Power
  Mobile Continuous Vision}. In \bibinfo{booktitle}{\emph{2018 ACM/IEEE 45th
  Annual International Symposium on Computer Architecture (ISCA)}}. IEEE
  Computer Society, \bibinfo{pages}{547--560}.
\newblock


\end{thebibliography}
